\documentclass{jfm}
\usepackage{graphicx}
\usepackage{dsfont}
\usepackage{bm}
\usepackage{amsmath}
\usepackage{comment} 
\usepackage{physics}
\usepackage{multirow}
\usepackage[normalem]{ulem}

\newcommand\redsout{\bgroup\markoverwith{\textcolor{red}{\rule[0.5ex]{2pt}{2.4pt}}}\ULon}

\newcommand{\stkout}[1]{\ifmmode\text{\redsout{\ensuremath{#1}}}\else\redsout{#1}\fi}

\usepackage{stmaryrd}
\usepackage{makecell}
\usepackage{subcaption}
\usepackage{xcolor} 
\usepackage{epstopdf, epsfig}
\usepackage{gensymb}

\shorttitle{Guidelines for authors}
\shortauthor{O. C. Marchand, S. Ramananarivo, C. Duprat, C. Josserand}

\title{Three-dimensional flow around and through a porous screen}

\author{Olivier C. Marchand
  \corresp{\email{olivier.marchand@ladhyx.polytechnique.fr}},
  Sophie Ramananarivo,
  Camille Duprat, 
  \& Christophe Josserand
 }

\affiliation{Laboratoire d'Hydrodynamique (LadHyX), CNRS, École polytechnique, Institut Polytechnique de Paris, 91120 Palaiseau, France}

\begin{document}

\maketitle

\begin{abstract}
We investigate the three-dimensional flow around and through a porous screen for various porosities at high Reynolds number $Re = \order{10^4}$. Historically, the study of this problem has been focused on two-dimensional cases and for a screen spanning completely or partially a channel. Since many recent problems have involved a porous object in a three-dimensional free flow, we present a three-dimensional model for porous screens initially based on \cite{Koo} and \cite{Steiros}, accounting for viscous effects in the vicinity of the screen, from which we can derive velocities, pressure distribution as well as aerodynamic forces. We characterize experimentally the aerodynamic drag coefficient for a porous square screen composed of fibers, immersed in a laminar air flow with different angles of attack. We test various fiber diameters to explore the effect of the space between the pores on the drag force. The drag prediction from the model is in good agreement with our experimental results.
Our theoretical and experimental results suggest that for high solidity, a homogeneous porous screens composed of fibers can have a higher drag coefficient than a flat plate with the same dimensions.
We also show that local viscous effects are important: at the same solidity and with the same air flow, the drag coefficient strongly depends on the Reynolds number based on the fiber diameter. 
The model, taking into account three-dimensional effects and the shape of the porous screen, may have many applications including the prediction of water collection efficiency for fog harvesters.
\end{abstract}


\section{Introduction}

The flow around porous structures has been largely investigated throughout the recent decades and has many engineering applications. It can be applied to parachute problems for the determination of drag and stability (\cite{Johari,Sarpkaya}), to vertical axis wind turbines (\cite{Ayati}) as well as to blockage correction in wind tunnel (\cite{SteirosTurbines}). \cite{Laws} highlighted the possibility of using screens in flow to control the velocity distribution and change the flow direction. The understanding and improvement of the water collection of fog harvesters in arid regions require a quantitative description of the flow in the vicinity of the net \citep{regalado_design_2016,moncuquet2022collecting}. Furthermore, such a quantitative description may provide a first step in the physical understanding of respiratory flows in the presence of a face mask, as used to reduce the propagation of airborne virus such as SARS-CoV-2 \citep{mittal2020flow,bourrianne2021quantifying}. In these cases, the flow can either pass through the porous net or mask, or is deviated around or through the leaks.
More generally, there may be an interest in reducing the constraints exerted on high panels or masts exposed to a flow for safety reasons, leading to an increased interest in large porous structures as mentioned by \cite{Giannoulis}. In some buildings, a permeable layer is added at a certain distance from the façade for energy efficiency reasons or to block a part of the sun rays. Also, windbreak panels are usually used in industry and power plant to control wind and dust pollution. In these cases an estimation of the cladding wind load is a useful information for architects and engineers (\cite{Pomaranzi}).
Furthermore, the modelling of the interaction of flow with arrays of fibers and the prediction of the corresponding drag can be helpful to understand the mechanism of filter feeding for numerous marine organisms for which arrays of bristles move in water to capture food particles (see \cite{Hood,Cheer}).

The main physical characteristics of the screen involved in flow resistance is the porosity and the permeability. For very thin porous screens, the porosity can be represented by the solidity which is the ratio between the solid surface area and the total surface area of the screen. The permeability is defined as the parameter relating the pressure gradient within a porous media to the local velocity of the flow, depending on the geometry of the pores.  
The porosity effects on pressure drop across porous screens and on drag force has largely been investigated, both theoretically and experimentally, whereas the permeability effect is much less understood, especially for very thin porous structures. However, some recent work at low Reynolds number has been conducted highlighting its influence: in particular, \cite{Ledda} has shown that the permeability has a strong effect on the wake characteristics, and \cite{Pezzula} has demonstrated that the drag coefficient of the screen depends on the permeability.
On the other hand, \cite{Steiros} have developed a model to predict the drag coefficient for two-dimensional perforated plates as a function of the solidity only. Although they obtained a good agreement with experimental data, this approach does not take explicitly into account the Reynolds number based on the scale of the pores, while it is known that the behavior of the flow in the vicinity of the screen depends on the specific geometry of the pores as well as the material and thickness of the screen (see section \ref{sec:pressure-Re}).

Several approaches have been adopted to model the flow. For instance \cite{Carvajal} used three-dimensional numerical simulations to access the aerodynamic characteristics during fog collection and model the net as a porous medium using Darcy's law, whereas \cite{Rivera} used the superposition principle applied to a flow passing around a solid plate and a flow forced to pass through the net to find an approximation of the velocity at the screen. The difficulty of the problem lies in the multiscale physical phenomena, from the characteristic scale of the flow around the screen that is of order  $0.1$ to $1\hspace{0.1cm}m$ to that of the flow through the screen (i.e. the pore size) which is of order $\order{10^{-3}}$ to $\order{10^{-6}}\hspace{0.1cm}m$; the scale of the Reynolds number thus varies from low Reynolds numbers at the pore scale (of order $\order{10^{-1}}$) to large ones at the net scale (of order $\order{10^{6}}$). Therefore the physical mechanisms of the whole system cannot be easily captured by a numerical simulation resolving all scales as noticed by \cite{Shklyar}, and the method generally used consists in modeling the porous surface as an imaginary interface where transfers of mass and momentum occur. The macroscopic jump laws for the velocities and pressure at the interface are deduced from a microscopic model at the vicinity of the screen where the fluid is generally governed by the steady Stokes equations. This can be obtained by periodic homogenization theory and has been recently used for porous surface by \cite{Ledda2}. However, these methods introduce some parameters like the permeability of the porous surface that are difficult to measure experimentally, although they could be obtained using pore-scale simulation. Our motivation is to obtain a model of the flow for porous surfaces based only on the porosity, the large-scale geometry of the screen and the Reynolds number at the scale of the holes (instead of the permeability), which are easy to access.

We also aim at predicting the drag coefficient of elevated porous panels of arbitrary shape placed in a laminar flow. We focus on porous screens composed of fibers, but the model can be applied without major changes to other kind of porous screens. Experimentally, we consider rectangular meshes of woven fibers, with fiber diameters between $6.0\hspace{0.1cm}\mu m$ and $1.9\hspace{0.1cm}mm$ and with typical pore sizes of the order of $10\hspace{0.1cm}\mu m$ to $1.0\hspace{0.1cm}cm$, placed in a uniform laminar flow of velocity varying from $0.5$ to $13\hspace{0.1cm}m.s^{-1}$ with three orientation angles. We measure the drag coefficient and perform Particle image velocimetry measurements of the flow around the screen.
Theoretically, we focus on the model first proposed by \cite{Taylor} and used by numerous authors (\cite{Neill}), which consists in considering the screen as a distribution of sources. This approach has been adopted by \cite{Koo} who proposed a two-dimensional mathematical model for a screen confined in a channel with two parallel boundaries. Recently, \cite{Steiros} derived the drag coefficient of a porous plate based partially on \cite{Koo} and 2D potential flow. Since their prediction showed a good agreement with their experimental data, here we propose to extend the model of \cite{Koo} to a three-dimensional free flow, keeping the same main hypothesis but taking into account the shape of the screen, three-dimensional effects, the base suction effect considered by \cite{Steiros}, as well as the viscous effects, i.e. finite pore Reynolds number. The model assumes a steady wake, and thus is not applicable in the presence of vortex shedding. We then derive the equations to predict the drag coefficient following the method proposed by \cite{Steiros}, and discuss the limits of this approach. First, the pressure jump used is based on clear physical assumptions derived following the method of \cite{Taylor} and \cite{Steiros}. Second, in order to take into account the viscous effects observed experimentally, we also incorporate in the model the empirical law of \cite{Brundrett} for the pressure jump. We finally compare the prediction of the drag coefficient for various porosities and Reynolds number based on the fibers' diameter to our experimental results (both with the theoretical and empirical law of the pressure jump). As far as we know, only \cite{Letchford} and \cite{Prandtl1932} performed measurements on the drag coefficient for elevated panels with different angles of attack. We therefore also study the flow for arbitrary angle of attacks and compare our theoretical prediction with experimental data for two angles of attack ($65^\circ$ and $43^\circ$). Finally, a measurement of the proportion of the flow that is deviated is presented for different solidities.

\begin{figure}
  \centerline{\includegraphics[width=12cm, trim = 0cm 0cm 0cm 0cm, clip]{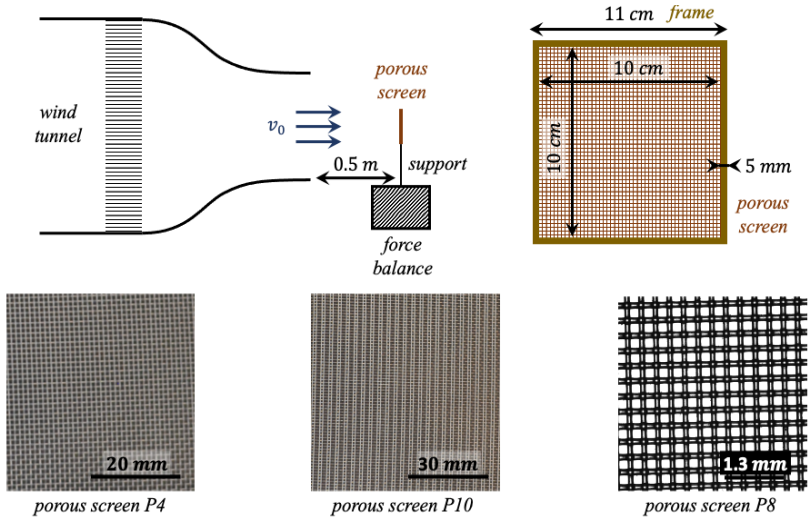}}
  \caption{Diagram of the the experimental set-up for the drag coefficient measurement, and examples of porous screens essentially in nylon woven mesh.}
\label{fig:set-up}
\end{figure}

\section{Experiments}\label{sec:types_paper}

We measure experimentally the drag coefficient for a series of porous planar structures consisting of regularly woven nylon yarns in a square mesh and other types of meshes like rod screens (parallel fibers), of size $L=10\hspace{0.1cm}cm\times10\hspace{0.1cm}cm$. The characteristics of the porous screens as well of their solidity are detailed in the appendix \ref{appC}, and we also present a synthesis in table \ref{tab:screens-article}. The solidity $s$ is defined as the ratio between the solid surface area of the screen to its total area $S_p$. We also characterize the drag coefficient for a classical surgical facemask for which the physical characteristics such as the fiber diameter and the solidity are taken from \cite{Monjezi} and \cite{Du} (screen P9 in table \ref{tab:screens-article}).  The porous structure is held in a planar configuration by a square frame with a width of $0.5\hspace{0.1cm}cm$ (that is $2.1\hspace{0.1cm}\%$ of the surface area of the porous structure), representing a small portion ($7.6\hspace{0.1cm}\%$) of the total cross-section of the laminar flow generated by the wind tunnel. Following \cite{Letchford}, the square frame is fixed on a $21.5\hspace{0.1cm}cm$ high mast to avoid boundary layer effects. The set-up is shown in figure \ref{fig:set-up}.

\renewcommand{\arraystretch}{1.2}
\begin{table}
  \begin{center}
\def~{\hphantom{0}}
  \begin{tabular}{cccc|cccc}
      \makecell{Screen \\ number}  & \makecell{Reynolds \\ number \\ $Re_d$} & Solidity $s$ & \makecell{Drag \\ coefficient \\ $C_D$} & \makecell{Screen \\ number}  & \makecell{Reynolds \\ number \\ $Re_d$} & Solidity $s$ & \makecell{Drag \\ coefficient \\ $C_D$} \\[3pt]
       P1 & $186$ & $0.58$ & $0.872$ & P17 & $173$ & $0.15$ & $0.198$ \\
       P2 & $705$ & $0.41$ & $0.566$ & P18 & $173$ & $0.28$ & $0.358$\\
       P3 & $1218$ & $0.87$ & $0.957$ & P19 & $167$ & $0.52$ & $0.803$ \\
       P4 & $167$ & $0.61$ & $0.922$ & P20 & $167$ & $0.42$ & $0.632$\\
       P5 & $64$ & $0.56$ & $0.941$ & P21 & $167$ & $0.32$ & $0.406$\\
       P6 & $115$ & $0.61$ & $0.935$ &  P22 & $32$ & $0.65$ & $0.956$\\
       P7 & $173$ & $0.45$ & $0.705$ & P23 & $16$ & $0.82$ & $0.951$\\
       P8 & $83$ & $0.70$ & $0.986$ &  P24 & $19$ & $0.75$ & $0.960$ \\
       P9 & \makecell{$0.6 - 12$ \\ mean $4$} & $0.26$ & $0.976$ & P25 & $24$ & $0.70$ & $0.985$\\
       P10 & $173$ & $0.11$ & $0.146$ & P26 & $282$ & $0.405$ & $0.560$\\
       P11 & $173$ & $0.37$ & $0.596$ & P27 & $38$ & $0.115$ & $0.185$\\
       P12 & $173$ & $0.31$ & $0.453$ & P28 & $141$ & $0.114$ & $0.141$\\
       P13 & $173$ & $0.17$ & $0.210$ & P29 & $6$ & $0.080$ & $0.281$\\
       P14 & $173$ & $0.24$ & $0.328$ & P30 & $282$ & $0.080$ & $0.082$\\
       P15 & $173$ & $0.24$ & $0.336$ & P31 & - & $1.00$ & $0.939$\\
       P16 & $173$ & $0.24$ & $0.343$ &  &  &  & \\
  \end{tabular}
 \caption{Porous screen characteristics. The Reynolds number $Re_d$ is calculated with the fiber diameter $d$ as characteristic size and with a velocity $v_0 = 10 \hspace{0.1cm}m.s^{-1}$ and a kinematic viscosity $\nu = 15.6 \times 10^{-6}\hspace{0.1cm}m^2.s^{-1}$, except for the screens P27, P28 and P29 where the velocity is $v_0 = 5 \hspace{0.1cm}m.s^{-1}$. The uncertainties as well as the fiber diameters, material type and geometry can be found in the appendix \ref{appC}.}
  \label{tab:screens-article}
  \end{center}
\end{table}

\begin{figure}
     \centering
     \hspace{-0.5cm}
     \begin{subfigure}[b]{0.45\textwidth}
         \centering
         \includegraphics[width=6.3cm,trim = 0cm 0cm 0cm 0cm, clip]{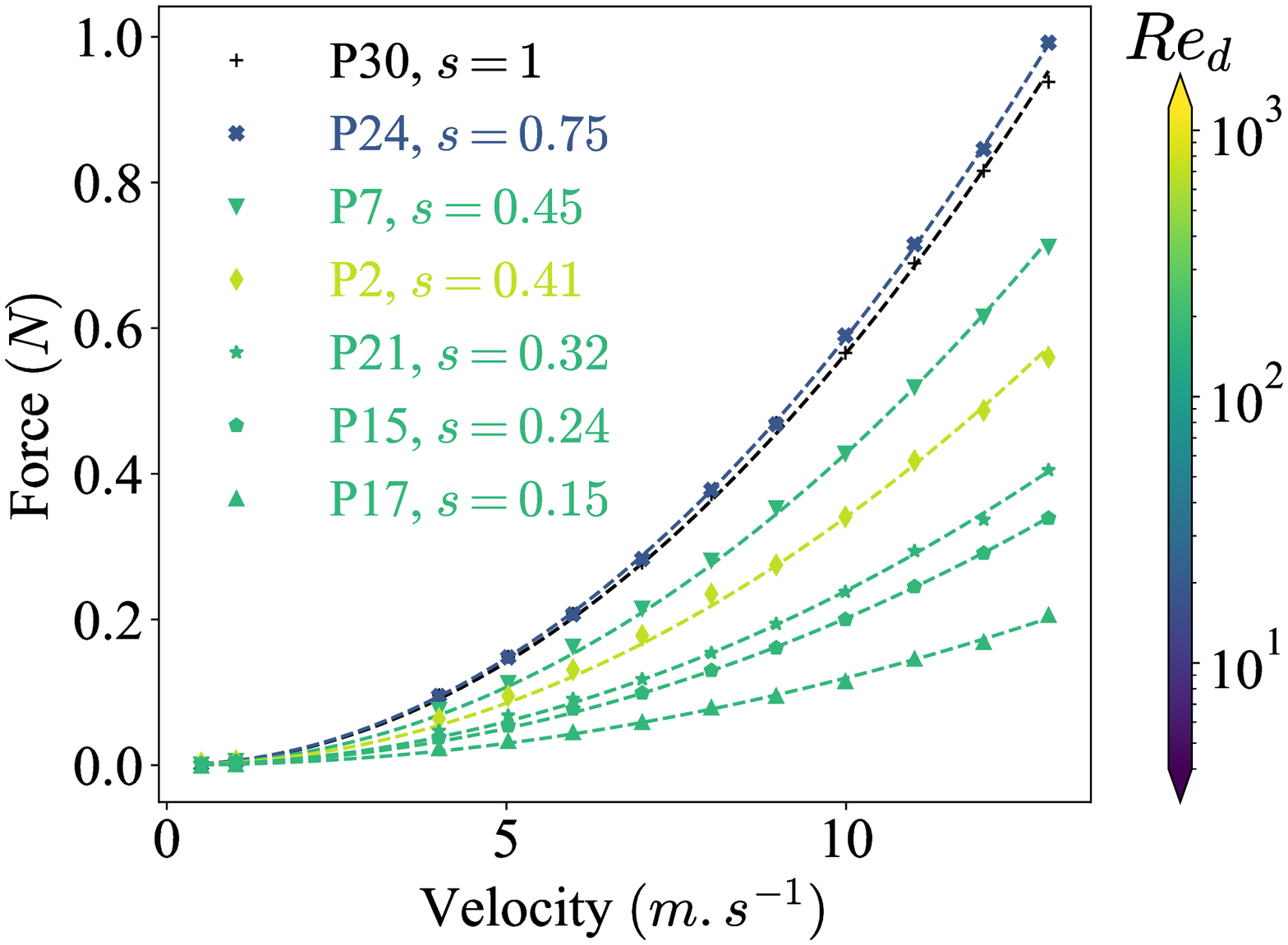}
         \caption{Drag force for different normal porous screens.}
         \label{fig:force-porous}
     \end{subfigure}
     \hspace{0.4cm}
     \begin{subfigure}[b]{0.45\textwidth}
         \centering
         \includegraphics[width=6.3cm,trim = 0cm 0cm 0cm 0cm, clip]{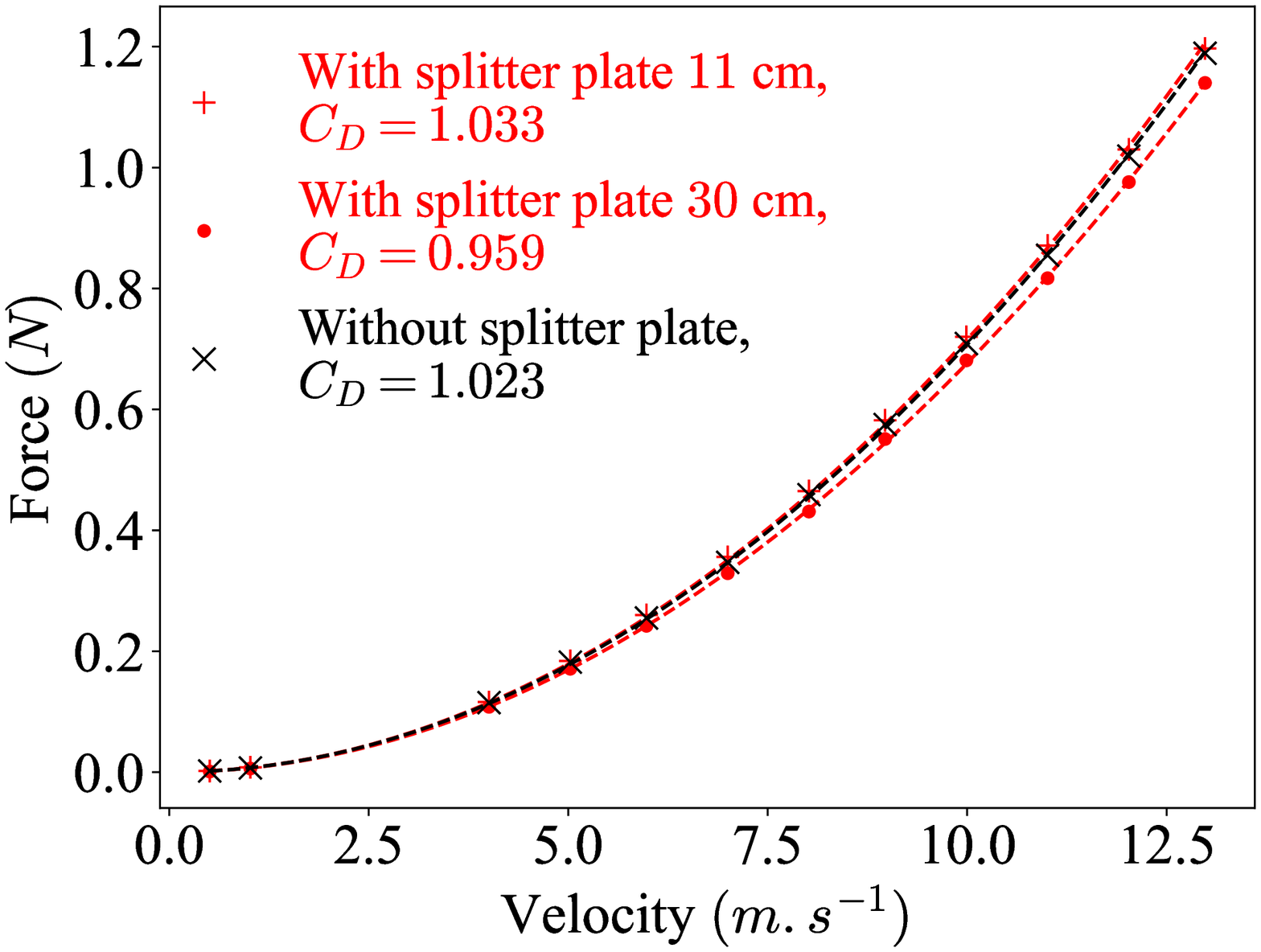}
         \caption{Drag force with an without a splitter plate}
         \label{fig:force-shedding}
     \end{subfigure}
     \caption{Drag force $(F_D)$ measured for different square porous screens at normal incidence (a). Comparison of the drag force with and without a splitter plate for a solid plate (without frame support) of dimensions $11.0\times 11.0\hspace{0.1cm}cm^2$ and thickness $3\hspace{0.1cm}mm$ (b). The fitting curves are obtained using a quadratic law.}
     \label{fig:force-drag}
\end{figure}

A force balance (SIXAXES, FX2.6, N°1026, $\pm 5\hspace{0.1cm}N$, sensitivity of $\pm 0.001\hspace{0.1cm}N$) is used to measure the force applied to the whole system. The laminar airflow is generated by an open jet wind tunnel with a square test section of width $40\hspace{0.1cm}cm$. To compute the drag coefficient at normal incidence, $12$ different velocities have been used as shown in figure \ref{fig:force-porous} where we plotted the force as a function of velocity for several screens; the data are fitted with a quadratic law to obtain the drag coefficient. We also perform experiments on inclined porous screens as will be detailed in section \ref{sec:results} ; in that case $8$ different velocities have been used. For both cases, the velocities vary from $0.5$ to $13\hspace{0.1cm}m.s^{-1}$. We can thus define the (global) Reynolds number of the problem as
\begin{equation}\label{eq:Re}
{\rm Re}=\frac{L_0v_0}{\nu},
\end{equation}
where $v_0$ is the uniform velocity of the flow far upstream of the screen, $L_0$ its typical size and $\nu$ the kinematic viscosity of the fluid (the air for the configurations considered here, so $\nu = 15.6 \times 10^{-6} \hspace{0.1cm}m^2.s^{-1}$). For a screen of size of few tenth of centimeters with a velocity of the order of one meter per second, we obtain ${\rm Re} \approx \order{10^4} \gg 1$. In addition we also define a local Reynolds number for the flow around each fiber denoted $Re_d$, for each screen.
This Reynolds number $Re_d$ is calculated with the diameter $d$, following:
\begin{equation}\label{eq-Red}
\begin{aligned}
   Re_d = \frac{v_0 d}{\nu}.
\end{aligned}
\end{equation}

We perform measurements with screens made of different fiber radii and pore sizes while keeping the solidity almost constant, which allows to probe the local effect of viscosity with $Re_d$ ranging from 0.6 to 1218. The details of the wind tunnel as well as the characteristics of the flow can be found in the thesis of Du \cite{Pontavice}. The system is placed at a distance of approximately $50\hspace{0.1cm}cm$ from the outlet of the wind tunnel in order to obtain a free flow.

Temperature and humidity were taken either from meteorological data of the site or local instruments placed upstream to reduce some of the uncertainty in the air density value. All the uncertainties estimates can be found in the Appendices.

\begin{figure}
  \centerline{\includegraphics[width=14cm, trim = 0cm 0cm 0cm 0cm, clip]{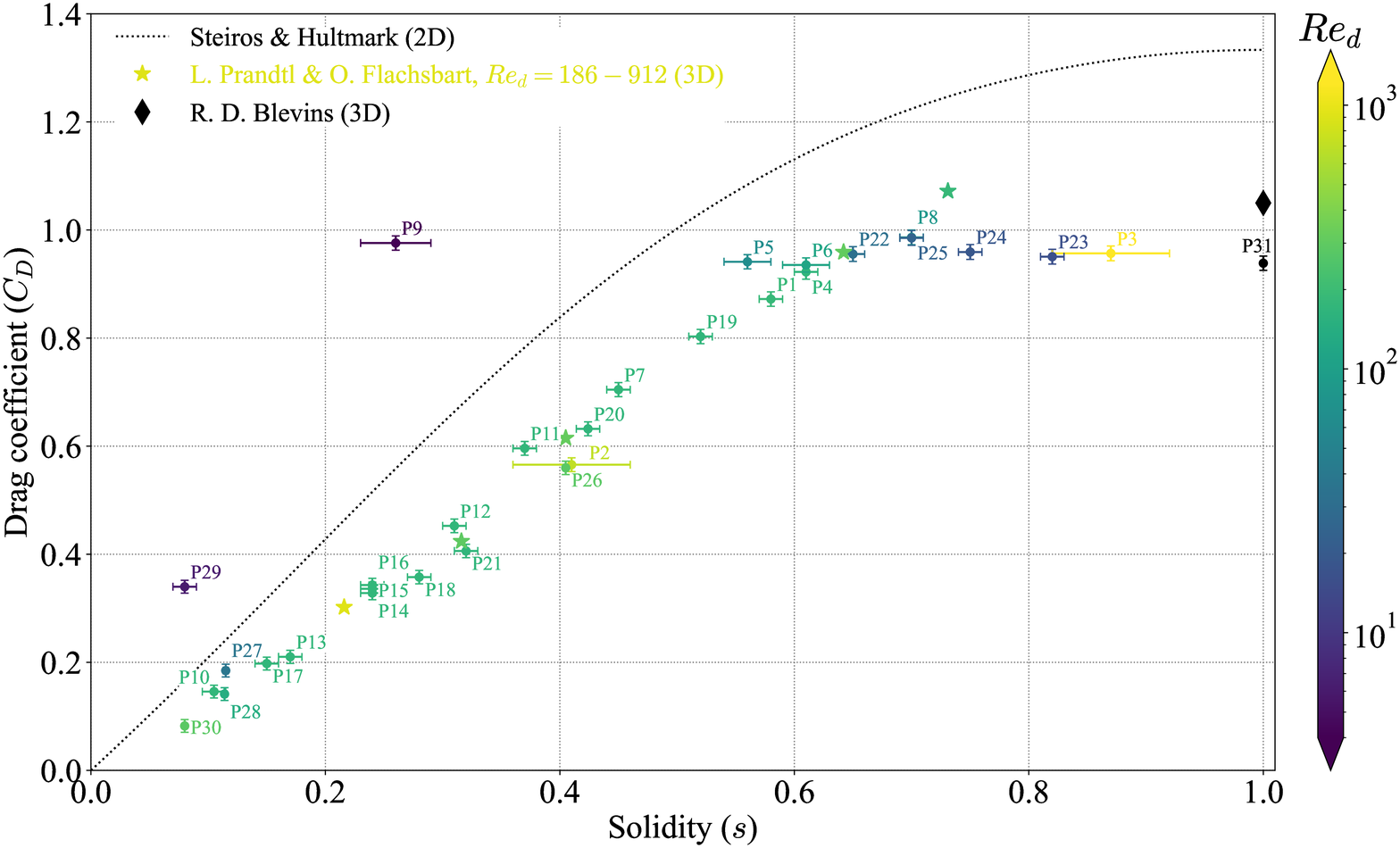}}
  \caption{Drag coefficient as a function the solidity for various square porous screens normal to the free flow and comparison with the two dimensional model of \cite{Steiros}. The color of the points indicates the value of the local Reynolds number following the color scale on the right. The data from \cite{Prandtl1932} and the value at solidity $s=1$ from \cite{Blevins} are also plotted.}
\label{fig:drag-solidity-exp}
\end{figure}

Figure \ref{fig:force-porous} shows the measured drag force $F_D$ as a function of the velocity $v_0$ for different screens at different Reynolds number $Re_d$, after subtraction of the contribution of the frame and the mast detailed in appendix \ref{appC}. The drag force is, as expected, proportional to the square of the fluid velocity $v_0$ upstream from the screen. 
Note that the vortex shedding occurring within a large range of Reynolds number can increase the drag significantly and reduce the base pressure, as seen in drag measurements in 2D by \cite{Steiros}. To evaluate the influence of the vortex shedding on our measurements, we used two splitter plates of different length located in the wake as done by \cite{Steiros}. Following \cite{Apelt}, for Reynolds number in the range $10^4<Re<5\times 10^4$, which corresponds to our case, the use of a splitter plate 3 times longer than the plate already suppresses the vortex shedding. In our experiments the splitter plate has the same height as the solid plate on which drag is measured but with length $11\hspace{0.1cm}cm$ and $30\hspace{0.1cm}cm$ and thickness of $3\hspace{0.1cm}mm$.
Figure \ref{fig:force-shedding} shows the measured drag force for a solid plate (solidity equal to $1$, square of $11.0\times 11.0 \hspace{0.1cm}cm^2$, $3\hspace{0.1cm}mm$ thickness) with and without a splitter plate. We see a reduction of the drag coefficient of $0.07$ with a splitter plate three times longer than the plate. This difference is expected to be less and less significant with decreasing solidity. We conclude that the vortex shedding has little influence on our experimental results.
From the force curves, we can then deduce a drag coefficient $C_D$ defined as
\begin{equation}
    C_D=\frac{F_D}{\frac12 \rho v_0^2 S_p},
    \label{CD_1}
\end{equation}
where $\rho$ is the fluid (air) density and $S_p$ the surface area of the screen. We note that here the surface $S_p$ is the total surface area of the screen and not its projected area along the far field stream direction $\bm{v_0}$.
Figure \ref{fig:drag-solidity-exp} shows the drag coefficient of the square screens as function of the solidity $s$.
The drag coefficient increases with increasing solidity, until it reaches a constant value $C_D\simeq 1$ at high solidities (for $s\leq 0.7$). Furthermore, we observe that for a given solidity, the drag coefficient increases for decreasing $Re_d$; this effect is particularly important at small solidities. The evolution of the drag coefficient is qualitatively consistent with previous experiments in particular \cite{Prandtl1932}, as well as with the 2D model derived by \cite{Steiros}, although the dependance on $Re_d$ is not considered there. Moreover, this model overestimates the drag coefficient obtained experimentally at moderately high $Re_d$ ($\sim 10^2$), in particular at high solidities, which highlights the importance to take into account 3D effects. The model also  underestimates the drag coefficient at low $Re_d$ where viscous effects must be included.  In the following, we aim at developing a model that takes into account both 3D and viscous effects. We then use this model to describe the flow around inclined rectangular screens and predict the corresponding drag coefficient.

To characterize the influence of the solidity on the flow deviation, we measure the flow field with a \textit{Particle image velocimetry} (PIV) method. We use a wind tunnel with square test section of width $22\hspace{0.1cm}cm$ at constant velocity $2.84 \pm 0.02\hspace{0.1cm}m.s^{-1}$. The fluid is seeded with micro-droplets of water of diameter $3.0 \pm 2.0\hspace{0.1cm}\mu m$. The Stokes number is defined as 
\begin{equation}
St=\frac{1}{18} \frac{\rho_p d_p^2 v_0}{\mu L},
\label{St}
\end{equation}
where $\rho_p$ and $v_p$ are the particles density and diameter respectively and $\mu= \rho \nu$ is the viscosity of the fluid. The Stokes number is of the order of $10^{-3} \ll 1$ and we can consider that the water droplets act as passive tracers of the flow. A $1\hspace{0.1cm}mm$-thick laser sheet is used to highlight the particles in a plane parallel to the flow. The laser (Elforlight LTD. model FCHPG-3000) has a wavelength $\lambda = 532\hspace{0.1cm}nm$ and maximum power of $6.0\hspace{0.1cm}W$.
A high speed camera PHOTRON was used at a frame rate of $4000\hspace{0.1cm}fps$ to record successive images that have been analysed with PIVlab in MATLAB (version 2.62).

\begin{figure}
     \centering
     \hspace{-0.5cm}
     \begin{subfigure}[b]{0.3\textwidth}
         \centering
         \includegraphics[width=4cm,trim = 0cm 0cm 0cm 1.5cm, clip]{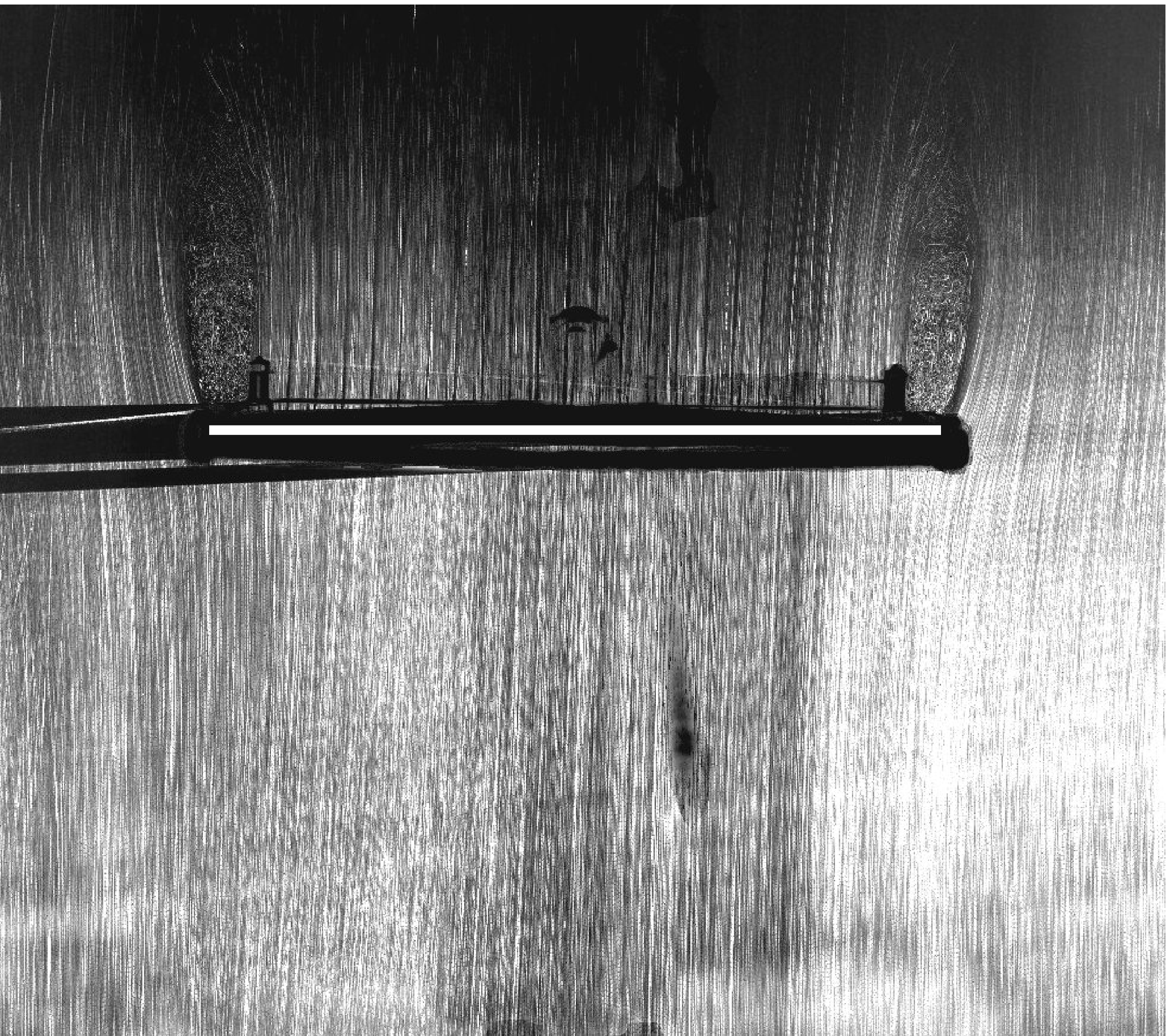}
         \caption{P14, $s=0.24$.}
         \label{fig:P14-streamline}
     \end{subfigure}
     \hspace{0.4cm}
     \begin{subfigure}[b]{0.3\textwidth}
         \centering
         \includegraphics[width=4cm,trim = 0cm 0cm 0cm 0cm, clip]{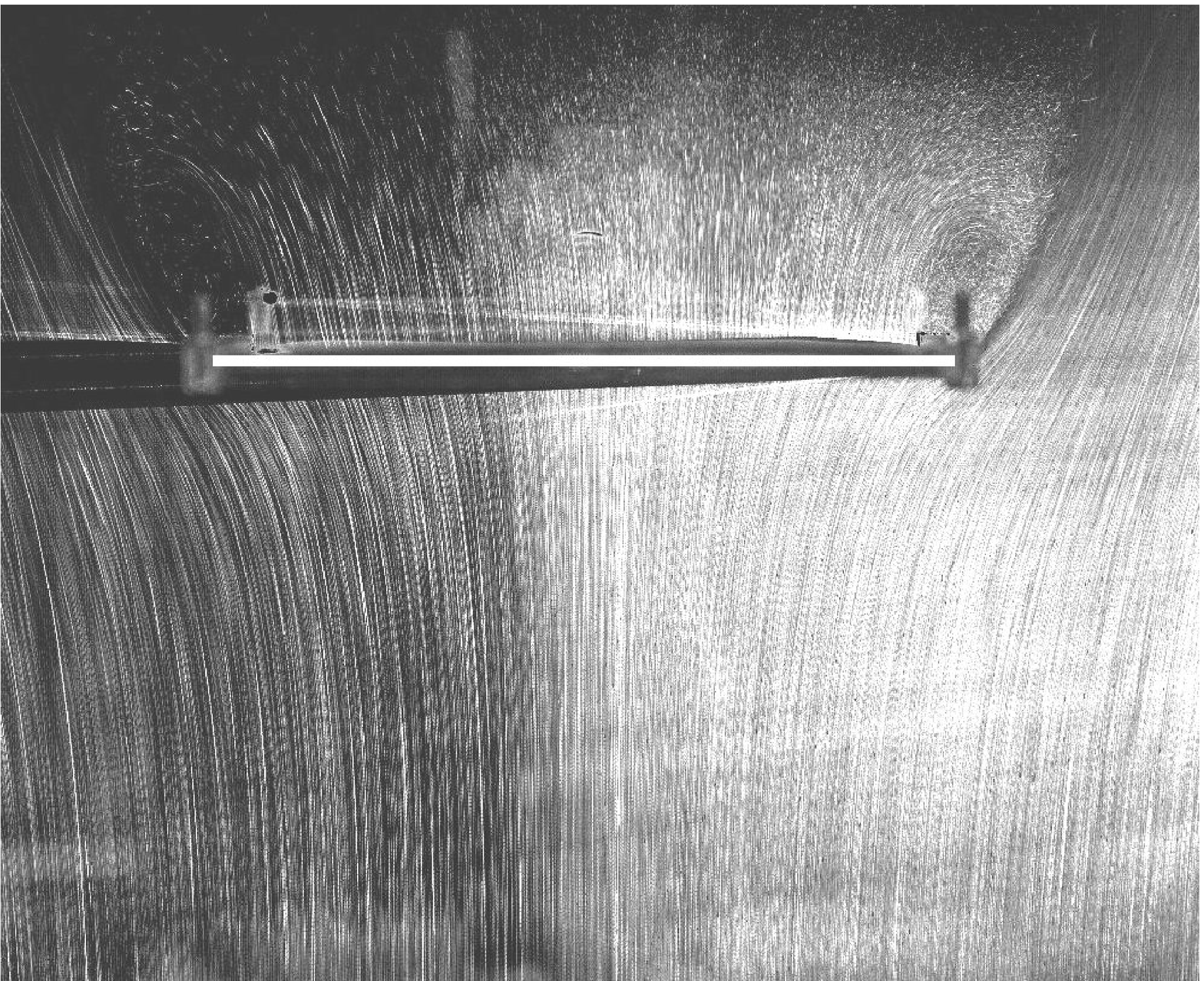}
         \caption{P6, $s=0.61$.}
         \label{fig:P6-streamline}
     \end{subfigure}
     \hspace{0.4cm}
     \begin{subfigure}[b]{0.3\textwidth}
         \centering
         \includegraphics[width=4cm,trim = 0cm 0cm 0.2cm 0.2cm, clip]{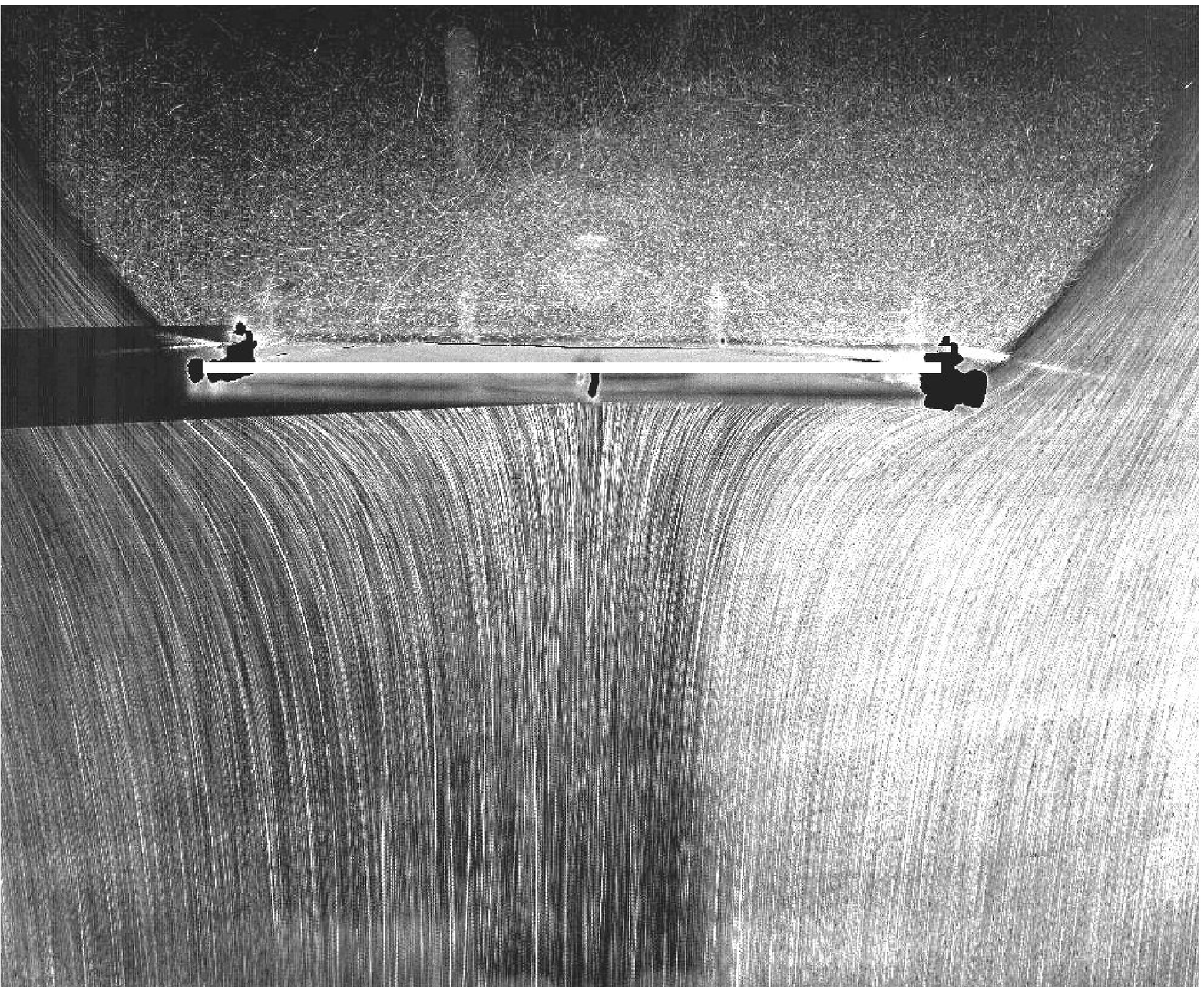}
         \caption{P23, $s=0.82$.}
         \label{fig:P26-streamline}
     \end{subfigure}
     \caption{Experimental evidence of the deviation of the flow and streamlines around a square porous screen at different solidities $s$ under normal uniform upstream flow $v_0 = 2.84\pm 0.02\hspace{0.1cm}m.s^{-1}$. Top view of the screen. The white bar ($110\hspace{0.1cm}mm$) corresponds to the screen embedded in the frame, the screen alone is $100\hspace{0.1cm}mm$ long.}
     \label{fig:streamline-exp}
\end{figure}

In figure \ref{fig:streamline-exp} we show the trajectories of the particles in a plane orthogonal to the screen at mid-height (i.e. at the middle of the screen), obtained by the superposition of the maximum intensity of $2000$ successive images. The upstream region is at the bottom of the figures and the downstream region at the top. We observe that the flow deviation around the screen increases as the solidity increases. We can further observe that the velocity decreases as the solidity increases, as shown by the variation of the length of the bright lines, shorter in figure \ref{fig:P6-streamline} than in \ref{fig:P14-streamline}. Furthermore, in figure \ref{fig:P26-streamline} we barely observe any particles crossing the screen, while some particles appear to be mixed by the recirculation in the wake.

\newpage

\section{Model}\label{sec:rules_submission}
In this section, we derive the equations describing the three-dimensional flow around and through a porous screen with arbitrary shape and solidity. In a second step, we take into account the small scale Reynolds number $Re_{d}$ (defined above in equation \ref{eq-Red}), using an empirical law for the pressure jump across the screen. First, we adopt the method used by \cite{Koo} which showed a good agreement with experimental results for 2D flow in a channel except at high solidity. This discrepancy at high solidity may come from the lack of base pressure and vortex shedding in their theory, as suggested by \cite{Steiros} (see section \ref{sec:pressure} and \ref{sec:drag}). We extend the model to the 3D case for free flow, i.e. a case where there are no boundaries constraining the flow, which is one of the major differences with the model of \cite{Koo}. We also take into account the effect of the base pressure. In section \ref{sec:pressure-Re} we incorporate local viscous effects by using an empirical formulation for the pressure jump that depends on the local Reynolds number $Re_d$, as first proposed by \cite{Brundrett}. We first obtain a general formulation of the equations for a porous screen of arbitrary shape. We then apply our equations to the case of a rectangular plate inclined in a laminar flow for which an analytical solution can be found.

\subsection{General formulation}

\begin{figure}
  \centerline{\includegraphics[width=13cm, trim = 0cm 0cm 0cm 0cm, clip]{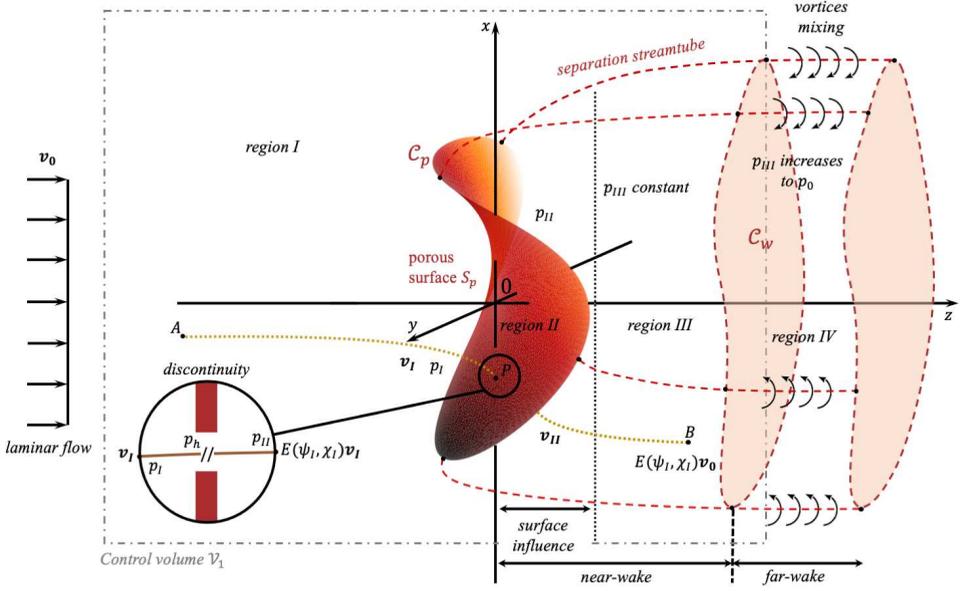}}
  \caption{Diagram of the model for a three-dimensional potential flow around and through a porous screen. The dashed lines are the separation streamlines used as a boundary between the regions. The dotted lines are the streamlines used in the model to calculate the velocities in the regions I and II. The incoming flow is laminar and is extended over the entire height of the system. $\mathcal{C}_w$ denotes the section of the wake.}
\label{fig:model}
\end{figure}

The flow around bluff bodies is complex. In order to obtain analytical or semi-analytical description of the flow, a widely used approach consists in simplifying the governing equations using potential flow theory outside the wake while introducing free parameters such as base pressure to account for viscous and complex phenomena in the wake and near the solid structure \citep{Parkinson}. The model proceeds with the same idea. The system is separated into four regions delimiting four flow regimes, as shown in figure \ref{fig:model}. The flow is assumed to be stationary, incompressible and inviscid everywhere except through the porous structure where viscous effects cannot be neglected.

In region I, we assume that the flow is potential and the velocity is denoted $\mathbf{v}_I(x,y,z)$. The region I is located upstream of the structure as well as downstream outside the wake zone contained by a well defined streamtube attached to the contour of the porous surface as shown in figure \ref{fig:model}. Therefore, the velocity derives from a velocity potential denoted $\phi_I(x,y,z)$ for the region I. Using the method employed by \cite{Koo}, and initially suggested by \cite{Taylor}, we calculate the flow by modelling the screen with a continuous source distribution with strength $\Omega(x_s,y_s,z_s)$ where $(x_s,y_s,z_s)$ denotes a point on the surface. We obtain the potential flow in region I by superposing the resulting potential flow from the distribution of sources with the uniform laminar flow $\mathbf{v}_0$. The two streamfunctions $\psi_I$ and $\chi_I$ needed to describe general three-dimensional incompressible flows, can then be deduced using the following relation (for a definition of three-dimensional stream functions see \cite{Yih}):
\begin{equation}\label{eq0}
   \mathbf{v}_I(x,y,z) = \nabla \psi_I(x,y,z) \wedge \nabla \chi_I(x,y,z).
\end{equation}

The regions II and III are located downstream of the porous surface in the near-wake. In these regions the flow can be rotational so that we cannot use anymore a velocity potential to describe the flow. In region II, the pressure and the velocity are not constant since they are influenced by the surface. However, in region III the flow is sufficiently far from the screen so that the streamlines tend to be aligned with the uniform flow $\mathbf{v}_0$ as represented by the contour $\mathcal{C}_w$ in figure \ref{fig:model}. Therefore the pressure tends towards a base suction pressure $p_{III}$ which is {\it a priori} different (and lower) than the constant external pressure $p_0$. This region is mathematically at infinity (there are no finite separation between region II and III), however since the flow aligns rapidly with the uniform flow, we indicate a region III in the near wake in figure \ref{fig:model}. In two dimensions, the approach of constant pressure along separating streamlines has been successfully used in free-streamline theory by \cite{Wu}, \cite{Parkinson} and \cite{Roshko} to model the wake. In this model, since we consider three dimensions we can not adopt free-streamline theory, but $p_{III}$ can be considered as having the same role as the constant pressure used in such theory. The flow in region II and III is found with matching conditions as explained later. 

In figure \ref{fig:model}, we added a region IV which is located in the far wake where the mixing with the outer flow can not be ignored. In this region, the pressure should increase to reach again the pressure $p_0$ outside the wake. We assume that this region has little influence on the flow near the porous screen and on the aerodynamic forces, and therefore it is not included in the model. Consequently, in our model the pressure in the far wake will remain equal to $p_{III}$.

\subsection{Determination of the flow in region I}

In region I, the flow is potential, and we derive the velocity from the velocity potential. For simplicity, we set the reference frame so that the axis $(Oz)$ is aligned with the velocity $\mathbf{v}_0$ without loss of generality. Due to the linearity of the Laplacian, we first consider the potential flow $\phi(x,y,z)$ for the source distribution only, then we add the potential flow for the uniform flow. The velocity potential from the source distribution $\Omega$ located on a general regular surface $\mathcal{S}_p$ is the solution of the following equation
\begin{equation}\label{eq1}
    \Delta \phi(x,y,z) = \Omega(x,y,z)\mathds{1}_{\mathcal{S}_p},
\end{equation}
where $\mathds{1}_{\mathcal{S}_p}$ denotes the dirac function associated to the surface $\mathcal{S}_p$.
For a point source in three-dimensions centered at the origin, the Green function of the Laplacian is
\begin{equation}\label{eq2}
    \Gamma(x,y,z) = - \frac{1}{4\pi}\frac{1}{\sqrt{x^2+y^2+z^2}}.
\end{equation}
Therefore, if we assume $\bm{\xi}: U \subset \mathbb{R}^2\to\mathbb{R}^3$ to be a surface patch of a general regular surface $\mathcal{S}_p$ with coordinates 
\begin{equation}\label{eq3}
\bm{\xi} = 
\begin{pmatrix} 
      x_s(u,v)\\ 
      y_s(u,v)\\
      z_s(u,v)
\end{pmatrix},
\end{equation}
parametrized by two parameters $u$ and $v$, with $(u,v)\in \left(U = \left[a,b\right]\times\left[c,d\right]\right)$ with $(a,b,c,d)\in \mathbb{R}^4$, then, the velocity potential $\phi(x,y,z)$ is expressed for all $(x,y,z)\in \mathbb{R}^3\setminus \mathcal{S}_p$ as (\cite{Pressley})

\begin{equation}\label{eq4}
    \phi(x,y,z) = \iint_{U} \Omega(u,v)\Gamma(x-x_s(u,v),y-y_s(u,v), z-z_s(u,v))\left\| \frac{\partial \bm \xi}{\partial u} \wedge \frac{\partial \bm \xi}{\partial v}\right\| \mathrm{d}u \mathrm{d}v,
\end{equation}
and the total velocity potential can be written as the following sum
\begin{equation}\label{eq5}
    \phi_I(x,y,z) = {v_0}_z z + \phi(x,y,z)
\end{equation}

We deduce the velocity in region I with $\mathbf{v_I}(x,y,z)=\mathbf{grad}(\phi_I(x,y,z))$. Note that the Green function can be changed without other modifications in the model to study the situation of a flow in a confined environment or near a wall.

\subsection{Determination of the flow in region II and III}

In region II, the flow can be rotational and therefore is not necessarily potential. The flow is obtained from the stream functions by considering, as done by \cite{Koo}, that the streamlines in region II have the same pattern as if they were obtained by the stream functions from the superposition of the distribution of sources and the uniform flow $\mathbf{v}_0$. 
This can be formulated in a general way by writing the two stream functions for the flow in region II as functions of the stream functions of region I. Let $\psi_{II}$ and $\chi_{II}$ be the stream functions in the region II. As defined above, $\psi_I$ and $\chi_{I}$ are the stream functions deduced from the flow in region I, functions that can be considered in the whole space. Then, without loss of generality, we choose an analog formulation as the one proposed by \cite{Koo} to describe the flow in region II, using the functions $f$ and $g$:
\begin{equation}\label{eqSF}
\psi_{II}=f(\psi_I)\psi_I \;\; {\rm and} \;\; \chi_{II}=g(\chi_I)\chi_I .
\end{equation}
This means that the velocity $\mathbf{v}_{II}$ in region II, and the velocity $\mathbf{v}_{I}$ obtained from the velocity potential $\phi_I(x,y,z)$, are co-linear at any point. We indeed obtain
\begin{equation}\label{eq12}
    {\mathbf{v}_{II}}(x,y,z) = \left(\frac{\mathrm{d}f}{\mathrm{d}\psi_I}+f(\psi_I)\right)\left(\frac{\mathrm{d}g}{\mathrm{d}\chi_I}+g(\chi_I)\right){\mathbf{v}_I}(x,y,z).
\end{equation}

We define the attenuation function $E$ as
\begin{equation}\label{eq12-bis}
    E(\psi_I,\chi_I)=\left(\frac{\mathrm{d}f}{\mathrm{d}\psi_I}+f(\psi_I)\right)\left(\frac{\mathrm{d}g}{\mathrm{d}\chi_I}+g(\chi_I)\right),
\end{equation}
which is the crucial quantity to determine the flow in region II. Since $E$ is a function only of the stream functions, it is constant along a streamline and is therefore entirely defined by considering its value on the screen.

In addition to the equation (\ref{eq12}), the mass flow rate must be conserved when the fluid passes through the screen implying the continuity of the normal velocity at the screen between the region I and the region II
\begin{equation}\label{eq11}
    {v_{I}}_n(x_s,y_s,z_s) = {v_{II}}_n(x_s,y_s,z_s).
\end{equation}

At this point, our system contains thus two unknowns which are the attenuation function $E$ and the distribution of sources $\Omega$. We have one equation (\ref{eq11}), and another equation linking the velocities and pressures in the vicinity of the porous structure is required to close our system of equations. For this purpose, two streamlines are considered as shown in figure \ref{fig:model}: $(AP)$ and $(PB)$ where the point $P$ is on the screen taken as a surface from a macroscopic point of view. Along each of these streamlines, Bernoulli's equation can be applied, and we thus obtain for $(AP)$
\begin{equation}\label{eq13}
    \frac{1}{2}\rho v_{I}^2(x_A,y_A,z_A)+p_{I}(x_A,y_A,z_A) = \frac{1}{2}\rho v_{I}^2(x_s,y_s,z_s)+p_{I}(x_s,y_s,z_s),
\end{equation}
and for $(PB)$
\begin{equation}\label{eq14}
    \frac{1}{2}\rho v_{II}^2(x_s,y_s,z_s)+p_{II}(x_s,y_s,z_s) = \frac{1}{2}\rho v_{II}^2(x_B,y_B,z_B)+p_{II}(x_B,y_B,z_B).
\end{equation}

We consider that the points $A$ and $B$ are far enough from the screen so that we can take constant values of the velocities and pressures (see \cite{Fail} for flat plates normal to an air stream). Therefore, upstream we have $\bm{v}_{I}(x_A,y_A,z_A)= \bm{v}_0$ and $p_{I}(x_A,y_A,z_A)=p_0$; downstream we take the mean value of the velocity over a section of the wake orthogonal to the far-field stream direction ($\bm{v}_0$) $\mathbf{v}_{II}(x_B,y_B,z_B) = \overline{\lim\limits_{z \rightarrow +\infty}\bm{v}_{II}(x,y,z)} = \overline{E(\psi_{I},\chi_{I})}\bm{v}_0$. In the rest of the paper, $\overline{E(\psi_{I},\chi_{I})}$ will be denoted $\overline{E}$. \cite{Koo} considered a far-downstream constant pressure $p_{0}$, however it is known that the pressure in the wake is lower than the pressure outside the wake which contributes to aerodynamic forces. \cite{Steiros} therefore introduced a suction base pressure $p_{III}$ at the point B, which is assumed to be constant far enough from the screen in region III as explained above. Thus, we introduce a third free parameter $p_{III}$ that will be determined from conservation equations in section \ref{sec:drag}.
Note that since we assume that the pressure $p_{III}$ is constant, the pressure will be discontinuous across the wake boundaries (as in the model of \cite{Koo} in the two-dimensional case).
By combining the equations (\ref{eq13}) and (\ref{eq14}), decomposing the velocities according the tangential and normal components on the screen (respectively ${v_I}_t$ and ${v_{I}}_n$) and by using equation (\ref{eq11}), we obtain the pressure difference
\begin{equation}\label{eq16}
\begin{aligned}
    p_0 - p_{III} ={}& \frac{1}{2}\rho\left(1-E^2\left(\psi_{I},\chi_{I}\right)\right){v_I^2}_t(x_s,y_s,z_s) + \frac{1}{2}\rho\left(\overline{E}^2-1\right) v_0^2 \\
     & + p_I(x_s,y_s,z_s) - p_{II}(x_s,y_s,z_s)
\end{aligned}
\end{equation}

If we are able to determine the pressure differences $p_0 - p_{III}$ and $p_I(x_s,y_s,z_s) - p_{II}(x_s,y_s,z_s)$ independently from these equations, then we can use the equations (\ref{eq11}) to find the attenuation function $E$ and equation (\ref{eq16}) to find the distribution of sources $\Omega$. In order to determine the suction pressure $p_{III}$, we will follow the method of \cite{Steiros} in section \ref{sec:drag}, using the conservation law of the momentum in the control volume $V_1$ shown in figure \ref{fig:model}, and at the vicinity of the screen. Before addressing this problem, we focus in the following section \ref{sec:pressure} on the pressure jump $p_I(x_s,y_s,z_s) - p_{II}(x_s,y_s,z_s)$.

\subsection{Pressure jump across the screen}\label{sec:pressure}

\begin{figure}
  \centerline{\includegraphics[width=8cm, trim = 0cm 0cm 0cm 0cm, clip]{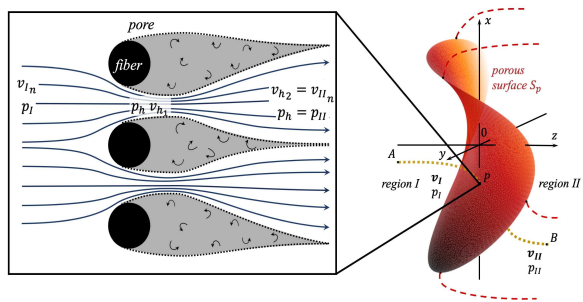}}
  \caption{Diagram of the flow at the scale of the pores in case of parallel fibers (left), and corresponding imaginary surface as used in the model (right).}
\label{fig:pressure-jump-screen}
\end{figure}

A summary of the models for the relation between the pressure jump and screen porosity can be found in \cite{Xu}. This problem has largely been discussed in many papers, raising the difficulties of a general formulation. In their model, \cite{Steiros} consider two streamlines passing through the screen in a hole where the velocity and the pressure are assumed to be uniform. Immediately upstream after the acceleration of the fluid, the characteristic pressure of the flow is denoted $p_h$ and is assumed to correspond physically to the mean pressure inside the hole, as shown in figure \ref{fig:pressure-jump-screen}. As noticed by several authors included \cite{TaylorDavies} and \cite{Wieghardt} the characteristic velocity immediately upstream should be regarded as the mean velocity after contraction of the flow within the holes, ${v_h}_1$, denoting the average velocity through the screen expressed as
\begin{equation}\label{eq-drag-5}
   {v_h}_1 = \frac{{v_I}_n(x_s, y_s, z_s)}{C(1-s)},
\end{equation}
where $C$ is the contraction coefficient taking into account the vena contracta within the pores. The velocity accelerates to ${v_h}_1$ in the hole and the pressure decreases to $p_h$ so that at this point there are no losses. Along this first streamline, Bernoulli's equation leads to
\begin{equation}\label{eq-drag-6-1}
\begin{aligned}
   p_{I}(x_s, y_s,z_s) + \frac{1}{2}\rho {v_I^2}(x_s, y_s,z_s)  & ={} p_h + \frac{1}{2}\rho {v_{h}^2}_1.
\end{aligned}
\end{equation}

Immediately downstream, considering also a homogenized velocity, the flow enlarges and the velocity reaches a  value equal to
\begin{equation}\label{eq-drag-6-1bis}
\begin{aligned}
  {v_h}_2 = {v_{II}}_n(x_s, y_s, z_s),
\end{aligned}
\end{equation}
taken as the characteristic velocity just after the hole. As formulated by \cite{TaylorDavies} and considered by \cite{Steiros}, the pressure immediately after the hole is assumed to be the same as inside the hole, i.e. $p_h$. It means that the pressure loss during the fluid acceleration is not recovered, because of the viscous effects. Bernoulli's equation along this second streamline leads to
\begin{equation}\label{eq-drag-6-2}
\begin{aligned}
   p_h + \frac{1}{2}\rho {v_{h}^2}_2 & ={} p_{II}(x_s, y_s, z_s) + \frac{1}{2}\rho {v_{II}^2}(x_s, y_s,z_s).
\end{aligned}
\end{equation}

We combine equations (\ref{eq12}), (\ref{eq11}) and (\ref{eq-drag-5}) to (\ref{eq-drag-6-2}) to obtain
\begin{equation}\label{eq-pressure}
\begin{aligned}
   {p_{II}}(x_s,y_s,z_s)-{p_{I}}(x_s,y_s,z_s) &={} \frac{1}{2}\rho {v_I}_t^2(x_s, y_s,z_s)\left(1-E^2(\psi_I,\chi_I)\right) \\
   & + \frac{1}{2}\rho{v_I}_n^2(x_s, y_s,z_s)\theta(s),
\end{aligned}
\end{equation}
with 
\begin{equation}\label{eq-8}
   \theta(s) = 1-\frac{1}{C^2(1-s)^2}.
\end{equation}

As far as we know, there are no measurements of vena contracta for screens constituted of fibers and especially in the three-dimensional case of free flow. \cite{Simmons} performed measurements of the velocity profile behind a porous screen made of a square mesh of woven material spanning a section of a channel and  suggest an estimate of the vena contracta assuming a uniform velocity ${v_h}_1 = v_0$ in the holes. In these experiments the screens are made of circular rods with diameter ranging from $0.112$ to $0.373\hspace{0.1cm}mm$ arranged in a square mesh. The velocity used was from $2.44$ to $10.36\hspace{0.1cm}m.s^{-1}$. The local Reynolds number $Re_d$ thus varies from $18$ to $248$, which is the same order of most of the porous screens used in our experiments.
For a solidity $s\approx 0.5$ the coefficient $C$ should be between $0.9$ and $1.0$. Note that with the formulation (\ref{eq-8}), if $C \neq 1$, the limit $s=0$ leads to a non-zero pressure difference, suggesting that $C$ may depend on the solidity $s$ at least for low solidities. For screens normal to the flow, \cite{Steiros} neglected the possible contraction of the flow within the holes, {\it i.e.} $C=1$. In order to compare the 3D model to their 2D model for the drag coefficient, we also take $C=1$. This will be discussed later.

Injecting the pressure difference $p_{I}-p_{II}$ obtained in equation \ref{eq-pressure} in equation (\ref{eq16}), we have
\begin{equation}\label{eq-pressure-pIII}
\begin{aligned}
     p_0 - p_{III} ={}& -\frac{1}{2}\rho{v_I^2}_n(x_s,y_s,z_s)\theta(s) + \frac{1}{2}\rho\left(\overline{E}^2-1\right) v_0^2.
\end{aligned}
\end{equation}

Assuming that $p_{III}$ is constant, the right-hand side of the above equation has to be constant, which leads to the condition
\begin{equation}\label{eq-cond-vn}
\begin{aligned}
     \mathbf{grad}_{\mathcal{S}_p}\left({v_I}_n^2(x_s,y_s,z_s)\theta(s)\right)={}& \bm{0},
\end{aligned}
\end{equation}

with $\theta(s)$ that can vary for non-homogeneous porous surfaces. Therefore, under the assumptions made so far, we are looking for a source strength $\Omega$ that satisfies this condition. It is possible to relax certain restrictions on the value of $\Omega$ by considering a variable base suction pressure $p_{III}$ or a variable far wake velocity $\mathbf{v}_{II}(x_B,y_B,z_B) = E(\psi_{I},\chi_{I})\bm{v}_0$. In that case, the problem is far more difficult, and should be solved numerically. Such a resolution is beyond the scope of present work.

\subsection{Drag coefficient}\label{sec:drag}

In this section, we extend the model of \cite{Steiros} in three-dimensions with a screen of arbitrary shape and orientation in the flow. It will give two expressions of the aerodynamic forces which leads to an equation determining the remaining free parameter $p_{III}$. An application of these equations will be given in the next section. 

A first expression is given by the momentum balance around the surface of the screen, using the drag coefficient, giving: 
\begin{equation}\label{eq-drag-1}
   C_D = \frac{1}{\frac{1}{2}\rho S_{p} v_0^3}\left(\iint_{\mathcal{S}_p}\left(p_I-p_{II}\right)\bm{v}_{0}\cdot \bm{n}_s\mathrm{d}S+\rho\iint_{\mathcal{S}_p} {v_I}_n \bm{v}_{0}\cdot \left(\bm{v}_{I}-\bm{v}_{II}\right)\mathrm{d}S\right),
\end{equation}

where the normal vector $\bm{n}_s$ points in the direction of the region II. 

The second expression of the drag coefficient is provided by the two equations (\ref{eq13}) and (\ref{eq14}) and the momentum balance in a control volume around the screen.
We consider the control volume in figure \ref{fig:model} and assume that the volume is large enough so that the velocities at the surfaces $\mathcal{S}_x$ and $\mathcal{S}_y$ on the sides of the block parallel to the z-axis are equal to $\bm{v}_0 + \bm{v}_{\epsilon}$ where $v_{\epsilon} \ll v_0$. With this approximation, we calculate the momentum balance leading to
\begin{equation}\label{eq-drag-2}
\begin{aligned}
        \bm{F}_D + \bm{F}_P ={} & \rho S_w \left(1-\overline{E(\psi_{I},\chi_{I})}^2\right){v_0}_z \bm{v}_0 + S_w\left(p_0 - p_{III}\right)\bm{e_z} \\
        & - \rho \bm{v}_0 \iint_{\mathcal{S}_x + \mathcal{S}_y} \bm{v}_I \cdot \bm{n}\mathrm{d}S
\end{aligned}
\end{equation}
where $\bm{F_P}$ is the lift, $\bm{F_D}$ is the drag, $S_w$ is the area of the section of the wake (orthogonal to the z-axis), and the bar over $E$ denotes the mean over the considered surface. The normal vector $\bm{n}$ points outwards from the control volume. The projection of the right-hand side of this equation onto the far-field stream direction ($\bm{v}_0$) gives an expression of the drag. Then, a mass balance in the same control volume gives
\begin{equation}\label{eq-drag-3}
\begin{aligned}
        \rho {v_0}_z S_w = \rho \iint_{\mathcal{S}_x + \mathcal{S}_y} \bm{v}_I \cdot \bm{n}\mathrm{d}S + \rho \overline{E(\psi_{I},\chi_{I})}{v_{0}}_zS_w.
\end{aligned}
\end{equation}

This equation allows us to find the value of the last term of the conservation of momentum equation (\ref{eq-drag-2}). Moreover, the section of the wake $S_w$ is determined with a mass balance through the screen
\begin{equation}\label{eq-drag-4}
\begin{aligned}
        \rho \iint_{\mathcal{S}_p} \bm{v}_I \cdot \bm{n}\mathrm{d}S = \rho \overline{E(\psi_{I},\chi_{I})}{v_{0}}_z S_w
\end{aligned}
\end{equation}

The equations (\ref{eq13}), (\ref{eq14}) and (\ref{eq-drag-2}) to (\ref{eq-drag-4}) are sufficient to obtain a second expression of the drag coefficient. This closes the problem and enables us to obtain the velocities and pressure at any location in the flow.

To summarize, in order to close their problem,  \cite{Koo} combined an equation for the pressure jump $p_I-p_{II}$ with the consideration of two streamlines, as we did here before for equation (\ref{eq16}), but considering $p_{III}=p_0$. Here, following \cite{Steiros}, after the addition of the base-suction effect $p_0 - p_{III}$, the problem is closed by first the momentum and mass conservation equations ((\ref{eq-drag-1}) to (\ref{eq-drag-4})), and second vy this screen-effect hypothesis on the pressure loss (\ref{eq-pressure}).

The final equation of the problem for the determination of the source-strength $\Omega$ then reads
\begin{equation}\label{eq-final-eq}
\begin{aligned}
        \overline{E}\iint_{\mathcal{S}_p} \left({\tilde{v}_{I_t}^2}\left(1-E^2\right) + {\tilde{v}_{I_n}^2}\theta(s)\right)\bm{n}_s\cdot\bm{e}_z + 2\tilde{v}_{I_n}\left(1-E\right){\bm{\tilde{v}}_{I_t}}\cdot \bm{e}_z \mathrm{d}S = \\
        \left(- \left(1-\overline{E}\right)^2 - \tilde{v}_{I_n}^2\theta(s)\right)\iint_{\mathcal{S}_p}\tilde{v}_{I_n} \mathrm{d}S,
\end{aligned}
\end{equation}

with dimensionless velocities $\tilde{v}=\frac{v}{v_0}$.
We give now the explicit expression for $v_{I_n}$ and $E$. Finding $v_{I_n}$ is a boundary surface potential problem. $v_{I_n}$ is found from the gradient of the velocity potential (\ref{eq5}) having therefore an integral term. This integral term is known as a harmonic double-layer potential with density $\Omega$ (\cite{Gunter}), which is defined on a sub-domain of $\mathbb{R}^3\setminus \mathcal{S}_p$. This integral term becomes singular if it is evaluated on the surface $\mathcal{S}_p$, however it can be continuously extended on the surface for each side and the value depends on the side by which we approach the surface. From the definitions of the regions I and II, for $\bm{m}=(x,y,z)\in \mathcal{S}_p$, we have
\begin{equation}\label{eq-final-eq-vn-1}
\begin{aligned}
        v_{I_n}(\bm{m}) = v_0\bm{e_z}\cdot\bm{n}_s + v_n^-(\bm{m}),
\end{aligned}
\end{equation}
and
\begin{equation}\label{eq-final-eq-vn-2}
\begin{aligned}
        v_{{II}_n}(\bm{m}) = E\left(\psi_I,\chi_I\right)\left(v_0\bm{e_z}\cdot\bm{n}_s + v_n^+(\bm{m})\right),
\end{aligned}
\end{equation}
with
\begin{equation}\label{eq-final-eq-vn-3}
\begin{aligned}
        v_n^{\pm}(\bm{m}) = \lim\limits_{\epsilon \to 0^+}\frac{\partial \phi}{\partial \bm{n}_s}\left(\bm{m} \pm \epsilon \bm{n}_s\right),
\end{aligned}
\end{equation}
with $\phi$ defined in equation (\ref{eq4}) and the directional outward normal derivative:
\begin{equation}\label{eq-final-eq-vn-4}
\begin{aligned}
        \frac{\partial \phi}{\partial \bm{n}_s}\left(\bm{m}\right) = \mathbf{grad}\left(\phi\left(\bm{m}\right)\right)\cdot \bm{n}_s.
\end{aligned}
\end{equation}

From a theorem which can be found in many books, for instance in \cite{Kress} p. 80, the value of the limits above can be expressed using an improper integral. For $\bm{m}=(x,y,z)\in \mathcal{S}_p$, this expression reads
\begin{equation}\label{eq-final-eq-vn-5}
\begin{aligned}
        v_{n}^{\pm}(\bm{m}) = \iint_{U} \Omega(u,v) \frac{\partial \Gamma}{\partial \bm{n}_s}\left(\bm{m}-\bm{m}_s(u,v)\right) \left\| \frac{\partial \bm \xi}{\partial u} \wedge \frac{\partial \bm \xi}{\partial v}\right\| \mathrm{d}u\mathrm{d}v \pm \frac{1}{2}\Omega(\bm{m}),
\end{aligned}
\end{equation}
with $\bm{m}_s(u,v)=(x_s(u,v),y_s(u,v),z_s(u,v))$.
Finally, $E$ is found by applying the continuity equation (\ref{eq11}) for the normal velocities across the surface, with the expressions given in equation (\ref{eq-final-eq-vn-1}) and (\ref{eq-final-eq-vn-2}). For $\bm{m}=(x,y,z)\in \mathcal{S}_p$, we have
\begin{equation}\label{eq-final-eq-vn-6}
\begin{aligned}
        E\left(\bm{m}\right) = \frac{v_0\bm{e_z}\cdot\bm{n}_s + v_n^-(\bm{m})}{v_0\bm{e_z}\cdot\bm{n}_s + v_n^+(\bm{m})}.
\end{aligned}
\end{equation}

In the next section, we solve the problem analytically in a basic but very common geometry.

\subsection{Application to an inclined rectangular screen in free flow}

\begin{figure}
  \centerline{\includegraphics[width=10cm, trim = 0cm 0cm 0cm 0cm, clip]{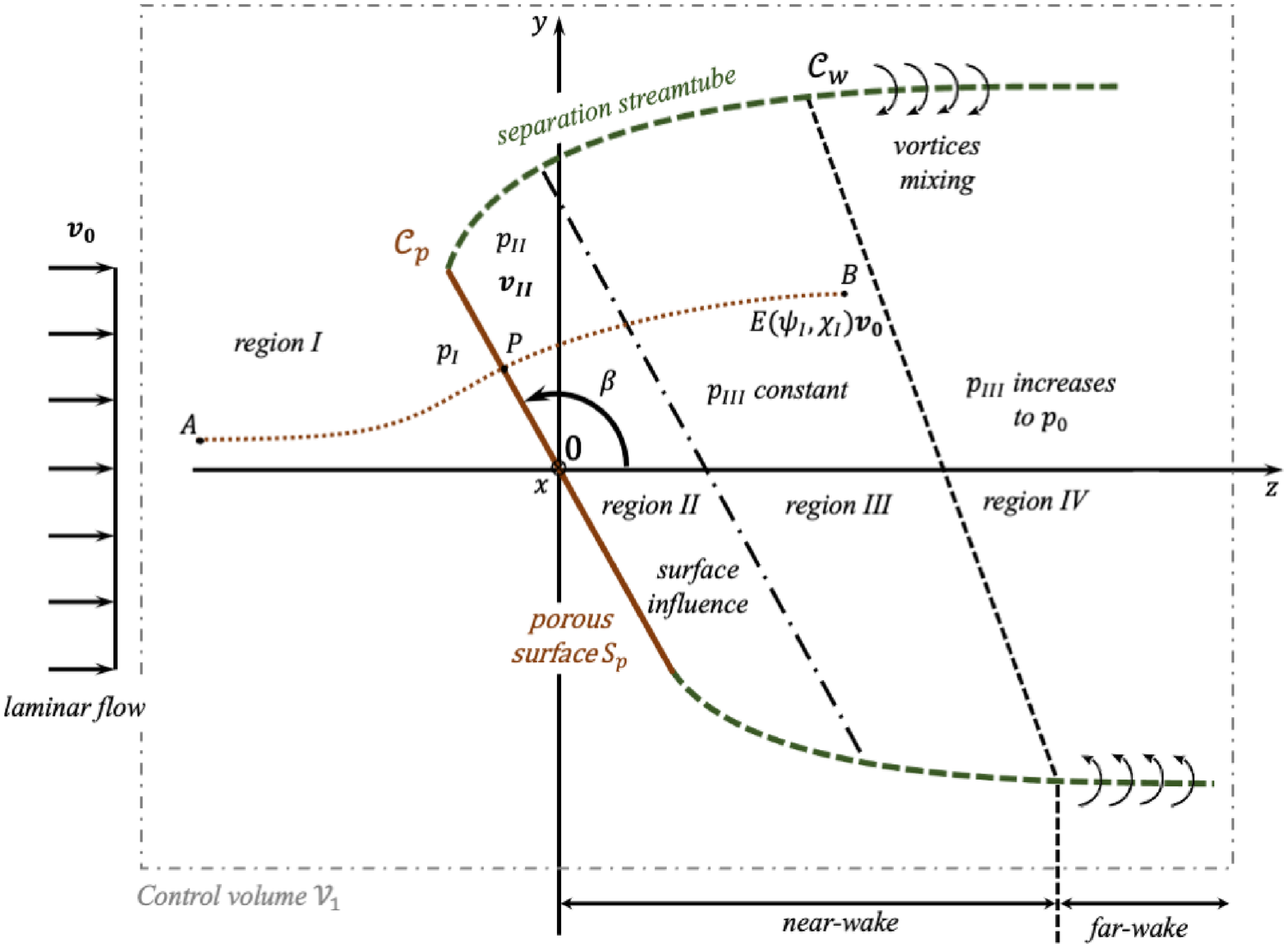}}
  \caption{Diagram of the inclined rectangular porous screen in the $(Oyz)$ plane in a three-dimensional free flow with an angle $\beta$. The dashed lines are the separation streamlines used as a boundary between the regions of the model. The dotted line between point A and B is the streamline used in the model to calculate the velocities in the regions I and III. The incoming flow is laminar.}
\label{fig:model-inclined-plate}
\end{figure}

For a rectangular geometry and homogeneous solidity it is possible to obtain without major difficulty an analytical solution of the equations of our model. We therefore apply the three-dimensional model to the simple case of a rectangular screen (centered at $z=0$ as shown in the figure \ref{fig:model-inclined-plate}) in a free laminar flow in order to find the flow and the drag coefficient as a function of the solidity. In that case, the velocity potential (\ref{eq5}) becomes
\begin{equation}\label{eq-rect-1}
    \phi_I(x,y,z) = v_0z + c - \frac{1}{4\pi}\iint_{\mathcal{S}_p}\frac{\Omega(u,v)\mathrm{d}u\mathrm{d}v}{\sqrt{(x-v)^2+(y-u\sin{(\beta)})^2+(z-u\cos{(\beta)})^2}}
\end{equation}

We then calculate the velocities at the screen. The objective of the calculus is to obtain $\Omega$ as a function of the solidity, which will determine all the other variables of the problem.

The normal component of the velocity in region I at the arbitrary position $(w,t)$ on the surface is
\begin{equation}\label{eq-rect-2}
    {v_{I}}_n^{\pm}(w,t) = v_0\sin{(\beta)} + \lim\limits_{\epsilon \to 0^\pm}\frac{1}{4\pi}\iint_{\mathcal{S}_p} f_{\epsilon}(u,v) \Omega(u,v) \mathrm{d}u\mathrm{d}v,
\end{equation}
with
\begin{equation}\label{eq-rect-2}
   f_{\epsilon}(u,v) = \frac{\epsilon\cos{(\beta)}\sin{(\beta)}}{\left((t-v)^2+(w-u)^2+(2(w-u) + \epsilon)\epsilon\cos{(\beta)}\right)^{\frac{3}{2}}}
\end{equation}

The details of the calculations are given in the appendix \ref{appB-bis}. At the screen, depending on the direction from which we approach the screen ($\epsilon\to0^{\pm}$), the magnitude of the normal component of the velocity is constant and is equal to
\begin{equation}\label{eq-rect-3}
    {v_{I}}_n(w,t) = {v}_n^{-}(w,t)  = v_0\sin{(\beta)} - \frac{1}{2} \Omega(w,t),
\end{equation}
\begin{equation}\label{eq-rect-4}
    {v_{II}}_n(w,t) = E(\psi_I,\chi_I){v}_n^{+}(w,t) = E(\psi_I,\chi_I)\left(v_0\sin{(\beta)} + \frac{1}{2} \Omega(w,t)\right).
\end{equation}

At this point, for an homogeneous screen ($s$ constant on the surface), the equation (\ref{eq-cond-vn}) leads to $\mathbf{grad}\left(\Omega(w,t)\right)={}\bm{0}$, i.e. the source strength is a constant.

Thus, the equation (\ref{eq11}) leads to a constant attenuation function
\begin{equation}\label{eq-rect-5}
    E = \frac{v_0\sin{(\beta)} - \frac{1}{2} \Omega}{v_0\sin{(\beta)} + \frac{1}{2} \Omega}.
\end{equation}

If we found constant normal velocity, it is not the case of the tangential velocity for which the magnitude varies on the surface of the screen.
The tangential component of the velocity at the surface is
\begin{equation}\label{eq-tang-princ}
\begin{aligned}
   {v_I}_t(x,w) &= \left(\Omega^2 \mathcal{I}_x^2(x,w) + \left(\Omega\left(\sin{(\beta)}\mathcal{I}_y(x,w)+\cos{(\beta)}\mathcal{I}_z(x,w)\right) \right.\right.\\
   & \left.\left. + \cos{(\beta)}v_0\right)^2\right)^{\frac{1}{2}},
\end{aligned}
\end{equation}
with $\mathcal{I}_x$, $\mathcal{I}_y$ and $\mathcal{I}_z$ the following surface integrals that are calculated in appendix \ref{appA}:
\begin{equation}\label{eq-tang-6-princ}
\begin{aligned}
   \mathcal{I}_x(x,w) &= -\frac{1}{4\pi}\iint_{\mathcal{S}_p} \frac{x-v}{\left(\left(x-v\right)^2+\left(w-u\right)^2\right)^{\frac{3}{2}}}
  \mathrm{d}u \mathrm{d}v
\end{aligned}
\end{equation}
\begin{equation}\label{eq-tang-7-princ}
\begin{aligned}
   \mathcal{I}_y(x,w) &= -\frac{1}{4\pi}\iint_{\mathcal{S}_p} \frac{(w-u)\sin{(\beta)}}{\left(\left(x-v\right)^2+\left(w-u\right)^2\right)^{\frac{3}{2}}}
  \mathrm{d}u \mathrm{d}v
\end{aligned}
\end{equation}
\begin{equation}\label{eq-tang-8-princ}
\begin{aligned}
    \mathcal{I}_z(x,w) &= -\frac{1}{4\pi}\iint_{\mathcal{S}_p} \frac{(w-u)\cos{(\beta)}}{\left(\left(x-v\right)^2+\left(w-u\right)^2\right)^{\frac{3}{2}}}
  \mathrm{d}u \mathrm{d}v
\end{aligned}
\end{equation}

For the sake of simplicity, we approximate the magnitude of the tangential component of the velocity as its root mean square. We note that since in equation  (\ref{eq-final-eq}), $v_{I_t}$ appears both in linear and quadratic form, this approximation becomes exact for a surface normal to the mean flow direction ($\beta=\frac{\pi}{2}$). We thus take
\begin{equation}\label{eq-tang-0}
\begin{aligned}
   {v_I}_t &= \left(\Omega^2 \gamma_0 + v_0^2\cos^2{(\beta)} \right)^{\frac{1}{2}}
\end{aligned}
\end{equation}

where $\gamma_0$ can be considered as a shape factor (see appendix \ref{appA}). The equation (\ref{eq-pressure}) becomes
\begin{equation}\label{eq-pressure-ex}
\begin{aligned}
   {p_{II}}-{p_{I}} &={} \frac{1}{2}\rho \left({v_I}_t^2\left(1-E^2\right)+{v_I}_n^2\theta(s)\right)
\end{aligned}
\end{equation}

The equation (\ref{eq16}) leads to : 
\begin{equation}\label{eq-pressure-2-ex}
\begin{aligned}
   {p_{III}}-{p_{0}} &={} \frac{1}{2}\rho \left(1-E^2\right)\left(v^2_0-{v_I}_t^2\right)  + p_{II} - p_{I} \\
   &={} \frac{1}{2}\rho\left(\left(1-E^2\right)v^2_0 + {v_I}_n^2\theta(s)\right)
\end{aligned}
\end{equation}

Using the equations (\ref{eq-drag-2}) to (\ref{eq-drag-4}) we obtain a first expression of the magnitude of the drag force $F_D$
\begin{equation}\label{eq-force-1-ex}
\begin{aligned}
        F_D & ={} \rho v_0\left(1-E\right){v_n} {S_p} + \frac{1}{v_0}\left(p_0 - p_{III}\right)\frac{{v_n}}{E} {S_p}.
\end{aligned}
\end{equation}

The second expression for the drag force $F_D$ is obtained with the equation (\ref{eq-drag-1})
\begin{equation}\label{eq-force-2-ex}
   F_D ={} \left({p_I}-{p_{II}}\right)\sin{(\beta)}{S_p} + \rho {v_n} v_0\cos(\beta)\left(1-E\right){S_p}.
\end{equation}

Note that for a rectangular screen orthogonal to the free flow the second term of equation (\ref{eq-force-2-ex}) vanishes since $\beta=\frac{\pi}{2}$ and we obtain a drag coefficient proportional to the pressure difference ${p_I}-{p_{II}}$. Now, by denoting $\omega = \frac{\Omega}{v_0}$ and combining the equations (\ref{eq-force-1-ex}) and (\ref{eq-force-2-ex}), we obtain the following equation that we have to solve to find the value of the source strength
\begin{equation}\label{eq-x-ex}
\begin{gathered}
  -\frac{1}{8} \omega^4\theta(s) + \omega^2 \sin^2{(\beta)}\Big(8\gamma_0 + \theta(s) - 2\Big) - \\
  4\omega\sin{(\beta)} - 2\sin^4{(\beta)}\theta(s) = 0
\end{gathered}
\end{equation}

\begin{figure}
     \centering
     \begin{subfigure}[b]{0.45\textwidth}
         \includegraphics[width=6.5cm,trim = 0.2cm 0cm 0cm 0cm, clip]{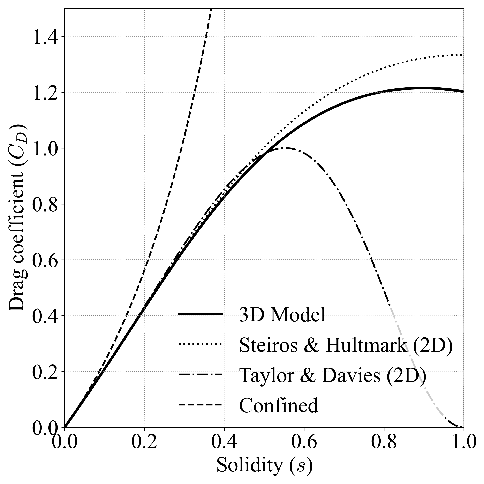}
         \caption{Theoretical drag coefficient.}
         \label{fig:drag-solidity-theor}
     \end{subfigure}
     \hspace{0.3cm}
     \begin{subfigure}[b]{0.45\textwidth}
         \includegraphics[width=6.5cm,trim = 0.2cm 0cm 0cm 0cm, clip]{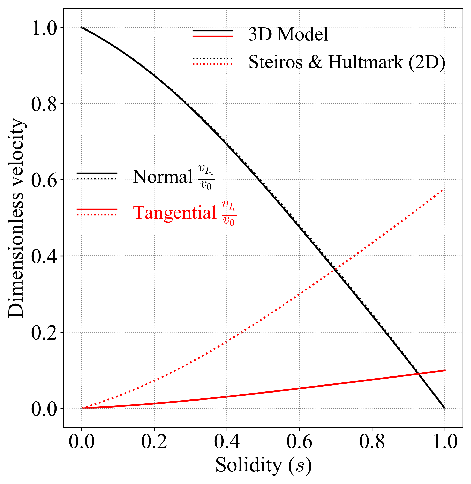}
         \caption{Theoretical velocity at the screen.}
         \label{fig:velocity-solidity-theo}
     \end{subfigure}
     \caption{Theoretical prediction of the drag coefficient and velocities on the surface as a function the solidity obtained with the three dimensional model applied to a square homogeneous screen normal to the flow ($\beta = \frac{\pi}{2}$), and comparison with different two-dimensional models. The full lines are obtained with the three-dimensional model developed in this article using the same pressure jump law across the screen than \cite{Steiros} (but taking account geometric 3D effects). (a) Drag coefficient. (b) Normal velocity and quadratic mean of the tangential velocity as defined respectively in equation (\ref{eq-rect-3}) and (\ref{eq-tang-0}).}
     \label{fig:drag-solidity-final}
\end{figure}

The solution for a square porous plate at normal incidence with $\beta=\frac{\pi}{2}$ is plotted in figure \ref{fig:drag-solidity-final} (a). At low solidity, our model for a square plate is close to the prediction of \cite{Steiros} and \cite{TaylorDavies}. Above a solidity  $s=0.4$ the drag coefficient becomes slightly different from the prediction of \cite{Steiros} and the difference increases with increasing solidity. Three dimensional effects are therefore important at high solidity.

At high solidity ($s\lesssim 1$), the curve converges towards the drag coefficient of a flat solid plate. This value is predicted to be $1.2$ in our model, which is close to the experimental measurement at global Reynolds number $Re \approx 10^4$, giving approximately $1.05$ according to \cite{Blevins}, $1.17$ according to the synthesis of \cite{Hoerner} on various drag measurements, and $0.939$ in pour experiments.

Our model takes into account the shape of the porous surface through the parameter $\gamma_0$.  For a rectangular screen, the aspect ratio, taken into account in $\gamma_0$, has an influence on the result as detailed in the appendix \ref{appA}. According to \cite{Hoerner}, for a flat plate normal to the flow, the drag coefficient increases very slowly when the aspect ratio is reduced until a ratio of approximately $0.1$. Beyond this point the increase becomes more pronounced, until it reaches $C_D=2.0$ for an infinitely thin plate ($1.90$ according to \cite{Blevins}). In our model, we indeed observe an increase of the drag when the aspect ratio decreases (see appendix \ref{appA}) from $C_D=1.2$ for a square plate (with $\gamma_0=0.0998$), 1.29 for a rectangle with aspect ratio 1/10 (with $\gamma_0=0884$) to $C_D=1.33$ for infinitely thin plate. This value for infinitely thin plate is lower than the experimental value, that is $C_D \approx 2.0$ according to \cite{Blevins} and \cite{Hoerner}. This difference may be due in this case of very high aspect ratio to the vortex shedding that is not taken into account in the present wake model, as noticed by \cite{Steiros}.

In figure \ref{fig:drag-solidity-final} (b), we compare the normal and tangential velocities in the 2D and 3D cases. 
The tangential velocity is taken in both cases as the quadratic mean over the whole surface, in 3D it is defined in equation (\ref{eq-tang-0}), the normal velocity is constant on the surface in both cases also, in 3D it is defined in equation (\ref{eq-rect-3}).
While the normal velocity is the same in 2D and 3D, we clearly see an influence of 3D effects on the tangential velocity.

We note finally that if the contraction of the flow is not neglected in equation (\ref{eq-8}) there is an increase of the drag coefficient compared to the curve plotted in figure \ref{fig:drag-solidity-theor}, and thus we would still overestimate the drag compared to our data in figure \ref{fig:drag-solidity-exp}.

While taking into account the 3D effects, which improves the prediction of $C_D$ at high solidities, the present model does not improve the prediction at moderate and low solidities. Moreover, the model still shows discrepancies with the experimental results. In particular, the variations of the drag coefficient with the local Reynolds number observed in the experiments are still not accounted for. Indeed, the relation used to estimate the pressure jump $p_{I}-p_{II}$ (equation (\ref{eq-pressure})) does not  explicitly consider the viscosity, as we will discuss in the next subsection.

Finally, the differences between our theoretical prediction and the experimental values might also be due to the limitations of potential flow theory to describe the complex flow around bluff bodies, as we will discuss later. However, we expect these limitations to arise at very high solidity, and to be negligible at low and moderate solidity.

\subsection{Pressure jump dependency on the local geometry of the pores and viscous effects}\label{sec:pressure-Re}

As explained above, in the equation of the pressure jump (\ref{eq-pressure}) with equation (\ref{eq-8}), as formulated by \cite{Steiros}, we do not take into account the dependency of the pressure drop on the local Reynolds number $Re_d$, computed at the scale of the screen pores (and thus much smaller than Re), the geometry of the holes and other possible dependency like the energy transfer between the material of the screen and the fluid.
For instance, \cite{Ando} showed that a layer of flexible fibers can have a higher permeability than the same layer of rigid fibers due to a flow-induced deformation. 
Moreover, as shown by \cite{Schubauer}, the angle of the screen relative to the laminar upstream free flow has an impact on the pressure drop. \cite{Kalugin} explained also that for inclined perforated plates, the structure of the flow in the holes depends on (1) the distance of the hole on the plate from the leading edge, (2) the angle of attack of the plate. For low angle of attack, the effective hole area can be significantly reduced due to the difficulty of the flow to deflect from its original direction mostly parallel to the surface of the plate. 
All these studies underline the current difficulty to obtain a general formulation of the pressure loss for arbitrary porous screen. Therefore, in what follows, we adopt another method based on empirical laws in order to test whether this would be sufficient to estimate the drag accurately.

Numerous experimental investigations have shown that the pressure drop can be reasonably considered proportional to the square of the velocity normal to the screen at the vicinity of it through the resistance coefficient $k$, especially \cite{TaylorDavies}. More recently, \cite{Ito} studied this problem in the two-dimensional case of a flow around and through a gauze for low resistance coefficient, while \cite{Eckert} studied the resistance coefficient for the case of a gauze spanning the entire section of a channel. In these cases, the pressure drop can be written:
\begin{equation}\label{eq6}
    \Delta p = \frac{1}{2} k \rho {v_{I}^2}_n(x_s,y_s,z_s),
\end{equation}

The resistance coefficient $k$ depends on the geometry of the holes, the material of the screen and the Reynolds number based on the scale of the holes (sometimes $k$ is directly related to what is called loss factor or friction factor). This has been mostly studied when the screen spans entirely a channel with normal incidence, and oblique incidence (\cite{Reynolds}, \cite{Schubauer}). Note that in equation (\ref{eq-pressure}), $\theta(s)$ can be interpreted as a resistance coefficient, the intervention of the tangential velocity in the equation (\ref{eq-pressure}) comes from the fact that in our case the fluid can pass around the porous structure.

To formulate the resistance coefficient dependency for porous screens, \cite{Pinker} has shown that the resistance coefficient $k$ can reasonably be considered as a product of a function of the solidity, $G(s)$, and a function $f(Re_n)$ of the Reynolds number $Re_n = Re_d \frac{v_{I_n}}{v_0}$ based on the scale of holes and the approach velocity that corresponds in our formulation to $v_{I_n}$. Among several fitted expressions for $G$ with respect to the solidity, \cite{Pinker} found that $G(s) = -\theta(s)$, which exhibits the best agreement with their data.

The pressure drop depending also on the inclination of the surface, we have to consider in fact the function $f(Re_n,\beta)$, as first proposed by \cite{Schubauer}. Since the geometry of our porous screen is arbitrary, $\beta$ should be considered as a local characteristic of the inclination of the surface of the screen relative to the direction of the laminar free flow in the far upstream. $\beta = 0$ means that the surface is parallel to the flow $\bm{v}_0$. We  thus propose the following relation for the pressure jump:
\begin{equation}\label{eq-pressure-final}
\begin{aligned}
   {p_{II}}(x_s,y_s,z_s)-{p_{I}}(x_s,y_s,z_s) &={} \frac{1}{2}\rho {v_I}_t^2(x_s, y_s,z_s)\left(1-E^2(\psi_I,\chi_I)\right) \\
   & + \frac{1}{2}\rho{v_I}_n^2(x_s, y_s,z_s)\theta(s)f(Re_n,\beta).
\end{aligned}
\end{equation}

The expression of $f$ is thus expected to be found experimentally. As far as we know, there is no general physical formulation of the pressure drop through the holes at the microscopic level that can cover all type of screens, and there is no general demonstration of an analytical expression of $f$ for arbitrary porous screen shape. Therefore it is expected that some modifications are required for particular porous structures taking into account for instance the geometry of the holes. For screens composed of fibers, with an angle of attack $\beta$ with the upstream flow, and for $10^{-4} < Re_n < 10^4$, \cite{Brundrett} gives the following empirical expression of f:
\begin{equation}\label{eq-pressure-f}
\begin{aligned}
   f(Re_n,\beta) &={} \sin^2{(\beta)}\left(\frac{c_1}{Re_n\sin{(\beta)}}+\frac{c_2}{\ln{(Re_n\sin{(\beta)}+1.25)}}+c_3\ln{(Re_n\sin{(\beta)})}\right),
\end{aligned}
\end{equation}
where $c_1$, $c_2$ and $c_3$ are real constant, \cite{Brundrett} obtain a good fitting with his data for wire mesh screens by taking $c_1\approx 7.125$, $c_2\approx 0.88$, $c_3\approx 0.055$. \cite{Bailey} found also a good fitting with their data by taking $c_1\approx 18$, $c_2\approx 0.75$, and $c_3\approx 0.055$. For the following sections of the article, we take the geometric mean of the different values $c_1 \approx 11$, $c_2 \approx 0.8$ and $c_3 \approx 0.055$.
For high $Re_n$ the function $f$ behaves as a logarithmic function of the Reynolds number $Re_n$, while for low $Re_n$, the variation of $f$ is much more pronounced (inverse function of the Reynolds number $Re_n$). We thus expect the viscous effects to be important for the typical flow speeds and mesh sizes considered in this work, and in applications such as fog harvesting and facemasks. In this model, equation \ref{eq-pressure-f}, the first term corresponds to the laminar contribution, the second one to the turbulent friction and the last one is valid at large Reynolds number \citep{Brundrett}.

For all mesh Reynolds number $Re_d$, we compute thus $Re_n=Re_d \frac{v_{In}}{v_0}=Re_d (\sin{\left(\beta\right)}-\frac{1}{2}\omega)$ using the new implicit relation for $\omega$:
\begin{equation}\label{eq-x-ex-bis}
\begin{gathered}
  -\frac{1}{8} \omega^4\theta(s)f(Re_n,\beta) + \omega^2 \sin^2{(\beta)}\Big(8\gamma_0 + \theta(s)f(Re_n,\beta) - 2\Big) - \\
  4\omega\sin{(\beta)} - 2\sin^4{(\beta)}\theta(s)f(Re_n,\beta) = 0.
\end{gathered}
\end{equation}

It should be emphasized that this expression which takes into account the local Reynolds number at the pore scale remains empirical, while that of Steiros and Hulmark is based on clear physical assumptions originating from Taylor and Davies (1944), but neglecting the influence of the viscosity on the flow. We will thus keep and compare both formulations ((\ref{eq-x-ex}) and (\ref{eq-x-ex-bis})) in the rest of the paper.

\begin{figure}
  \centerline{\includegraphics[width=14cm, trim = 0cm 0cm 0cm 0cm, clip]{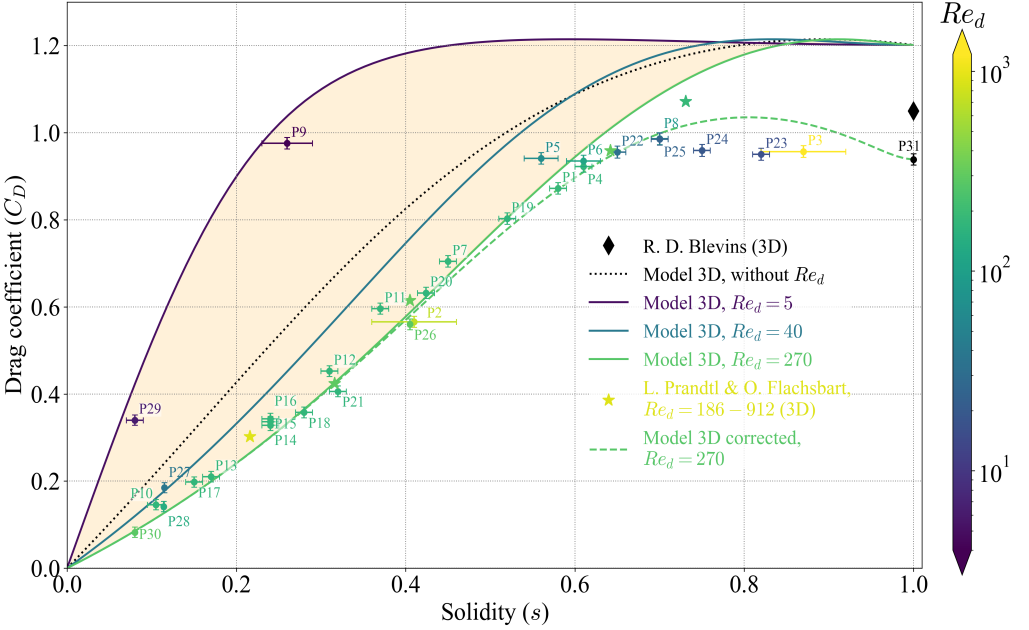}}
  \caption{Drag coefficient as a function of the solidity for various square porous screens normal to the free flow with different $Re_d$, and comparison between the three-dimensional model. The full lines are obtained with the three-dimensional model developed in this article, solution of equation (\ref{eq-x-ex-bis}). The black doted curve is the solution of equation (\ref{eq-x-ex}), thus without taking into account $Re_d$. The slashed green curve is obtained using the correction (\ref{eq-asymp-5}) of the model explained in section \ref{asymptotic}, with $C_D^0 = 0.939$.}
\label{fig:drag-solidity-exp-final}
\end{figure}

For high solidity, and for a certain range of the two Reynolds number $Re$ and $Re_d$, the term $\theta(s)f(Re_n,\beta)$ should behave as the inertial term of the Darcy-Forchheimer equation for porous media. For a porous screens composed of square fiber meshes the inertial term of the Darcy-Forchheimer equation calculated with the method of \cite{Wang} has a value reasonably close to $\theta(s)$ when $s \approx 0.9$. For very thin porous surface, as discussed by \cite{Teitel}, the concept of permeability for porous media involved in the equation of Darcy, and Darcy-Forshheimer, may not always hold for the pressure loss through screens depending on the regime of the flow. According to \cite{Brundrett} and \cite{Bailey} the expression (\ref{eq-pressure-f}) seems to be valid over a larger flow regime.


\section{Results}\label{sec:results}

\subsection{Three-dimensional and viscous effects on the drag coefficient}\label{sec:results-1}

As seen previously, our experiments suggest a strong effect of the local Reynolds number on the drag coefficient. We now compare in figure \ref{fig:drag-solidity-exp-final} our experimental results with the prediction of the 3D model, including local viscous effects using the function $f(Re_n,\beta)$. For all mesh Reynolds number $Re_d$, the trend of the curve remains globally the same: there is an approximately linear increase of the drag coefficient at low solidity before the curve flattens and reaches a plateau, which is well represented by the 3D model derived using the method proposed by \cite{Steiros}, i.e. without  $f(Re_n,\beta)$ (equation (\ref{eq-x-ex})). However, the slope of the initial linear part strongly depends on $Re_d$: it first decreases with increasing mesh Reynolds number $Re_d$, then increases after a critical Reynolds number. This non-monotony is directly related to the non-monotony of the function $f$. 

\begin{figure}
  \centering
  \includegraphics[width=6.5cm,trim = 0.0cm 0cm 0cm 0cm, clip]{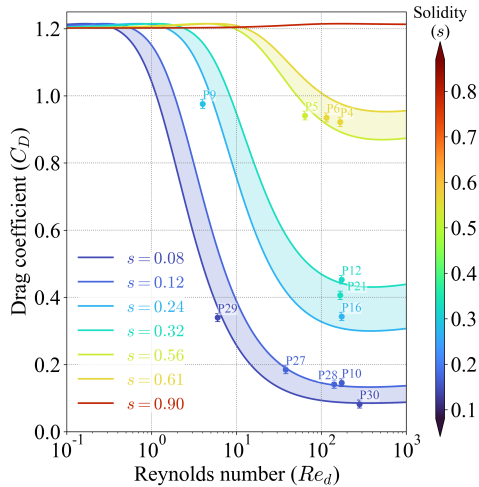}
  \caption{Drag coefficient as a function of the Reynolds number $Re_d$ for three different narrow ranges of solidities: $s=0.1 \pm 0.02$, $s=0.28 \pm 0.04$, $s=0.585 \pm 0.025$. The bullets represent the experimental measurements, and the plain lines the results of our model (equation (\ref{eq-x-ex-bis}) taking into account the effect of mesh Reynolds number $Re_d$).}
  \label{fig:drag-Red}
\end{figure}

Using our model, the equation (\ref{eq-pressure-final}) with the empirical formulation of \cite{Brundrett}, we obtain a good fitting with our data for solidity $s\leq 0.6$, for different Reynolds numbers. This shows the importance of both the 3D effects and the viscous effects through the local Reynolds number $Re_d$. In particular, for a given solidity, the drag coefficient strongly decreases for increasing $Re_d$, as are plotted in figure \ref{fig:drag-Red}.

We indeed observe a strong effect of the mesh Reynolds number, that also depends on the solidity. We observe a decrease of drag coefficient with increasing $Re_d$ that is well captured by the empirical formulation (equation (\ref{eq-x-ex-bis})). This effect has also been observed for perforated plates by \cite{Bray}. At low $Re_d$, all curves tend towards the value of the drag coefficient of a flat solid plate, i.e. no fluid is passing through the screen and most of it is deviated around. We then note rapid variations for intermediate values of $Re_d$ ($5<Re_d<50$ depending on the solidity), to finally converge towards a constant value at high $Re_d$. This reduction of drag decreases with increasing solidity. As expected, for high solidity ($s=0.9$), there are no effects of $Re_d$, as the flow through the screen is weak. We obtain a similar drag coefficient using the values for the constants $c_1$, $c_2$ and $c_3$ from either \cite{Brundrett} or \cite{Bailey}. Indeed we see that for all solidity and moderate Reynolds number $Re_d=\mathcal{O}(200)$ which corresponds to most of the screens we tested, the difference is small ($1\leq\frac{C_D(Bailey)}{C_D(Brundrett)}\leq 1.05$), and the difference is even lower with increasing Reynolds number $Re_d$. For low Reynolds number $Re_d=\mathcal{O}(5)$, the difference is however higher.

To further check the validity of our model, we made screens that have the same solidity but different hole size and number, keeping however $Re_d$ constant (see Table \ref{tab:screens-article} for $s=0.24$ (P14, P15 and P16) and $s=0.7$ (P8 and P25)), but still with a periodic distribution. As expected from our model, we do not observe any difference in the drag coefficient within the bounds of the measurement uncertainty, demonstrating that for these regular screens, the friction coefficient depends only on $s$ and $Re_d$.

\subsection{Drag coefficient for low angle of attack}\label{sec:results-2}

\begin{figure}
     \centering
     \hspace{-0.5cm}
     \begin{subfigure}[b]{0.45\textwidth}
         \centering
         \includegraphics[width=6.3cm,trim = 0cm 0cm 0cm 0cm, clip]{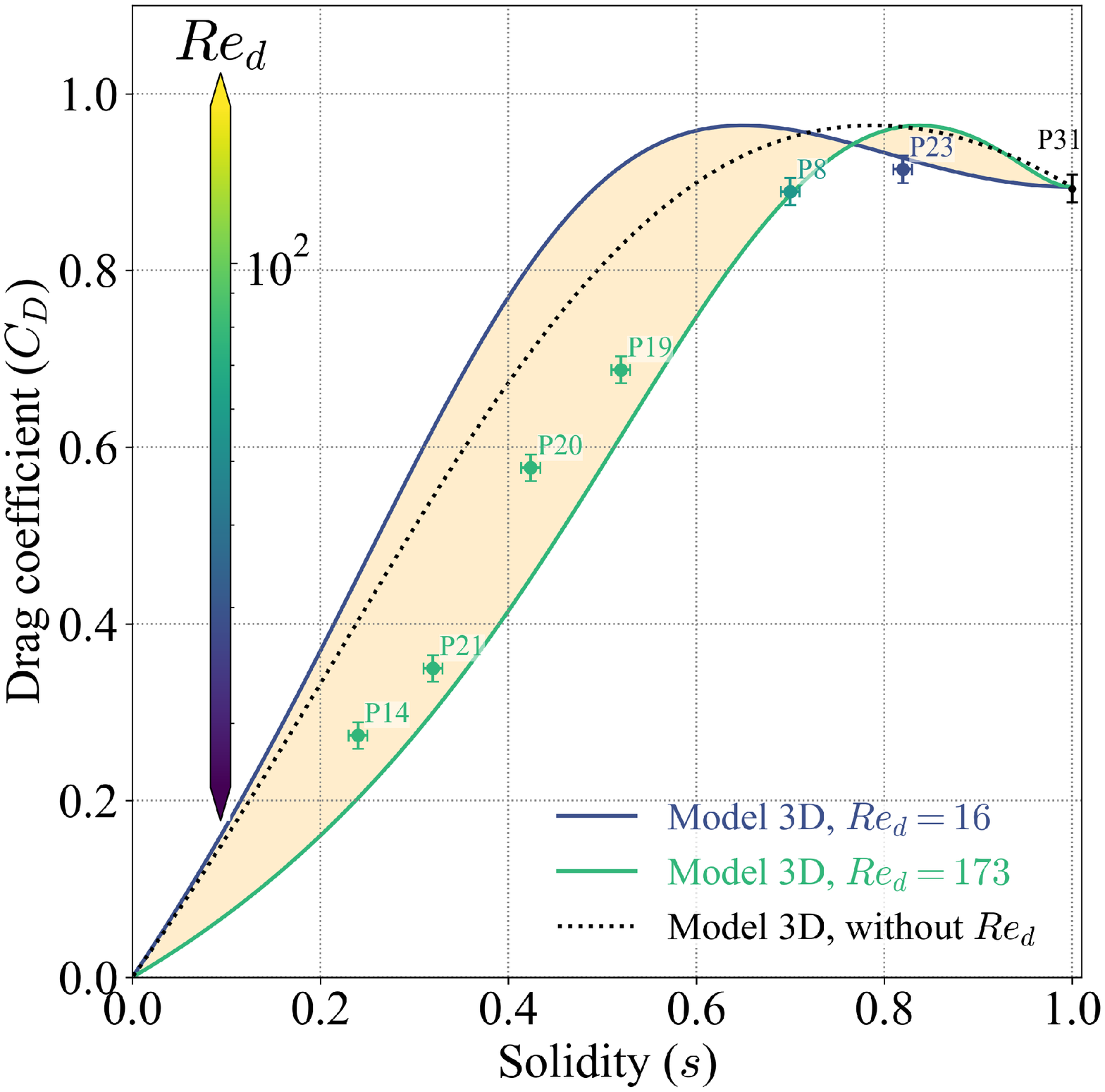}
         \caption{Drag coefficient at $\beta = 65^\circ$.}
         \label{fig:angle60}
     \end{subfigure}
     \hspace{0.4cm}
     \begin{subfigure}[b]{0.45\textwidth}
         \centering
         \includegraphics[width=6.3cm,trim = 0cm 0cm 0cm 0cm, clip]{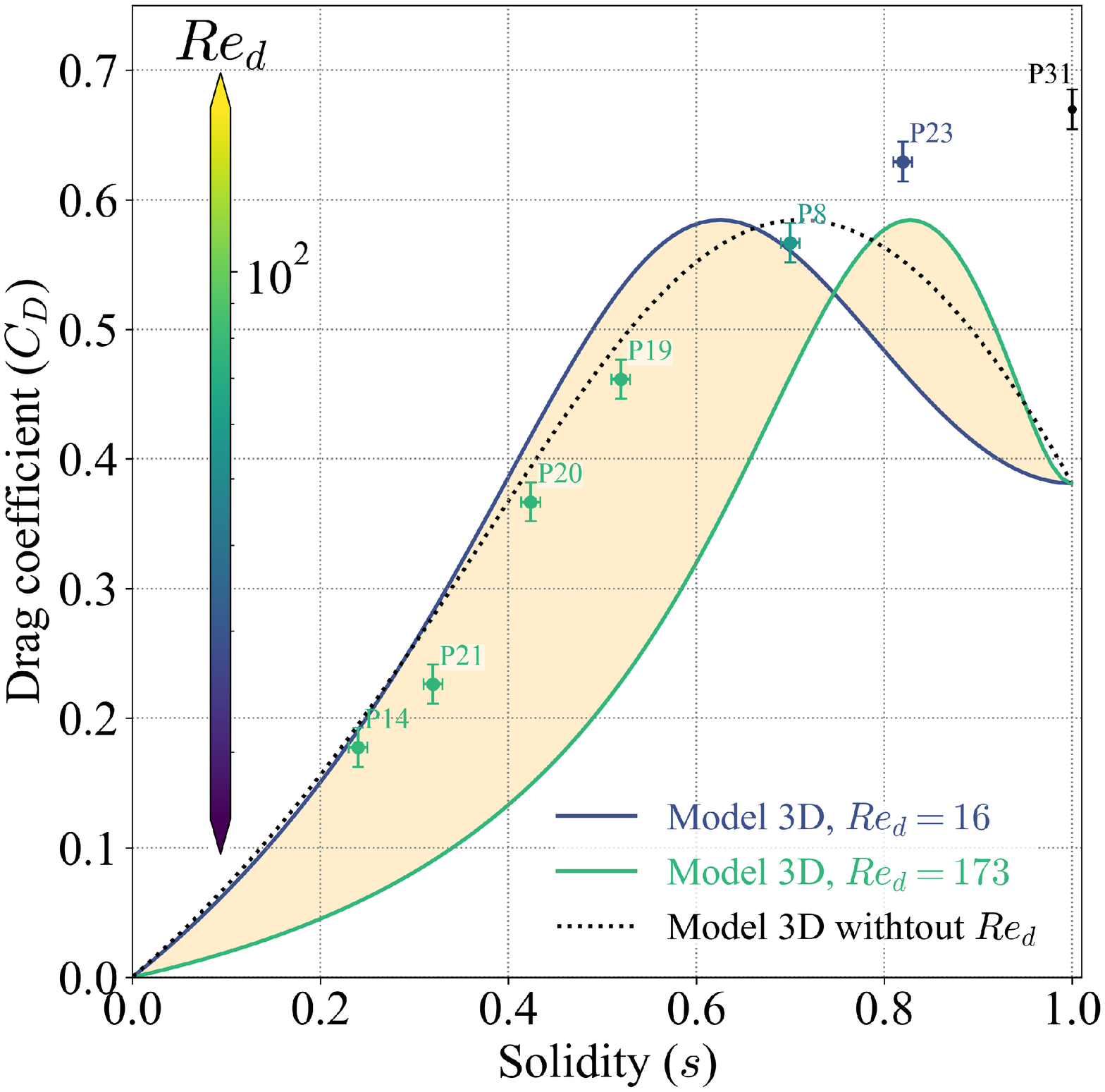}
         \caption{Drag coefficient at $\beta = 43^\circ$.}
         \label{fig:angle40}
     \end{subfigure}
     \caption{Comparison between the three-dimensional model and our experiments for the prediction of the drag coefficient at two angles of inclination for various porous square screens (a) $\beta = 65^\circ$ (b) $\beta = 43^\circ$ using equation (\ref{eq-x-ex}). The black dotted line is the theoretical result obtained with equation (\ref{eq-x-ex-bis}).}
     \label{fig:force-drag-angle}
\end{figure}

We then vary the orientation angle of the screen in the flow $\beta$. The experimental results are compared with our model for the two angles $\beta=65^\circ$ and $\beta=43^\circ$ on figure \ref{fig:force-drag-angle}.
As for panels at a normal incidence, the drag coefficient increases with increasing solidity. Our model is in good agreement with the experimental data at $\beta=65^\circ$ (figure \ref{fig:angle60}). However, it underestimates the drag at orientations further from the normal incidence ($\beta=43^\circ$ Fig \ref{fig:angle40}). In the case of an inclined porous screen composed of fibers, the effective solidity should increase as the angle of attack decreases. The use of this effective solidity would result in shifting our data closer to the green curve ($Re_d = 173$) in figure \ref{fig:angle40}.
We finally observe that both the slope of the linear part and the final value at $s=1$ depend on the angle and that the drag coefficient decreases with decreasing angle.

In addition, our model predicts a maximum of the drag coefficient at high solidity, as shown in figure \ref{fig:drag-solidity-exp} (or \ref{fig:drag-solidity-exp-final}) and more pronounced in figure \ref{fig:force-drag-angle}. For a square screen normal to the flow, several of the porous screens have indeed (on average) a higher drag coefficient than a solid screen (e.g. for P3, P5, P8, P9, P22, P23, P24 and P25). However, due to the uncertainty and the interference drag with the frame used for the support of the screens (see Appendix \ref{appC}), the difference is not significant enough to draw a clear conclusion. We do not observe such a maximum in our measurements for inclined plates at an angle $43\degree$. However, for an angle of $65\degree$ the drag coefficient of the screen $P23$ is on average higher than the solid screen. It is also possible that in our experiments we were outside the flow regime for such a non-monotonic behavior in the drag coefficient.

\subsection{Flow visualisation}\label{sec:results-3}

\begin{figure}
     \centering
     \begin{subfigure}[b]{0.32\textwidth}
         \centering
         \includegraphics[width=4.5cm,trim = 0cm 0cm 0cm 0cm, clip]{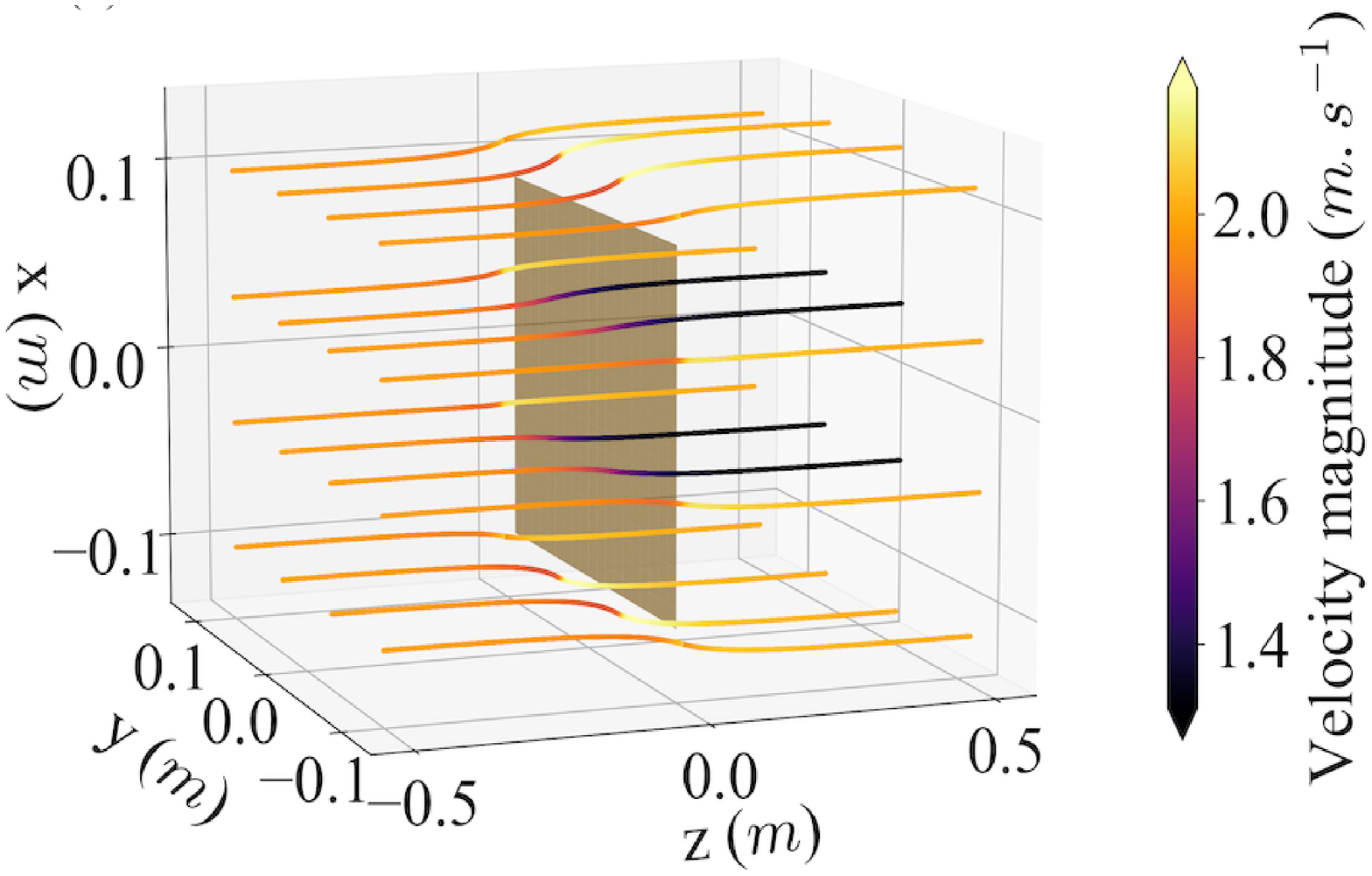}
         \caption{$s=0.3$ and $\beta = 90\degree$}
         \label{fig:30}
     \end{subfigure}
     \hfill
     \begin{subfigure}[b]{0.32\textwidth}
         \centering
         \includegraphics[width=4.5cm,trim = 0cm 0cm 0cm 0cm, clip]{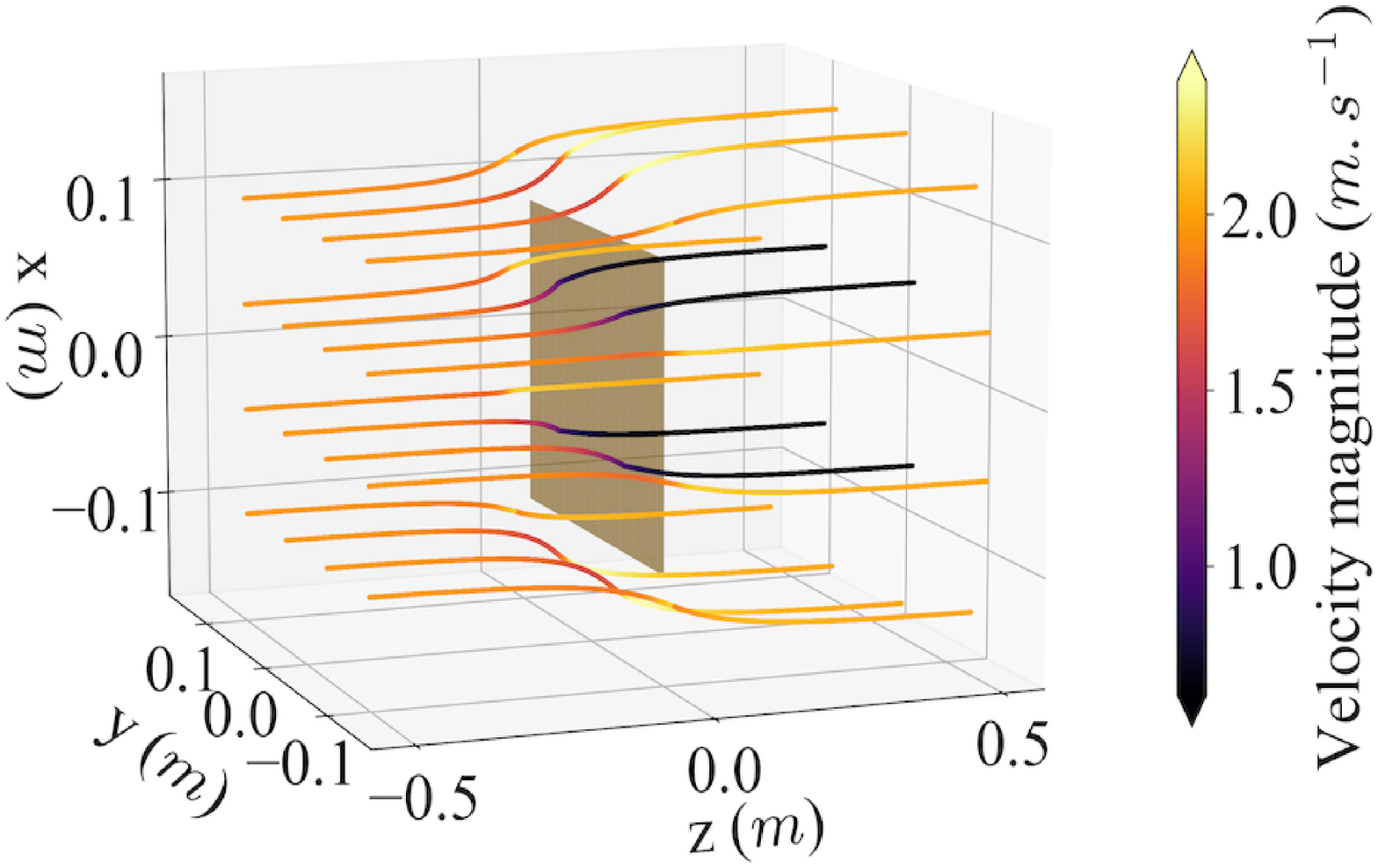}
         \caption{$s=0.6$ and $\beta = 90\degree$}
         \label{fig:60}
     \end{subfigure}
     \hfill
     \begin{subfigure}[b]{0.32\textwidth}
         \centering
         \includegraphics[width=4.5cm,trim = 0cm 0cm 0cm 0cm, clip]{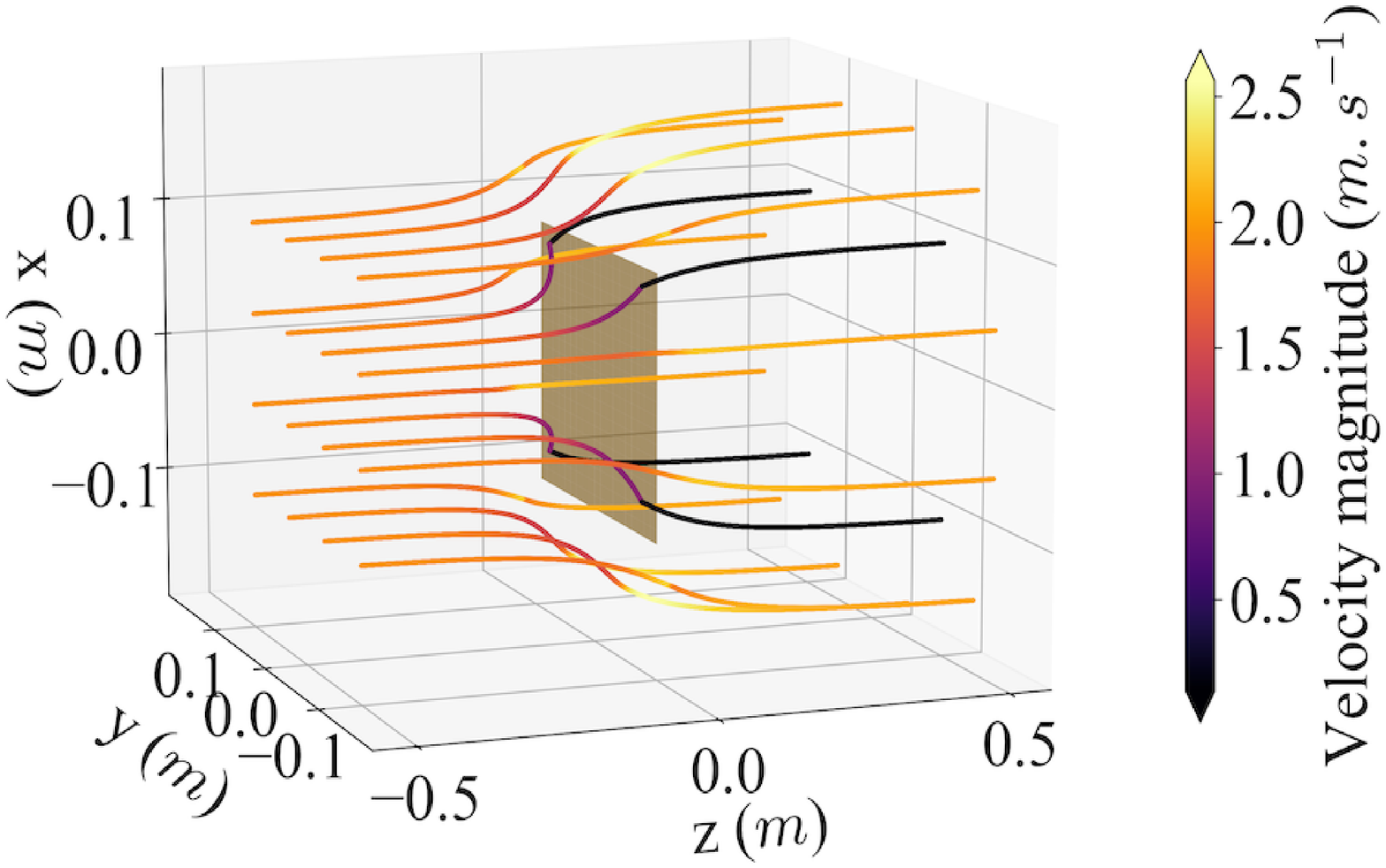}
         \caption{$s=0.9$ and $\beta = 90\degree$}
         \label{fig:90}
     \end{subfigure}
     \hfill
          \begin{subfigure}[b]{0.32\textwidth}
         \centering
         \includegraphics[width=4.5cm,trim = 0cm 0cm 0cm 0cm, clip]{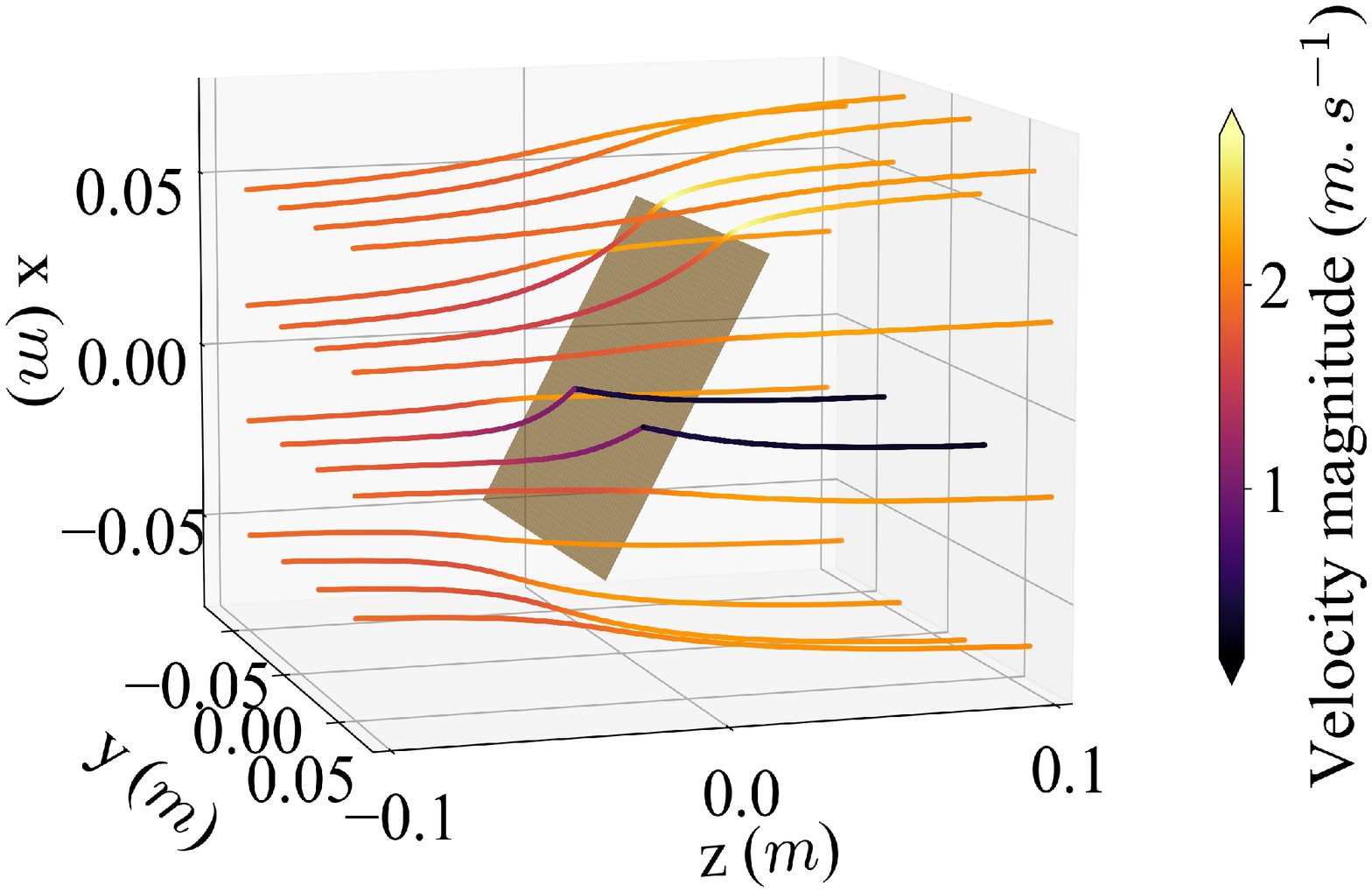}
         \caption{$s=0.8$ and $\beta = 60\degree$}
         \label{fig:80-2}
     \end{subfigure}
     \hfill
     \begin{subfigure}[b]{0.32\textwidth}
         \centering
         \includegraphics[width=4.5cm,trim = 0cm 0cm 0cm 0cm, clip]{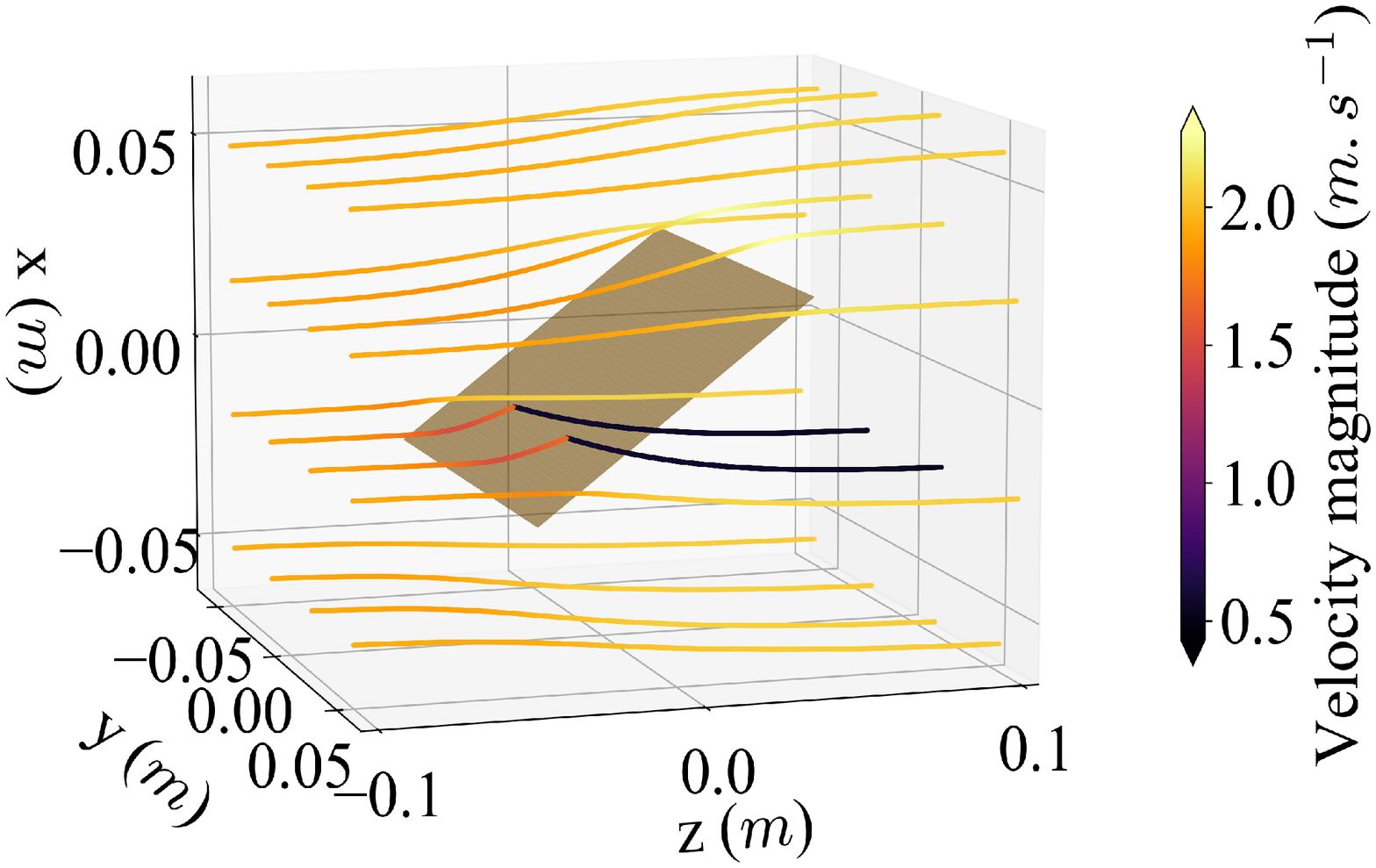}
         \caption{$s=0.8$ and $\beta = 30\degree$}
         \label{fig:80-1}
     \end{subfigure}
     \hfill
     \begin{subfigure}[b]{0.32\textwidth}
         \centering
         \includegraphics[width=4.5cm,trim = 0cm 0cm 0cm 0cm, clip]{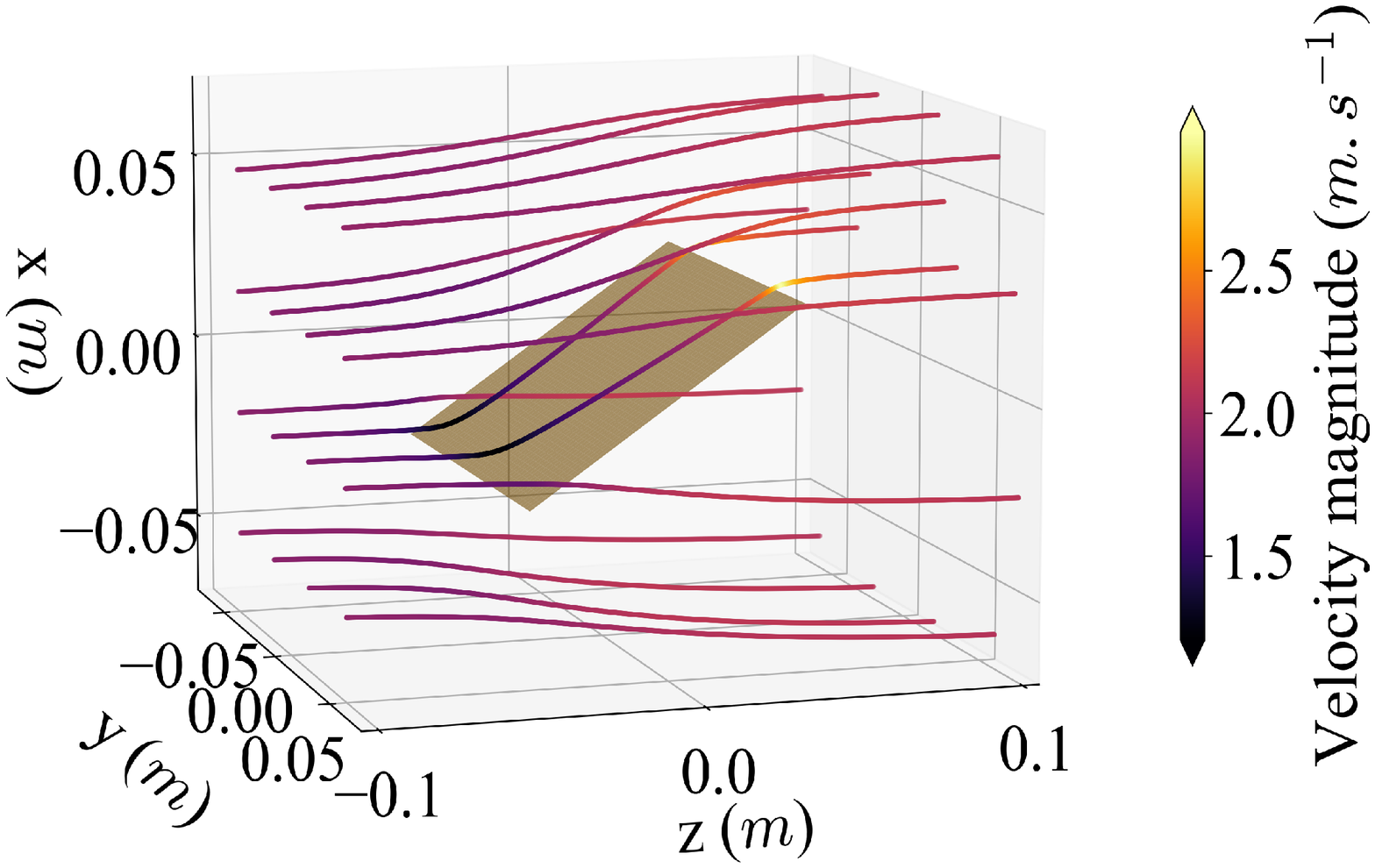}
         \caption{$s=1$ and $\beta = 30\degree$}
         \label{fig:100}
     \end{subfigure}
        \caption{Streamlines and velocity magnitude obtained with the three-dimensional model for a square screen normal and inclined to a flow at $v_0=2.0\hspace{0.1cm}m.s^{-1}$ for different solidities and the pressure jump defined in equation $\ref{eq-pressure}$ (thus without dependency on the mesh Reynolds number $Re_d$).}
        \label{fig:three graphs}
\end{figure}

We can gain some insights into those behaviours by plotting the streamlines and velocity magnitude of the flow around and through the screen with our model (figure \ref{fig:three graphs}).
For a screen normal to the flow, we observe that as the solidity increases, a larger part of the flow is deviated around the screen and that the flow is strongly slowed down in front of the screen, consistently with the measured increase of drag coefficient (figure \ref{fig:30}-\ref{fig:90}). When the orientation angle is increased away for the normal incidence, the deviation of the streamlines is less important and the velocity is slightly higher (figure \ref{fig:80-2}-\ref{fig:80-1}). For the extreme case of a solid plate in figure \ref{fig:100}, two streamlines are deviated along the plate, the flow slows down first and increases again with a peak value above the plate as the fluid particle leaves it and is re-entrained in the surrounding flow. 
However, if the asymptotic behavior of the model may provide some indications about the global flow and the aerodynamic forces, it is expected to be outside of the assumptions of the model as discussed in the next section.

\begin{figure}
     \centering
     \begin{subfigure}[b]{0.45\textwidth}
         \centering
         \includegraphics[width=6.3cm,trim = 0cm 0cm 0cm 0cm, clip]{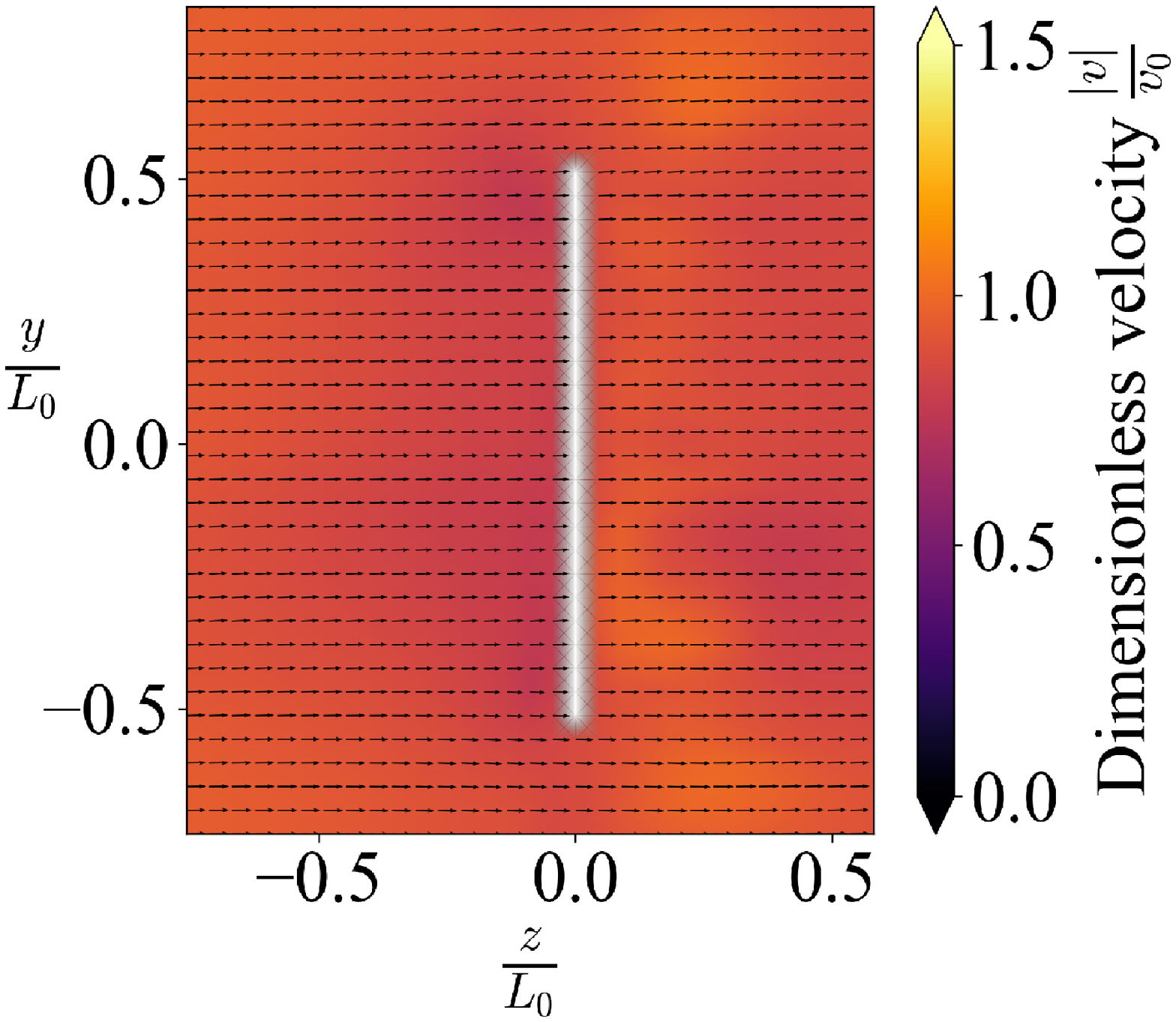}
         \caption{P14, $s=0.24$, experimental.}
         \label{fig:vel-field-14-exp}
     \end{subfigure}
     \begin{subfigure}[b]{0.45\textwidth}
         \centering
         \includegraphics[width=6.3cm,trim = 0cm 0cm 0cm 0cm, clip]{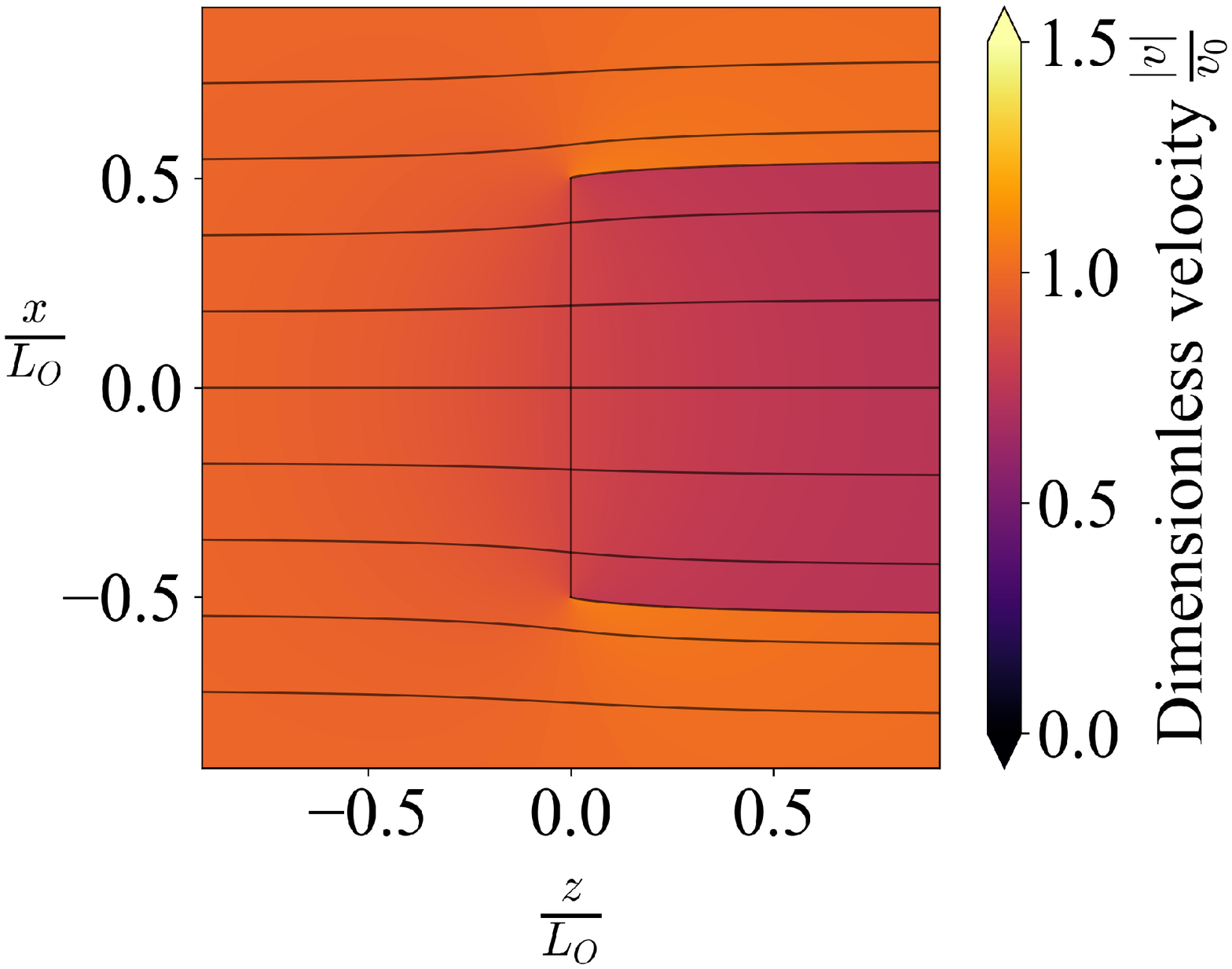}
         \caption{P14, $s=0.24$, theoretical.}
         \label{fig:vel-field-14-theor}
     \end{subfigure}
     \begin{subfigure}[b]{0.45\textwidth}
         \centering
         \includegraphics[width=6.3cm,trim = 0cm 0cm 0cm 0cm, clip]{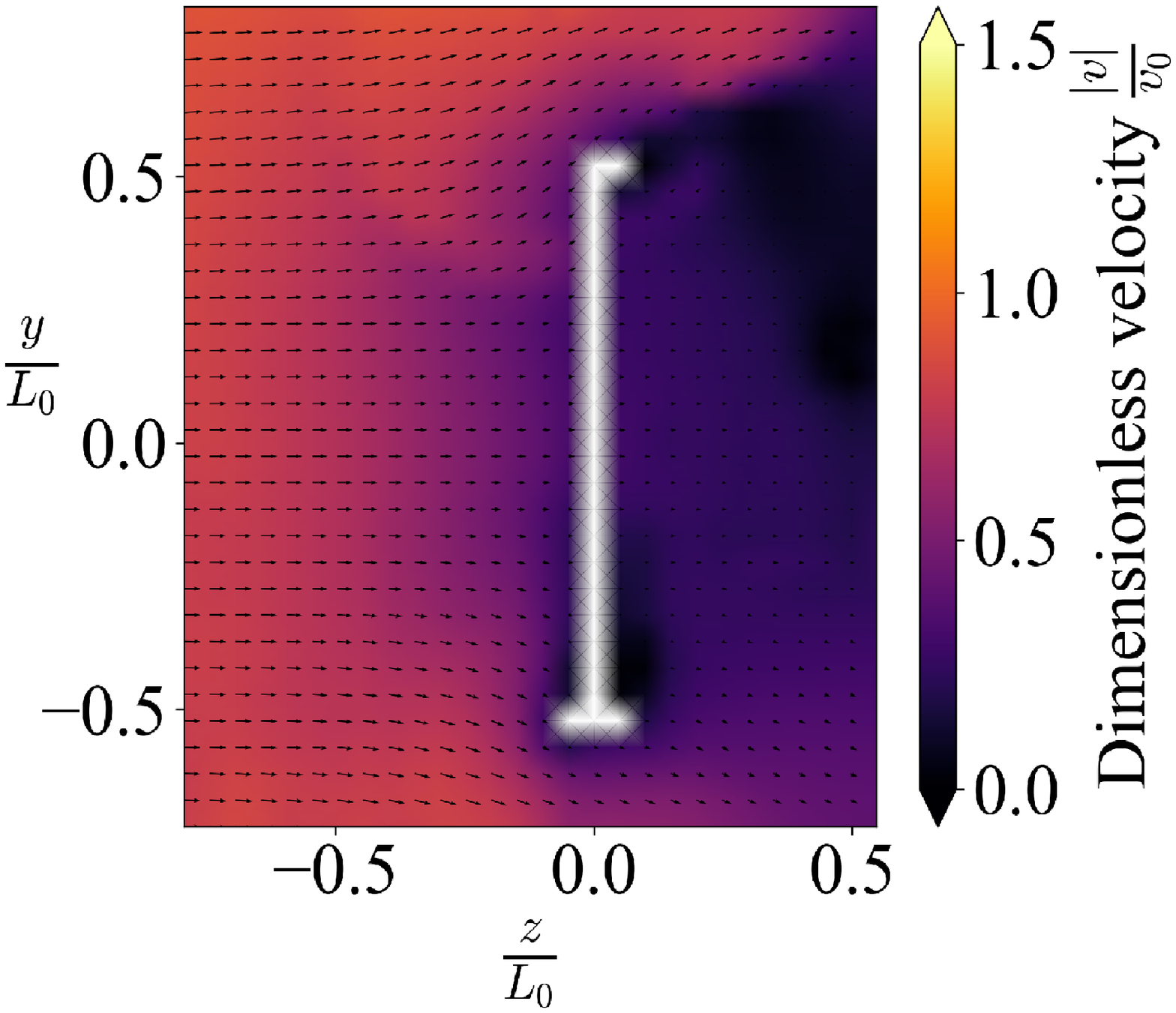}
         \caption{P6, $s=0.61$, experimental.}
         \label{fig:vel-field-6-exp}
     \end{subfigure}
     \begin{subfigure}[b]{0.45\textwidth}
         \centering
         \includegraphics[width=6.3cm,trim = 0cm 0cm 0cm 0cm, clip]{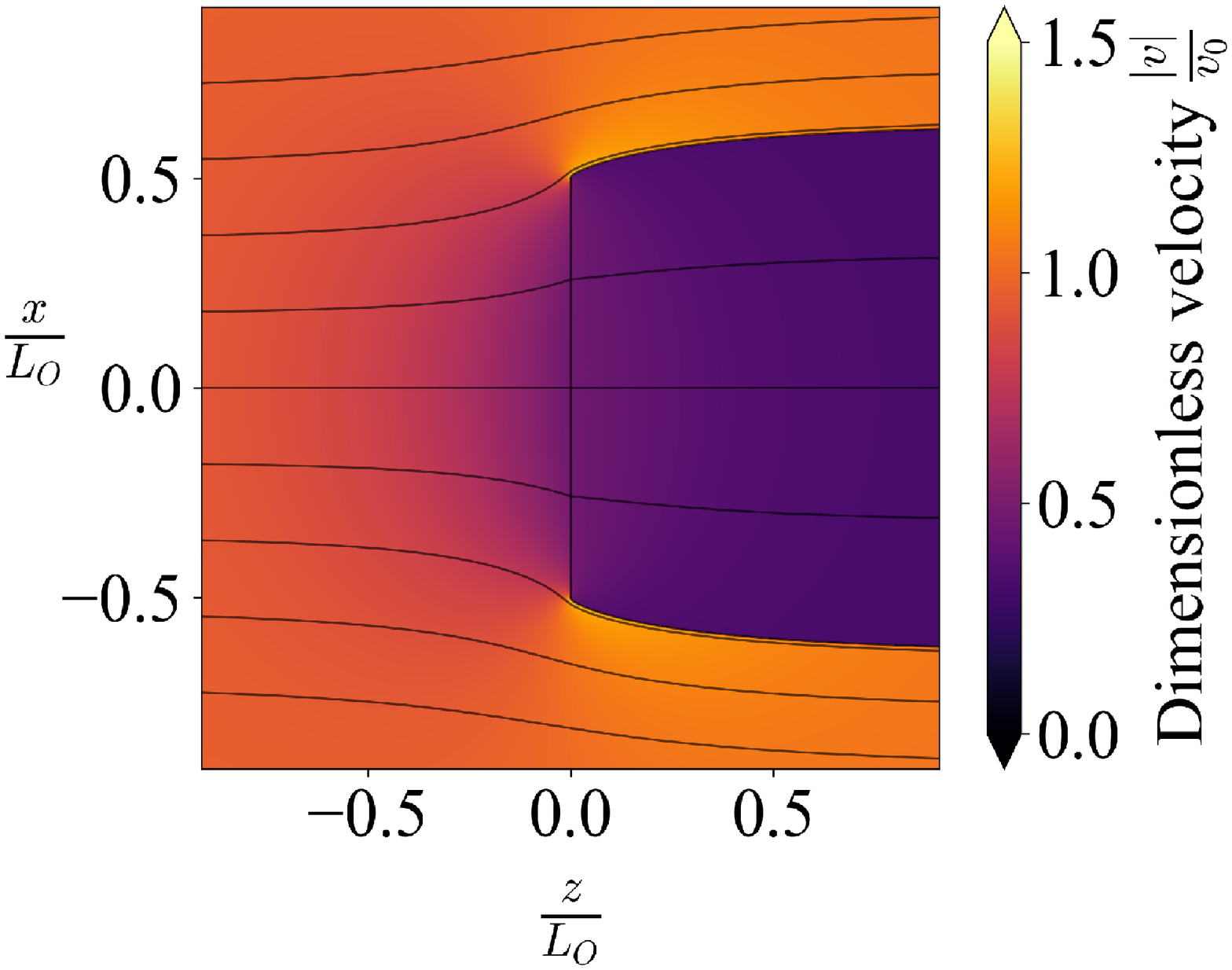}
         \caption{P6, $s=0.61$, theoretical.}
         \label{fig:vel-field-6-theor}
     \end{subfigure}
     \begin{subfigure}[b]{0.45\textwidth}
         \centering
         \includegraphics[width=6.3cm,trim = 0cm 0cm 0cm 0cm, clip]{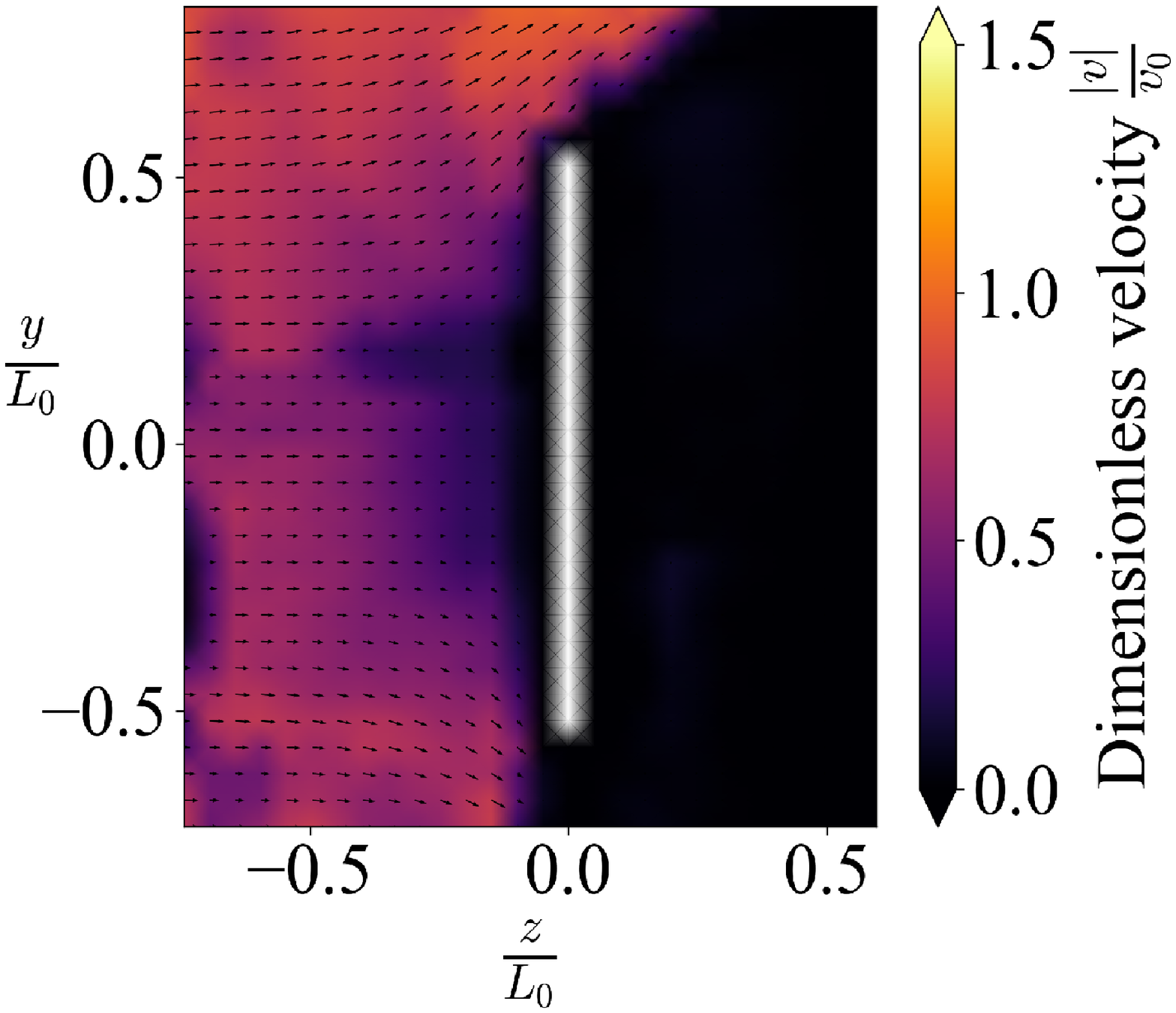}
         \caption{P23, $s=0.82$, experimental.}
         \label{fig:vel-field-26-exp}
     \end{subfigure}
     \begin{subfigure}[b]{0.45\textwidth}
         \centering
         \includegraphics[width=6.3cm,trim = 0cm 0cm 0cm 0cm, clip]{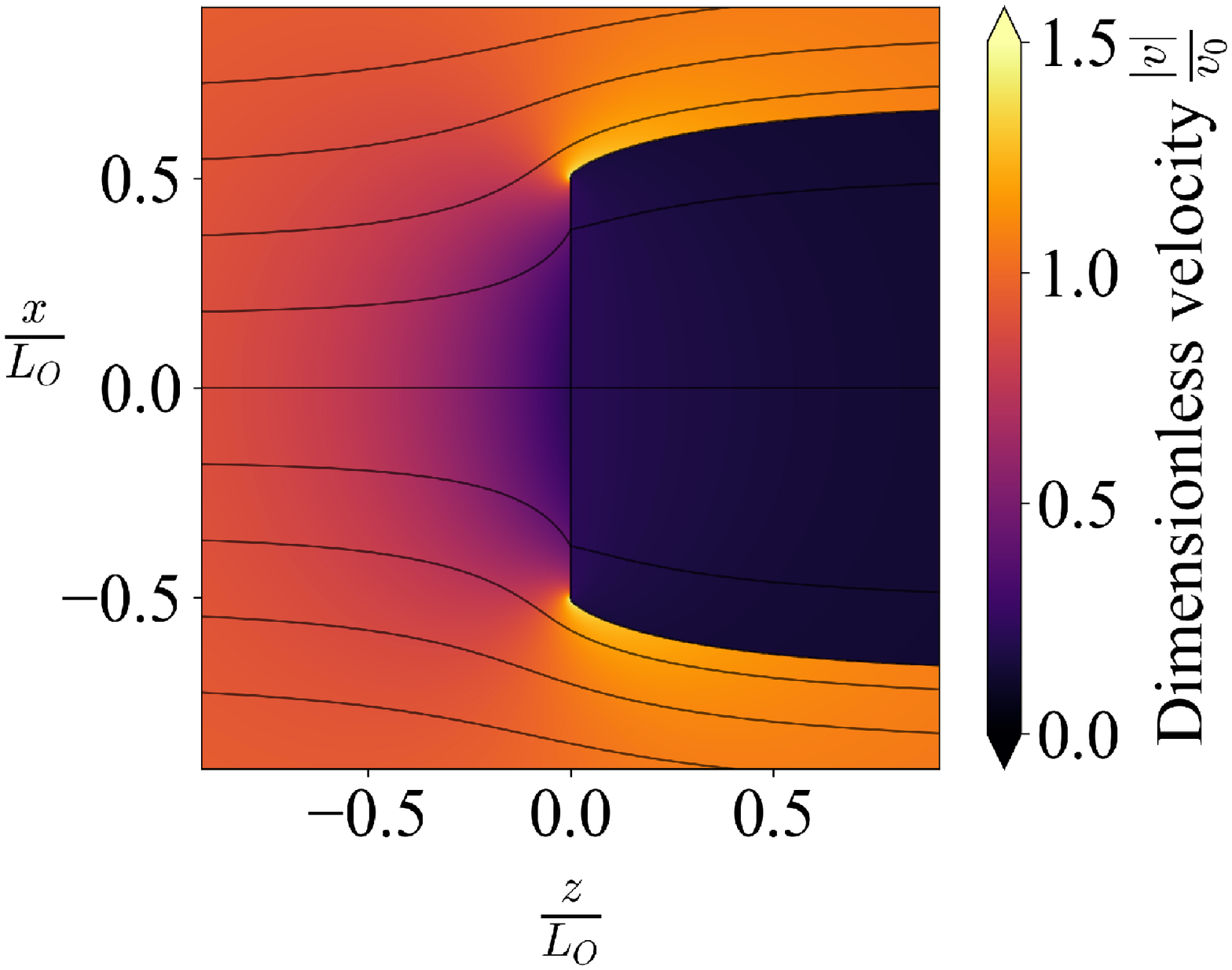}
         \caption{P23, $s=0.82$, theoretical.}
         \label{fig:vel-field-26-theor}
     \end{subfigure}
     \caption{Comparison between the three-dimensional theoretical model and our experiments for the prediction of the velocity field in the $(z,x)$ plane with $y=0$ for three different solidities (low, moderate and high). The theoretical velocity field has been obtained using the theoretical pressure jump law (equation (\ref{eq-pressure}) and (\ref{eq-8})). Uncertainties on the velocities are estimated to be around $0.1\hspace{0.1cm}m.s^{-1}$.}
     \label{fig:vel-field-comparison}
\end{figure}

This analysis is taken further by plotting in figure \ref{fig:vel-field-comparison} both the PIV measurements (as described in section \ref{sec:types_paper})  in the symmetry plane of the screen, as well as the theoretical predictions using the pressure jump equation (\ref{eq-pressure}) for the sake of simplicity (thus not accounting for mesh Reynolds number $Re_d$ effect). As the solidity increases, we can observe a stronger attenuation of the velocity behind the screen, a larger deviation of the flow around the screen, as well as the apparition of a slower region upstream of the screen. We note that the theoretical prediction of the velocity magnitudes are in good agreement with the measured velocities even for the velocity attenuation downstream in the wake. We also observe that the attenuation of the velocity upstream is well captured by the model. However, for high solidity, the width of the wake appears larger in experiments than in the model, which might also be due to the presence of a thicker frame around the mesh in the experiments. Indeed, as observed in the streamlines in figure \ref{fig:streamline-exp}, there are vortices attached to the edges of the frame that may impact both the normal velocity and the shape of the wake. 

\begin{figure}
  \centering
  \includegraphics[width=6.5cm,trim = 0.2cm 0cm 0cm 0cm, clip]{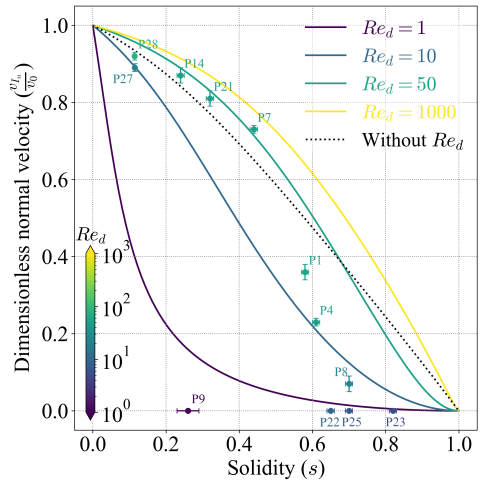}
  \caption{Dimensionless normal velocity $\frac{v_{I_n}}{v_0}$ obtained with equation (\ref{eq-x-ex-bis}) for the solid lines, and with equation (\ref{eq-x-ex}) for the dotted black line. The bullets represent the experimental measurements obtained using a hot wire anemometer.}
  \label{fig:velocity-Red}
\end{figure}

\renewcommand{\arraystretch}{1.2}
\begin{table}
  \begin{center}
\def~{\hphantom{0}}
  \begin{tabular}{c|cccccccccccccc}
      \makecell{Screen number} & P1 & P4 & P7 & P8 & P9 & P14 & P19 & P21 & P22 & P23 & P25 & P27 & P28 & P32\\
      \makecell{Reynolds \\ number $Re_d$} & $53$ & $47$ & $49$ & $24$ & \makecell{$0.16-3.5$ \\ mean $0.75$} & $49$ & $47$ & $47$ & $9.1$ & $5.0$ & $6.7$ & $22$ & $80$ & $80$
  \end{tabular}
 \caption{Reynolds number $Re_d$ calculated with the fiber diameter $d$ as characteristic size and with a velocity $v_0 = 2.84 \hspace{0.1cm}m.s^{-1}$ and a kinematic viscosity $\nu = 15.6 \times 10^{-6}\hspace{0.1cm}m^2.s^{-1}$.}
  \label{tab:Red}
  \end{center}
\end{table}

The proportion of the incoming fluid that goes through the screen is directly given by the dimensionless normal velocity $\frac{v_{I_n}}{v_0}$. In figure \ref{fig:velocity-Red} we plot this ratio $\frac{v_{I_n}}{v_0}$ as a function of the solidity $s$ for different meshes, i.e. different $Re_d$ as presented in table \ref{tab:Red}. The velocity has been measured using a constant temperature anemometer from \textit{Dantec Dynamics} (MiniCTA, with probe 55P11, tungsten wire with diameter $5\hspace{0.1cm}\mu m$ and length $1.25\hspace{0.1cm} mm$, precision of $0.01\hspace{0.1cm}m.s^{-1}$, minimum velocity of $0.20\hspace{0.1cm}m.s^{-1}$). For high mesh Reynolds number and low solidity the model exhibits good agreements with the data. However, we observe a strong effect of $Re_d$ on the normal velocity: notably, for small mesh Reynolds number $Re_d$ the normal velocity drops more rapidly than predicted by the model. The model also overpredicts the normal velocity for high solidity. For screens commonly used for fog harvesting ($Re_d = \order{100}$), the normal velocity in the case where $Re_d$ is taken into account can be up to $21 \%$ greater than the normal velocity in the case where $Re_d$ is not taken into account (according to our theoretical results, this is true for $0.25 < s < 0.75$), and is thus not negligible. These data are coherent with the PIV measurements.

\section{Discussion: asymptotic behaviour at large solidities}\label{asymptotic}

Although the model is built for porous surfaces, it is interesting to explore its asymptotic behavior when the solidity tends to $1$. In this limit, one can question whether a porous description of the surface is still valid.

When the screen tends to a solidity equal to one, we can obtain an expression of the drag coefficient. Taking the equation (\ref{eq-x-ex-bis}), dividing by $\theta(s)$ which tends to $+\infty$ when $s \to 1$, we have
\begin{equation}\label{eq-asymp-1}
   \left(\omega^2-4\sin^2{(\beta)}\right)^2=0.
\end{equation}

Therefore (excluding the case when $\omega$ is negative which would not corresponds to the type of flow we study in this paper) we obtain
\begin{equation}\label{eq-asymp-2}
   \omega = 2\sin{(\beta)},
\end{equation}
which gives ${v_I}_n=0$ as expected. Introducing this value into the expression of the drag coefficient gives
\begin{equation}\label{eq-asymp-3}
\begin{aligned}
   C_D &={} 2\sin^3{(\beta)}\Big(1-4\gamma_0\Big).
\end{aligned}
\end{equation}

Note that the solid drag coefficient does not depend on the assumption of the pressure loss across the screen through the expression of $\theta(s)f(Re_n,\beta)$ (in equation (\ref{eq-pressure-final})) as expected.

\begin{figure}
  \centerline{\includegraphics[width=13cm, trim = 0cm 0cm 0cm 0cm, clip]{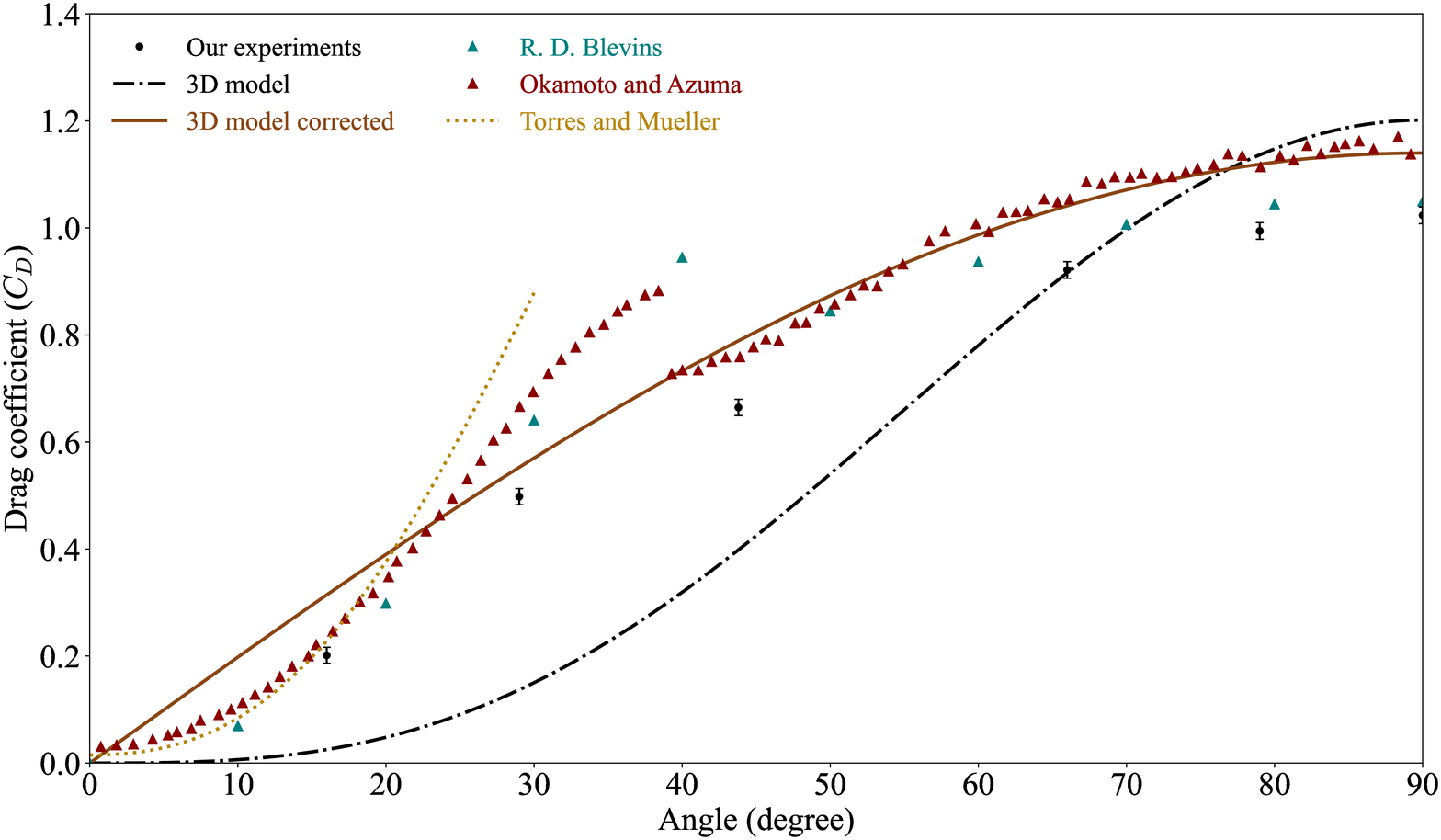}}
  \caption{Comparison between the drag coefficient prediction of the three-dimensional model and the experimental measurement for a square plate in a free flow for different angle of inclination. Data are added from \cite{Blevins}, \cite{Okamoto}, \cite{Torres}. The correction plotted with the solid line is obtained by using the equation \ref{eq-asymp-5} with $C_D^0 = 1.14$, which corresponds to the drag coefficient at angle of attack $90\degree$ from \cite{Torres}.}
\label{fig:flat-plate-drag}
\end{figure}

We measured the drag coefficient as a function of the angle of attack for a square plate at solidity $1$. We plotted the result in figure \ref{fig:flat-plate-drag}, and as we can see, there is a gap between the prediction (\ref{eq-asymp-3}) and the experimental value of the drag coefficient especially at low angle of attack.
Several explanations for this gap can be listed. First, for solid plate, assuming a no slip boundary condition, the tangential component of the velocity on the surface is equal to zero and increases gradually in the boundary layer. Here, this component of the velocity is not equal to zero but takes a value which corresponds to the conservation of the pressure head along the streamlines in region I. Moreover, it is clearly possible that the pressure difference $p_I-p_{II}$ is not always well determined through the assumptions of the model, especially for low angle of attacks (or particular shapes) where detachment of the boundary layer and the formation of separation bubbles can occur which strongly affects the pressure distribution around the surface (\cite{Crompton}). 

For engineering purposes, by the Bernouilli's equation, this over or under-estimate of the pressure distribution may be integrated from the potential model through a correction of the tangential component of the velocity on the surface. Indeed, we can introduce a correction $\sigma$ of the parameter $\gamma_0$ so that the tangential component of the velocity at the surface in the model can be interpreted as a homogenized tangential velocity in the boundary layer due to friction effects with the solid parts of the surface and the suction increase or decrease due to the detachment of the boundary layer. This value should depend on the solidity (since it is well known that the decrease of the solidity can detach the recirculation bubble behind the solid plate as well as the detachment phenomenon of the boundary layer), on the shape of the surface and the inclination $\beta$. In the case of a rectangular plate inclined in a free flow, this consideration can be written down by enforcing the expression of the tangential velocity:
\begin{equation}\label{eq-asymp-4}
\begin{aligned}
   {v_I}^2_t &={} \Big(\omega^2\gamma_0\sigma(s,\beta)+\cos^2{(\beta)}\Big)v_0^2.
\end{aligned}
\end{equation}
instead of equation (\ref{eq-tang-0}).
Although $\sigma$ could be determined directly by comparing the model with the experiments for all solidity and angle, we have observed that taking the value of $\sigma$ at $s=1$ is sufficient to improve the model. As shown on figure \ref{fig:flat-plate-drag}, the dependency on $\beta$ is complicated, particularly at low angle of attack where a discontinuity is observed around $\beta=40\degree$. $\sigma(\beta)$ is chosen to ensure that the drag coefficient (equation \ref{eq-asymp-4}) is equal to the experimental value for $\beta \approx \frac{\pi}{2}$ which gives:
\begin{equation}\label{eq-asymp-5}
\begin{aligned}
   \sigma(\beta) &={} \frac{1}{4\gamma_0}\left(1-\frac{C_D^0}{2\sin^2{(\beta)}}\right),
\end{aligned}
\end{equation}
with $C_D^0$ the drag coefficient of the solid screen at angle of attack $90°$.
The corrected drag coefficient obtained with the correction (\ref{eq-asymp-5}) is plotted on figure \ref{fig:flat-plate-drag} (3D model corrected curve), using $C_D^0$ at $90\degree$ from \cite{Okamoto}. This rescaling can greatly improve the prediction of our model at the other solidities without addition of an independent parameter, as shown in figures \ref{fig:drag-solidity-exp} and \ref{fig:force-drag-angle} for three various $\beta$ (for these figures, the drag coefficient used is $C_D^0 = 0.939$ at $s=1$ and $\beta = 90\degree$ from our experimental data).

Remarkably, this rescaling has no rigorous physical formulation, but it enables us to fit the drag coefficient with respect to the angle of attack shown in figure \ref{fig:flat-plate-drag}.

At this stage, no skin friction drag effects are modeled at the surface, since it is generally very small and neglected. Indeed, an estimate of the skin friction drag coefficient based on the Blasius boundary layer approximation can be done using the parameters of our model (assuming a characteristic length $l$ and a characteristic velocity ${v_I}_t$), leading to:
\begin{equation}\label{eq-asymp-3-1}
\begin{aligned}
   C_f &={} 0.664\cos{(\beta)}s\sqrt{\frac{\nu}{l {v_I}_t}}.
\end{aligned}
\end{equation}
Taking from our experiments ${v_I}_t = v_0\approx 0.1\hspace{0.1cm}m.s^{-1}$, $s=1$, $l=0.1\hspace{0.1cm}m$ and $\beta=0$, we obtain approximately $C_f \approx 0.01$. It is in good agreement with the experimental observation of \cite{Torres}, who found a mean skin friction drag of $0.015$ for flat plates of different shapes including rectangular plates of aspect ratio between $0.5$ and $2$.

Finally, it might be interesting to develop in further studies a more complete model by introducing a tensor rather than a function $\sigma$ in order to specify the influence of the friction with the surface according to the direction for non-homogeneous porous surface (e. g. aligned fibers).

\section{Conclusion}
We have studied both experimentally and theoretically the flow around and through porous screens. 
Our three-dimensional model can be considered as an extension of the model of \cite{Koo} integrating the wake of \cite{Steiros}. In its simplest formulation, the model uses only three types of information, first the solidity, second the macroscopic geometry of the screen and third the Reynolds number based on the fiber diameter. This local Reynolds number $Re_d$ as well as three-dimensional effects have an impact which can be significant on the flow and aerodynamic forces. We performed experiments on more than $30$ porous screens composed of fibers to measure the drag force. The drag coefficient for square porous screens, either normal or with a high angle of attack, show a good agreement between the model and the experiments except at very high solidity for which however the prediction is improved compared to previous models. We show both theoretically and experimentally that for screens of identical solidity, the lower the local Reynolds number based on the fiber diameter is, the higher the drag coefficient is. Moreover, our experiments, supported by our model, suggest that a porous screen at high solidity or very low local Reynolds number can have a higher drag coefficient than a solid screen. For other types of screens, including perforated plates, once the pressure jump law across is known, it can be directly implemented in the model.

If a value of the drag coefficient of the screen at solidity equal to $1$ is known, which is generally easily available, we can then determine a value of the parameter $\sigma$ which is useful for the accuracy of the model for very high solidity. However for low solidity $\sigma$ has nearly no influence on the model and can be set equal to $1$. Our model might be also useful for non-homogeneous porous screen. Indeed, for high solidity perforated square plates, \cite{Bray} found that the drag coefficient depends on the distribution of the perforations, i.e. is slightly higher with outer holes than inner holes. It is worth to emphasize that it should be possible in our model to implement such a surface with non-homogeneous solidity and therefore to try to reproduce such difference in the drag coefficient. However, this consideration goes beyond the scope of this article and will be explored in a further study.

We note that this model may find an application to the wind tunnel blockage correction (\cite{SteirosTurbines}) and turbines modelling (\cite{Ayati}). According to \cite{SteirosTurbines}, the use of a porous plate potential model as the one we use in the present paper improves the blockage correction accuracy for moderate and high solidity compared to other models. However, these models are in two-dimensions (\cite{SteirosTurbines,Ayati}), while the turbine can have a circular three-dimensional structure. Our theoretical and experimental results show that the use of a three-dimensional model may improve the accuracy at moderate and high solidity for the drag compare to the two-dimensional models (see figures \ref{fig:drag-solidity-theor} and \ref{fig:drag-solidity-exp}). Moreover, our experimental results show that the normal velocity $v_{I_n}$ on the screen may be significantly overestimated at high solidity by both 2D and 3D models. Our results with different order of magnitude of the local Reynolds number $Re_d$ highlight the importance of taking into account the particular geometry at the pore scale that can lead to significant discrepancies with the results obtained using the traditional pressure jump law in equation (\ref{eq-pressure}).

Further study should focus (1) first on more complex shape with curvature, a straightforward formulation would be to use several small rectangular plates like the well-known panel method (adapting therefore the velocity potential $\phi$), (2) second on the pressure and velocity distribution around the porous screen, as there are very few - if any - data for free three-dimensional flow. We stresses again that our model assumes a steady wake, and thus is not applicable in the presence of vortex shedding that occurs at high solidity for a certain range of Reynolds number.

\vspace{1cm}

\noindent\textbf{Declaration of interests.} The authors report no conflict of interest.

\newpage
\appendix

\section{}\label{appA}
This appendix contains the details of the calculation of the velocities for a rectangular screen inclined with an angle $\beta$ (with respect to the $z$-axis) in a laminar flow.

Using the previous notations in the section 2, the velocity potential in region I is
\begin{equation}\label{eq-appB-0}
\begin{aligned}
    \phi_I(x,y,z) & ={} v_0z + c + \phi(x,y,z) \\
    & ={} v_0z + c - \frac{1}{4\pi}\iint_{\mathcal{S}_p}\frac{\Omega(u,v)\mathrm{d}u\mathrm{d}v}{\sqrt{(x-v)^2+(y-u\sin{(\beta)})^2+(z-u\cos{(\beta)})^2}}.
\end{aligned}
\end{equation}

\subsection{Normal component of the velocity}

From the velocity potential we deduce the velocity in region I:
\begin{equation}\label{eq-appB-1}
\begin{aligned}
     \mathbf{v_I}(x,y,z) ={} & \left(v_0 + \frac{\partial \phi}{\partial z}(x,y,z)\right)\mathbf{e_z} + \frac{\partial \phi}{\partial y}(x,y,z)\mathbf{e_y}  + \frac{\partial \phi}{\partial x}(x,y,z)\mathbf{e_x}.
\end{aligned}
\end{equation}

The vector normal to the surface with an angle $\beta$ is
\begin{equation}\label{eq-appB-2}
   \mathbf{e_n} = -\cos{(\beta)}\mathbf{e_y}+\sin{(\beta)}\mathbf{e_z}.
\end{equation}

Therefore the component of the velocity normal to the surface is
\begin{equation}\label{eq-appB-3}
\begin{aligned}
     \mathbf{{v_I}_n}(x,y,z) ={} & \left(\sin{(\beta)}\left(v_0 + \frac{\partial \phi}{\partial z}(x,y,z)\right) -\cos{(\beta)} \frac{\partial \phi}{\partial y}(x,y,z)\right)\mathbf{e_n},
\end{aligned}
\end{equation}
with
\begin{equation}\label{eq-appB-4}
\begin{aligned}
     \frac{\partial \phi}{\partial z}(x,y,z) ={} & -\frac{1}{4\pi}\iint_{\mathcal{S}_p} \frac{(z-u\cos{(\beta)})\Omega(u,v)}{\left((x-v)^2+(y-u\sin{(\beta)})^2+(z-u\cos{(\beta)})^2\right)^{\frac{3}{2}}}\mathrm{d}u\mathrm{d}v,
\end{aligned}
\end{equation}
\begin{equation}\label{eq-appB-5}
\begin{aligned}
     \frac{\partial \phi}{\partial y}(x,y,z) ={} & -\frac{1}{4\pi}\iint_{\mathcal{S}_p} \frac{(y-u\sin{(\beta)})\Omega(u,v)}{\left((x-v)^2+(y-u\sin{(\beta)})^2+(z-u\cos{(\beta)})^2\right)^{\frac{3}{2}}}\mathrm{d}u\mathrm{d}v.
\end{aligned}
\end{equation}

The normal velocity magnitude is therefore  
\begin{equation}\label{eq-appB-6}
\begin{aligned}
    {v_I}_n(x,y,z) ={} & \sin{(\beta)}v_0 -\frac{1}{4\pi}I(x,y,z),
\end{aligned}
\end{equation}
with
\begin{equation}\label{eq-appB-6}
\begin{aligned}
    I(x,y,z) ={} & \iint_{\mathcal{S}_p} \frac{(\sin{(\beta)}z - \cos{(\beta)}y)\Omega(u,v)}{\left((x-v)^2+(y-u\sin{(\beta)})^2+(z-u\cos{(\beta)})^2\right)^{\frac{3}{2}}}\mathrm{d}u\mathrm{d}v.
\end{aligned}
\end{equation}

When this component is evaluated on the surface, the integral becomes singular at the position of this evaluation. We have to calculate the value of this singularity. To simplify the calculation we introduce two parameters :
\begin{equation}\label{eq-appB-7}
\begin{aligned}
    h ={} & y\sin{(\beta)} + z\cos{(\beta)}, \\
    t ={} & y\cos{(\beta)} - z\sin{(\beta)}.
\end{aligned}
\end{equation}
Thus we have inversely
\begin{equation}\label{eq-appB-8}
\begin{aligned}
    y ={} & h\sin{(\beta)} + t\cos{(\beta)}, \\
    z ={} & h\cos{(\beta)} - t\sin{(\beta)}.
\end{aligned}
\end{equation}

Denoting $\Tilde{I}$ the integral $I$ with the new parameters and without the source strength $\Omega$ assumed to be continuous. We obtain
\begin{equation}\label{eq-appB-9}
\begin{aligned}
   \Tilde{I}(x,h,t) ={} & -\int_{v_a}^{v_b}\int_{u_a}^{u_b} \frac{t}{\left((x-v)^2+t^2+(h-u)^2\right)^{\frac{3}{2}}}\mathrm{d}u\mathrm{d}v\\
   ={} & F_{u_a,v_a}(x,h,t) - F_{u_b,v_a}(x,h,t) - F_{u_a,v_b}(x,h,t) + F_{u_b,v_b}(x,h,t),
\end{aligned}
\end{equation}
with 
\begin{equation}\label{eq-appB-10}
\begin{aligned}
    F_{u_a,v_a}(x,h,t) ={} & \arctan{\left(\frac{(h-u_a)(v_a-x)}{t\sqrt{(x-v_a)^2+t^2+(h-u_a)^2}}\right)}.
\end{aligned}
\end{equation}

We can introduce now
\begin{equation}\label{eq-appB-11}
\begin{aligned}
    y ={} & w\sin{(\beta)}, \\
    z ={} & (w+\epsilon)\cos{(\beta)}.
\end{aligned}
\end{equation}

The parameters become
\begin{equation}\label{eq-appB-11-bis}
\begin{aligned}
    h ={} & w, \\
    t ={} & -\epsilon\cos{(\beta)}\sin{(\beta)}.
\end{aligned}
\end{equation}

At the surface $\epsilon \to 0^{\pm}$ ($\pm$ depending on the direction from which we approach the surface, upstream or downstream), denoting $\Tilde{\Tilde{I}}$ the integral $I$ with the new parameters $w$ we have 
\begin{equation}\label{eq-appB-12-0}
\begin{aligned}
    {v_I}_n(x,w) ={} & v_n^-,
\end{aligned}
\end{equation}
\begin{equation}\label{eq-appB-12-1}
\begin{aligned}
    {v_{II}}_n(x,w) ={} & E v_n^+,
\end{aligned}
\end{equation}
with
\begin{equation}\label{eq-appB-12}
\begin{aligned}
    v_n^{\pm} ={} & \sin{(\beta)}v_0 -\frac{1}{4\pi} \lim_{\epsilon \to 0^{\pm}} \Tilde{\Tilde{I}}(x,w).
\end{aligned}
\end{equation}

Whatever are the constant integral bounds, we have the following limit
\begin{equation}\label{eq-appB-13}
\begin{aligned}
    \lim_{\epsilon \to 0^{\pm}} \Tilde{\Tilde{I}}(x,w) ={} & \pm 2\pi \Omega(x,w).
\end{aligned}
\end{equation}

Therefore the normal component of the velocity at the position $(x,w)$ on the porous surface is
\begin{equation}\label{eq-appB-14}
\begin{aligned}
    {v_I}_n(x,w) ={} & \sin{(\beta)}v_0 -  \frac{1}{2}\Omega(x,w),
\end{aligned}
\end{equation}
and
\begin{equation}\label{eq-appB-14-bis}
\begin{aligned}
    {v_{II}}_n(x,w) ={} & E\left(\sin{(\beta)}v_0 + \frac{1}{2}\Omega(x,w)\right).
\end{aligned}
\end{equation}

\subsection{Tangential component of the velocity}

We define the tangential vector on the porous surface as
\begin{equation}\label{eq-vect-t0}
\begin{aligned}
   \mathbf{{t}} &= \mathbf{{t_1}}  +  \mathbf{{t_2}},
\end{aligned}
\end{equation}
with
\begin{equation}\label{eq-vect-t1}
\begin{aligned}
   \mathbf{{t_1}} &= \mathbf{e_x},
\end{aligned}
\end{equation}
\begin{equation}\label{eq-vect-t2}
\begin{aligned}
   \mathbf{{t_2}} &= \sin{(\beta)}\mathbf{e_y}+ \cos{(\beta)}\mathbf{e_z}.
\end{aligned}
\end{equation}

Thus the magnitude of the tangential component of the velocity in region I is
\begin{equation}\label{eq-tang-1}
\begin{aligned}
   {v_I}_t(x,y,z) &= \left\|(\mathbf{v_I}.\mathbf{{t_1}})\mathbf{{t_1}} + (\mathbf{v_I}.\mathbf{{t_2}})\mathbf{{t_2}}\right\|\\
   &= \left(\left(\frac{\partial \phi}{\partial x}\right)^2 + \left(\sin{(\beta)}\frac{\partial \phi}{\partial y} + \cos{(\beta)}\left(v_0 + \frac{\partial \phi}{\partial z}\right)\right)^2\right)^{\frac{1}{2}}.
\end{aligned}
\end{equation}

If we consider a constant source strength $\Omega$ then we can write the tangential velocity at the point $(x,w)$ on the surface with $w$ defined in \ref{eq-appB-11} as
\begin{equation}\label{eq-tang-2}
\begin{aligned}
   {v_I}_t(x,w) &= \left(\Omega^2 \mathcal{I}_x^2(x,w) + \left(\Omega\left(\sin{(\beta)}\mathcal{I}_y(x,w)+\cos{(\beta)}\mathcal{I}_z(x,w)\right) + \cos{(\beta)}v_0\right)^2\right)^{\frac{1}{2}}.
\end{aligned}
\end{equation}

For simplicity, we use the root mean square of the magnitude of the tangential component of the velocity on the surface. We have
\begin{equation}\label{eq-tang-3}
\begin{aligned}
   {v_I}_t &= \frac{1}{\sqrt{S_p}}\left(\iint_{\mathcal{S}_p}  {v_I}_t^2(x,w) \mathrm{d}x\mathrm{d}w\right)^{\frac{1}{2}} \\
   &= \left(\Omega^2 \gamma(\beta) + \Omega \alpha(\beta)v_0 + v_0^2\cos^2{(\beta)} \right)^{\frac{1}{2}},
\end{aligned}
\end{equation}
with
\begin{equation}\label{eq-tang-4}
\begin{aligned}
   \gamma(\beta) &= \frac{1}{S_p}\iint_{\mathcal{S}_p}  \mathcal{I}_x^2(x,w) + \left(\sin{(\beta)} \mathcal{I}_y(x,w) + \cos{(\beta)} \mathcal{I}_z(x,w) \right)^2 \mathrm{d}x\mathrm{d}w,
\end{aligned}
\end{equation}
and
\begin{equation}\label{eq-tang-5}
\begin{aligned}
   \alpha(\beta) &= \frac{1}{S_p}\iint_{\mathcal{S}_p} 2\cos{(\beta)} \left(\sin{(\beta)} \mathcal{I}_y(x,w) + \cos{(\beta)} \mathcal{I}_z(x,w) \right) \mathrm{d}x\mathrm{d}w,
\end{aligned}
\end{equation}
with the following integrals:

\begin{equation}\label{eq-tang-6}
\begin{aligned}
   \mathcal{I}_x(x,w) &= -\frac{1}{4\pi}\iint_{\mathcal{S}_p} \frac{x-v}{\left(\left(x-v\right)^2+\left(w-u\right)^2\right)^{\frac{3}{2}}}
  \mathrm{d}u \mathrm{d}v,
\end{aligned}
\end{equation}
\begin{equation}\label{eq-tang-7}
\begin{aligned}
   \mathcal{I}_y(x,w) &= -\frac{1}{4\pi}\iint_{\mathcal{S}_p} \frac{(w-u)\sin{(\beta)}}{\left(\left(x-v\right)^2+\left(w-u\right)^2\right)^{\frac{3}{2}}}
  \mathrm{d}u \mathrm{d}v,
\end{aligned}
\end{equation}
\begin{equation}\label{eq-tang-8}
\begin{aligned}
        \mathcal{I}_z(x,w) &= -\frac{1}{4\pi}\iint_{\mathcal{S}_p} \frac{(w-u)\cos{(\beta)}}{\left(\left(x-v\right)^2+\left(w-u\right)^2\right)^{\frac{3}{2}}}
  \mathrm{d}u \mathrm{d}v.
\end{aligned}
\end{equation}

There is no particular difficulty to compute the integrals, if we define a function $F_{a,b}: (x,w)\mapsto F_{a,b}(x,w)$ as well as $G_{a,b}: (x,w)\mapsto G_{a,b}(x,w)$ where and $\vert a \vert \geq \vert x\vert$, $\vert b \vert \geq \vert w\vert$, 
\begin{equation}\label{eq-tang-9}
\begin{aligned}
        F_{a,b}(x,w) &= \frac{1}{4\pi} \ln{\left(\sqrt{\left(x-a\right)^2+\left(w-b\right)^2} + w - b\right)},
\end{aligned}
\end{equation}
and
\begin{equation}\label{eq-tang-10}
\begin{aligned}
        G_{a,b}(x,w) &= \frac{1}{4\pi} \ln{\left(\sqrt{\left(x-a\right)^2+\left(w-b\right)^2} + x - a\right)},
\end{aligned}
\end{equation}
then, if we integrate over the rectangular domain $[v_a,v_b]\times[u_a,u_b]$ we have
\begin{equation}\label{eq-tang-11}
\begin{aligned}
   \mathcal{I}_x(x,w) ={}& F_{v_a,u_a}(x,w) - F_{v_a,u_b}(x,w) - F_{v_b,u_a}(x,w) + F_{v_b,u_b}(x,w),
\end{aligned}
\end{equation}
\begin{equation}\label{eq-tang-12}
\begin{aligned}
   \mathcal{I}_y(x,w) ={}& \sin{(\beta)}\left(G_{v_a,u_a}(x,w) - G_{v_a,u_b}(x,w)\right. \\
   & \left.- G_{v_b,u_a}(x,w) + G_{v_b,u_b}(x,w)\right),
\end{aligned}
\end{equation}
\begin{equation}\label{eq-tang-13}
\begin{aligned}
   \mathcal{I}_z(x,w) ={}& \cos{(\beta)}\left(G_{v_a,u_a}(x,w) - G_{v_a,u_b}(x,w)\right. \\
   & \left.- G_{v_b,u_a}(x,w) + G_{v_b,u_b}(x,w)\right),
\end{aligned}
\end{equation}
and those expressions are integrable again over the rectangular domain $[v_a,v_b]\times[u_a,u_b]$.

For a screen normal to the free flow $\mathbf{v_0}$, $\beta = \frac{\pi}{2}$ and the tangential velocity is reduced to
\begin{equation}\label{eq-tang-14}
\begin{aligned}
   {v_I}_t &= \Omega\sqrt{\gamma\left(\frac{\pi}{2}\right)}.
\end{aligned}
\end{equation}

\subsection{Value of the shape parameters for a rectangular screen with different aspect ratio}

Starting from the equations (\ref{eq-tang-4}) and (\ref{eq-tang-5}) we can rewrite the expression of $\gamma(\beta)$ and $\alpha(\beta)$ in order to separate the shape terms and the inclination terms, we see for this particular case of a rectangular plate that these expressions can be simplified as
\begin{equation}\label{eq-tang-15}
\begin{aligned}
   \gamma(\beta) &= \gamma_0,
\end{aligned}
\end{equation}
and
\begin{equation}\label{eq-tang-16}
\begin{aligned}
   \alpha(\beta) &= 2\alpha_0\cos{(\beta)},
\end{aligned}
\end{equation}
with
\begin{equation}\label{eq-tang-17}
\begin{aligned}
  \mathcal{J}(x,w) &= -\frac{1}{4\pi}\iint_{\mathcal{S}_p} \frac{w-u}{\left(\left(x-v\right)^2+\left(w-u\right)^2\right)^{\frac{3}{2}}}
  \mathrm{d}u \mathrm{d}v \\
  &= G_{v_a,u_a}(x,w) - G_{v_a,u_b}(x,w)- G_{v_b,u_a}(x,w) + G_{v_b,u_b}(x,w).
\end{aligned}
\end{equation}
\begin{equation}\label{eq-tang-18}
\begin{aligned}
  \gamma_0 &= \frac{1}{S_p}\iint_{\mathcal{S}_p} \mathcal{I}^2_x(x,w) + \mathcal{J}^2(x,w)
  \mathrm{d}x \mathrm{d}w,\\
  \alpha_0 &= \frac{1}{S_p}\iint_{\mathcal{S}_p} \mathcal{J}(x,w)
  \mathrm{d}x \mathrm{d}w.
\end{aligned}
\end{equation}

For symmetry reasons, $\alpha_0 = 0$. Thus the tangential component of the velocity at the surface for a rectangular plate with inclined in the fow is actually
\begin{equation}\label{eq-tang-19}
\begin{aligned}
   {v_I}_t &= \left(\Omega^2 \gamma_0 + v_0^2\cos^2{(\beta)} \right)^{\frac{1}{2}}.
\end{aligned}
\end{equation}
We computed the values of the parameter $\gamma_0$ in the table \ref{tab:shape-param} for different aspect ratio ($v_b = - v_a = \frac{l_1}{2}$ and $u_b = -u_a = \frac{l_2}{2}$).

\renewcommand{\arraystretch}{1.2}
\begin{table}
  \begin{center}
\def~{\hphantom{0}}
  \begin{tabular}{ccccc}
      Aspect ratio & $l_1$ & $l_2$ & $\gamma_0$ \\[3pt]
      $1$ & $0.1$ & $0.1$ & 0.0998 \\
      $2$ & $0.1$ & $0.2$ & 0.0977 \\
      $4$ & $0.1$ & $0.4$ & 0.0934 \\
      $8$ & $0.1$ & $0.8$ & 0.0894 \\
      $10$ & $0.1$ & $1.0$ & 0.0884 \\
      $20$ & $0.1$ & $2.0$ & 0.0861 \\
  \end{tabular}
 \caption{Shape parameter for rectangular plate with different aspect ratio.}
  \label{tab:shape-param}
  \end{center}
\end{table}

\section{}\label{appB-bis}

This appendix contains the details of the calculation of the source strength $\Omega$ for the case of a rectangular screen inclined to the laminar free flow with an angle $\beta$ with respect to the $z$-axis. We start from the following equations.

The pressure difference expressed as
\begin{equation}\label{eq-appB-bis-1}
\begin{aligned}
   {p_{II}}-{p_{I}} &={} \frac{1}{2}\rho \left({v_I}_t^2\left(1-E^2\right)+{v_I}_n^2\theta(s)f(Re_n,\beta)\right),
\end{aligned}
\end{equation}
\begin{equation}\label{eq-appB-bis-2}
\begin{aligned}
   {p_{III}}-{p_{0}} &={} \frac{1}{2}\rho\left(\left(1-E^2\right)v^2_0 + {v_I}_n^2\theta(s)f(Re_n,\beta)\right).
\end{aligned}
\end{equation}

And the drag forces are expressed as
\begin{equation}\label{eq-appB-bis-3}
\begin{aligned}
    F_D & ={} \rho v_0\left(1-E\right){v_n} {S_p} + \frac{1}{v_0}\left(p_0 - p_{III}\right)\frac{{v_n}}{E} {S_p},
\end{aligned}
\end{equation}
\begin{equation}\label{eq-appB-bis-4}
   F_D ={} \left({p_I}-{p_{II}}\right)\sin{(\beta)}{S_p} + \rho {v_n} v_0\cos^2(\beta)\left(1-E\right){S_p}.
\end{equation}

In these expressions the normal and tangential components of the velocity at the surface of the screen are
\begin{equation}\label{eq-appB-bis-5}
\begin{aligned}
    {v_I}_n ={} & \sin{(\beta)}v_0 -  \frac{1}{2}\Omega,
\end{aligned}
\end{equation}
\begin{equation}\label{eq-appB-bis-6}
\begin{aligned}
   {v_I}_t &= \left(\Omega^2 \gamma_0 + v_0^2\cos^2{(\beta)} \right)^{\frac{1}{2}}.
\end{aligned}
\end{equation}

Then, by denoting $\omega = \frac{\Omega}{v_0}$ we have for the velocities
\begin{equation}\label{eq-appB-bis-7}
\begin{aligned}
    \frac{{v_I}_n}{v_0} ={} & \sin{(\beta)} -  \frac{1}{2}\omega,
\end{aligned}
\end{equation}
\begin{equation}\label{eq-appB-bis-8}
\begin{aligned}
   \frac{{v_I}_t}{v_0} &= \left(\omega^2 \gamma_0 + \cos^2{(\beta)} \right)^{\frac{1}{2}}.
\end{aligned}
\end{equation}

The attenuation coefficient $E$ is
\begin{equation}\label{eq-appB-bis-9}
\begin{aligned}
   E &= \frac{\sin{(\beta)} -  \frac{1}{2}\omega}{\sin{(\beta)} +  \frac{1}{2}\omega}.
\end{aligned}
\end{equation}

And for the pressure differences we obtain
\begin{equation}\label{eq-appB-bis-10}
\begin{aligned}
   \frac{{p_{II}}-{p_{I}}}{\frac{1}{2}\rho v_0^2} &={} \left(\omega^2 \gamma_0 +  \cos^2{(\beta)} \right)\frac{2\omega\sin{(\beta)}}{\left(\sin{(\beta)} +  \frac{1}{2}\omega\right)^2}+\left(\sin{(\beta)} -  \frac{1}{2}\omega\right)^2\theta(s)f(Re_n,\beta),
\end{aligned}
\end{equation}
\begin{equation}\label{eq-appB-bis-11}
\begin{aligned}
   \frac{{p_{III}}-{p_{0}}}{\frac{1}{2}\rho v_0^2} &={} \frac{2x\sin{(\beta)}}{\left(\sin{(\beta)} +  \frac{1}{2}x\right)^2}+\left(\sin{(\beta)} -  \frac{1}{2}x\right)^2\theta(s)f(Re,\beta).
\end{aligned}
\end{equation}

The first expression of the drag force is
\begin{equation}\label{eq-appB-bis-12}
   \frac{F_D}{\frac{1}{2}\rho v_0^2 S_p} ={} \frac{2\omega\left(\sin{(\beta)} - \frac{1}{2}\omega\right)}{\sin{(\beta)} + \frac{1}{2}\omega} + \left(\sin{(\beta)} + \frac{1}{2}\omega\right)\frac{{p_{0}}-{p_{III}}}{\frac{1}{2}\rho v_0^2}.
\end{equation}

The second expression of the drag force is
\begin{equation}\label{eq-appB-bis-13}
   \frac{F_D}{\frac{1}{2}\rho v_0^2 S_p} ={} \frac{{p_{I}}-{p_{II}}}{\frac{1}{2}\rho v_0^2}\sin{(\beta)} + \cos^2(\beta)\frac{2\omega\left(\sin{(\beta)} - \frac{1}{2}\omega\right)}{\sin{(\beta)}+ \frac{1}{2}\omega}.
\end{equation}

The two expressions of the drag coefficient are
\begin{equation}\label{eq-appB-bis-14}
\begin{aligned}
  C_D ={} \frac{-\omega^2}{\sin{(\beta)} + \frac{1}{2}\omega} - \left(\sin{(\beta)} + \frac{1}{2}x\right)\left(\sin{(\beta)} -  \frac{1}{2}x\right)^2\theta(s)f(Re_n,\beta),
\end{aligned}
\end{equation}
\begin{equation}\label{eq-appB-bis-15}
\begin{aligned}
  C_D ={}&  \cos^2(\beta)\frac{2x\left(\sin{(\beta)} - \frac{1}{2}\omega\right)}{\sin{(\beta)}+ \frac{1}{2}x} - \sin{(\beta)}\left(\sin{(\beta)} -  \frac{1}{2}\omega\right)^2\theta(s)f(Re_n,\beta) \\
  & - \left(\omega^2 \gamma_0 + \cos^2{(\beta)} \right)\frac{2\omega\sin^2{(\beta)}}{\left(\sin{(\beta)} +  \frac{1}{2}\omega\right)^2}.
\end{aligned}
\end{equation}

By combining the expression (\ref{eq-appB-bis-14}) and (\ref{eq-appB-bis-15}) we obtain the equation
\begin{equation}\label{eq-appB-bis-16}
\begin{gathered}
  -\frac{1}{8} \omega^4\theta(s)f(Re_n,\beta) + \omega^2 \sin^2{(\beta)}\Big(8\gamma_0 + \theta(s)f(Re_n,\beta) - 2\Big) - \\
  4\omega\sin{(\beta)} - 2\sin^4{(\beta)}\theta(s)f(Re_n,\beta) = 0.
\end{gathered}
\end{equation}

Note that for a rectangular screen normal to the free flow this equation is reduced to:
\begin{equation}\label{eq-appB-bis-17}
\begin{aligned}
  -\frac{1}{8} \omega^4\theta(s)f(Re_n,\beta) + \omega^2\left(8\gamma_0 + \theta(s)f(Re_n,\beta) - 2\right) - 4\omega - 2\theta(s)f(Re_n,\beta) = 0.
\end{aligned}
\end{equation}

If the length $L$ (or the width $D$) of the rectangular screen is infinitely long then we can show that:
\begin{equation}\label{eq-appB-bis-17}
\begin{aligned}
  \lim_{L\to\infty} \gamma_0 = \frac{1}{12},
\end{aligned}
\end{equation}
and we tend to the two-dimensional case studied by \cite{Steiros}.

\newpage

\section{}\label{appC}
This appendix contains the details of the experimental data and their processing.

\subsection{Screen samples}

\begin{figure}
  \centerline{\includegraphics[width=9cm, trim = 0cm 0cm 0cm 0cm, clip]{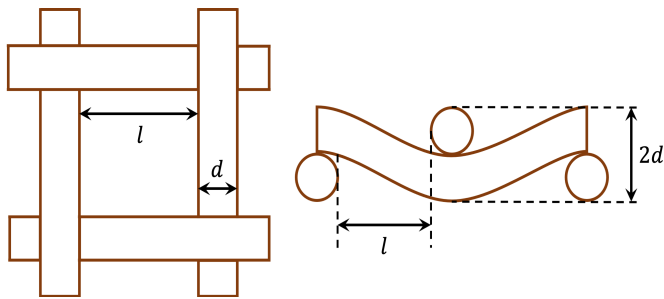}}
  \caption{Diagram of the fiber mesh.}
\label{fig:mesh}
\end{figure}

The characteristics of the different porous screens used for the measurement are presented in the table \ref{tab:screens}. They are mainly square wire mesh screens, as represented in figure \ref{fig:mesh}. In addition, $5$ other types of screen were used to test the robustness of the model.

\renewcommand{\arraystretch}{1.2}
\begin{table}
  \begin{center}
\def~{\hphantom{0}}
  \begin{tabular}{cccccc}
      \makecell{Screen \\ number}  & \makecell{Fiber diameter \\ $d$ (mm)} & \makecell{Reynolds \\ number \\ $Re_d$} & Solidity $s$ & Note \\[3pt]
       P1 & $0.29 \pm 0.01$ & $186$ & $0.58 \pm 0.01$ & \makecell{Square woven mesh, nylon fibers} \\
       P2 & $1.1 \pm 0.5$ & $705$ & $0.41 \pm 0.05$ & \makecell{Regular net} \\
       P3 & $1.9 \pm 1.0$ & $1218$ & $0.87 \pm 0.05$ & \makecell{Regular net} \\
       P4 & $0.26 \pm 0.01$ & $167$ & $0.61 \pm 0.01$ & \makecell{Square woven mesh, nylon fibers} \\
       P5 & $0.10 \pm 0.01$ & $64$ & $0.56 \pm 0.02$ & Square woven mesh, nylon fibers \\
       P6 & $0.18 \pm 0.02$ & $115$ & $0.61 \pm 0.02$ & Square woven mesh, metal fibers  \\
       P7 & $0.27 \pm 0.01$ & $173$ & $0.45 \pm 0.01$ & \makecell{Square woven mesh, nylon fibers} \\
       P8 & $0.13 \pm 0.01$ & $83$ & $0.70 \pm 0.01$ & \makecell{Square woven mesh, nylon fibers} \\
       P9 & \makecell{$0.0009 - 0.019$ \\ mean $0.006 \pm 0.003$} & \makecell{$0.6 - 12$ \\ $4$} & $0.26 \pm 0.03$ & \makecell{Surgical facemask, \\ physical characteristics according \\ to \cite{Monjezi} \\ and \cite{Du}}  \\
       P10 & $0.27 \pm 0.01$ & $173$ & $0.11 \pm 0.01$ & \makecell{Square woven mesh, nylon fibers} \\
       P11 & $0.27 \pm 0.01$ & $173$ & $0.37 \pm 0.01$ & \makecell{Square woven mesh, nylon fibers}\\
       P12 & $0.27 \pm 0.01$ & $173$ & $0.31 \pm 0.01$ & \makecell{Square woven mesh, nylon fibers} \\
       P13 & $0.27 \pm 0.01$ & $173$ & $0.17 \pm 0.01$ & \makecell{Square woven mesh, nylon fibers} \\
       P14 & $0.27 \pm 0.01$ & $173$ & $0.24 \pm 0.01$ & \makecell{Square woven mesh, nylon fibers} \\
       P15 & $0.27 \pm 0.01$ & $173$ & $0.24 \pm 0.01$ & \makecell{Square woven mesh, nylon fibers} \\
       P16 & $0.27 \pm 0.01$ & $173$ & $0.24 \pm 0.01$ & \makecell{Square woven mesh, nylon fibers} \\
       P17 & $0.27 \pm 0.01$ & $173$ &  $0.15 \pm 0.01$ & \makecell{Square woven mesh, nylon fibers} \\
       P18 & $0.27 \pm 0.01$ & $173$ & $0.28 \pm 0.01$ & \makecell{Square woven mesh, nylon fibers} \\
       P19 & $0.26 \pm 0.01$ & $167$ & $0.52 \pm 0.01$ & \makecell{Square woven mesh, nylon fibers} \\
       P20 & $0.26 \pm 0.01$ & $167$ & $0.42 \pm 0.01$ & \makecell{Square woven mesh, nylon fibers} \\
       P21 & $0.26 \pm 0.01$ & $167$ & $0.32 \pm 0.01$ & \makecell{Square woven mesh, nylon fibers} \\
       P22 & $0.050 \pm 0.002$ & $32$ & $0.65$ & \makecell{Square woven mesh, homogeneous, \\ polyamide fibers} \\
       P23 & $0.025 \pm 0.002$ & $16$ & $0.82$ & \makecell{Square woven mesh, homogeneous, \\ polyamide fibers} \\
       P24 & $0.030 \pm 0.002$ & $19$ & $0.75$ & \makecell{Square woven mesh, homogeneous, \\ polyamide fibers} \\
       P25 & $0.037 \pm 0.002$ & $24$ & $0.70$ & \makecell{Square woven mesh, homogeneous, \\ polyamide fibers} \\
       P26 & $0.44 \pm 0.01$ & $282$ & $0.405 \pm 0.002$ & Parallel nylon fibers \\
       P27 & $0.12 \pm 0.01$ & $38$ & $0.115 \pm 0.002$ & Parallel nylon fibers \\
       P28 & $0.44 \pm 0.01$ & $141$ & $0.114 \pm 0.002$ & Parallel nylon fibers \\
       P29 & $0.02 \pm 0.002$ & $6$ & $0.080 \pm 0.002$ & Parallel copper fibers \\
       P30 & $0.44 \pm 0.01$ & $282$ & $0.080 \pm 0.001$ & Parallel nylon fibers \\
       P31 & - & - & $1.00$ & Flat plate, thickness $0.1\hspace{0.1cm}mm$ \\
  \end{tabular}
 \caption{Porous screen characteristics. The Reynolds number $Re_d$ is calculated with a velocity $v_0 = 10 \hspace{0.1cm}m.s^{-1}$ and a kinematic viscosity $\nu = 15.6 \times 10^{-6}\hspace{0.1cm}m^2.s^{-1}$, except for the screens P27, P28 and P29 where the velocity is $v_0 = 5 \hspace{0.1cm}m.s^{-1}$.}
  \label{tab:screens}
  \end{center}
\end{table}

\subsection{Correction of the coupling drag}
Due to the mast and frame supporting the porous structure, to obtain the drag coefficient of the porous structure from the raw data composed of the measured forces denoted $F_{t+m}$, the contributions of each part must be decoupled. 

In our analysis of the data, we consider the most simple assumption that the coupling is negligible. Therefore the drag force of the porous screen is the drag force of the total system (mast, frame and porous screen) minus the drag force of the mast and the frame (without the porous screen, measured before the series of measurements). In order to justify this assumption, we estimated the coupling drag force (the additional term due to the interference between the frame and the screen) for what we assumed to be the worst case, that is for the solidity equal to $1$.

\begin{figure}
  \centerline{\includegraphics[width=11cm, trim = 0cm 0cm 0cm 0cm, clip]{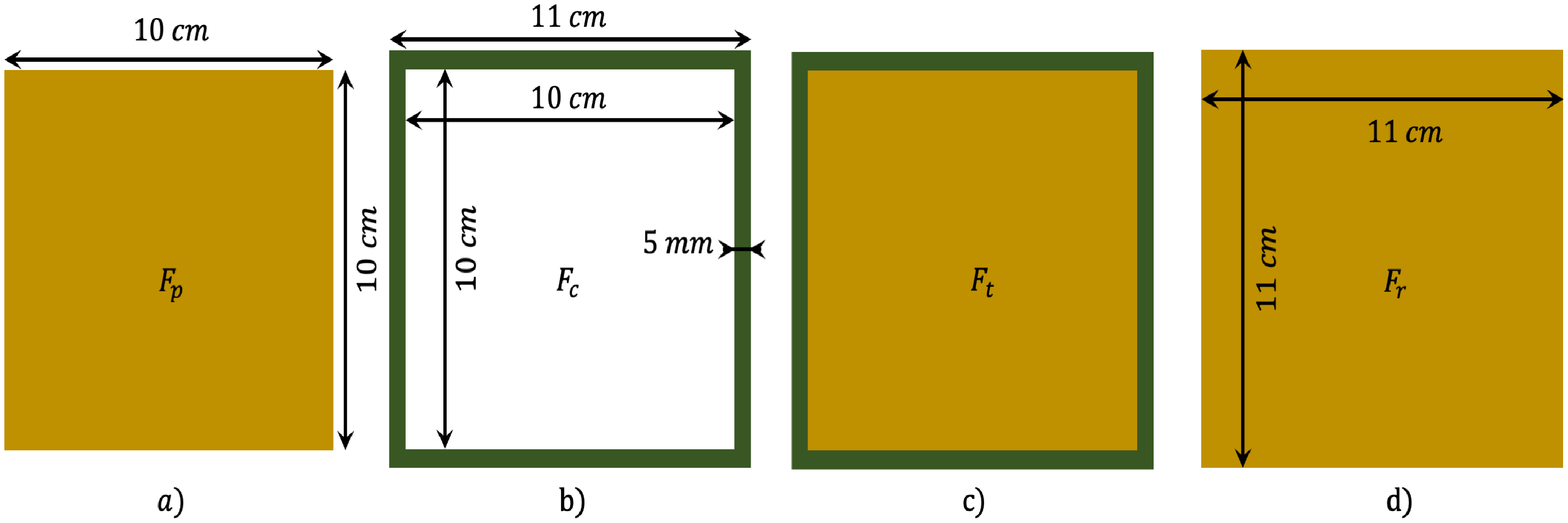}}
  \caption{Diagram of the screens (in dark gold) and the frame support (in green).}
\label{fig:frame}
\end{figure}

As far as we know, this coupling is non-linear and there is no general method. We adopt the approach we detail here, based on different measurements with and without the frame illustrated in figure \ref{fig:frame}. Due to the elongated shape of the mast and the way it is connected to the screen, it is reasonable to assume that the drag force of the mast $F_m$ and the rest of the system $F_t$ add up (giving what we denote $F_{t+m}$). For the frame, the coupling with the porous structure is expected to be more important. We measured the drag force of the mast alone $F_m$ and subtracted the value to $F_{t+m}$. We measured the drag force of the mast with the frame $F_{m+c}$ so that we obtain the drag force of the frame $F_c = F_{m+c} - F_m$. The drag force of the porous screen is denoted $F_p$ and the coupling term is denoted $\Delta F$. The total force measured of the frame and the porous screen $F_t$ can be written down
\begin{equation}\label{eq-corr-1}
   F_t = F_c + F_p + \Delta F.
\end{equation}

To determine the coupling (or interference) term $\Delta F$, we know exactly the value of the drag force of the porous structure for two points: the first at solidity $s=0$ where $F_p= 0$, and the second at solidity $s=1$ for which we can determine easily the drag force without a frame since we can use a solid plate with the same surface area and thickness.

Note that the value of the coupling is not independent on the solidity: indeed, if we assume a monotone dependency, then the lower the solidity, the lower the coupling will be, until it reaches a zero value at zero solidity. These two points allow us to estimate the coupling.

The frame has a thickness of $0.2\hspace{0.1cm}mm$ each side of the system. Therefore, to quantify the influence of this extrusion on the drag force, we measured for the solidity $s=1$ the drag force $F_r$ of a plate (diagram d) in figure \ref{fig:frame}-d) and the drag force $F_t$ of the system composed of the frame and the porous screen in figure \ref{fig:frame}-c), with the same width at the border.

\renewcommand{\arraystretch}{1.2}
\begin{table}
  \begin{center}
\def~{\hphantom{0}}
  \begin{tabular}{cccccccc}
      Experience & $v_0\hspace{0.1cm}(m.s^{-1})$  & $F_p\hspace{0.1cm}(N)$ & $F_c\hspace{0.1cm}(N)$ & $F_t\hspace{0.1cm}(N)$ & $F_r\hspace{0.1cm}(N)$ & \makecell{estimation of \\ $\Delta F\hspace{0.1cm}(N)$} & $F_t - F_c\hspace{0.1cm}(N)$\\[3pt]
      \hline
       \multirow{12}{*}{Screen at $90\degree$} & 0.51 & 0.002 & 0.002 & 0.004 & 0.003 & 0.000 & 0.002\\
       & 1.02 & 0.008 & 0.003 & 0.011 & 0.010 & 0.000 & 0.008\\
       & 4.00 & 0.101 & 0.022 & 0.118 & 0.122 & -0.005 & 0.096\\
       & 5.03 & 0.159 & 0.036 & 0.185 & 0.191 & -0.010 & 0.149\\
       & 5.98 & 0.221 & 0.052 & 0.260 & 0.272 & -0.013 & 0.208\\
       & 7.00 & 0.301 & 0.072 & 0.350 & 0.370 & -0.023 & 0.278\\
       & 8.02 & 0.398 & 0.095 & 0.467 & 0.490 & -0.026 & 0.372\\
       & 8.97 & 0.495 & 0.118 & 0.588 & 0.606 & -0.025 & 0.470\\
       & 9.99 & 0.620 & 0.144 & 0.710 & 0.750 & -0.054 & 0.566\\
       & 11.01 & 0.753 & 0.173 & 0.862 & 0.919 & -0.064 & 0.689\\
       & 12.03 & 0.897 & 0.208 & 1.024 & 1.092 & -0.081 & 0.816\\
       & 12.98 & 1.042 & 0.240 & 1.178 & 1.262 & -0.104 & 0.938\\
       \hline
       Drag coefficient $C_D$ & - & 0.993 & - & 0.973 & 0.986 & - & 0.939\\
       \hline
  \end{tabular}
 \caption{Summary of the values of the coupling drag force for a flat plate (solidity $s=1$) for different velocities. All the values have an uncertainty of approximately $0.020 \hspace{0.1cm}N$. For the measurement of $F_p$ and $F_r$ a solid flat plate with $4.0\hspace{0.1cm}mm$ thickness has been used, corresponding to the thickness of the frame used for measuring $F_c$ and $F_t$. The drag coefficients are obtained using the dimensions shown in figure \ref{fig:frame} with fluid density estimated from the measurement of temperature, pressure and humidity during the different experiments.}
  \label{tab:coupling}
  \end{center}
\end{table}

\begin{figure}
    \centering
    \includegraphics[width=8cm, trim = 0cm 0cm 0cm 0cm, clip]{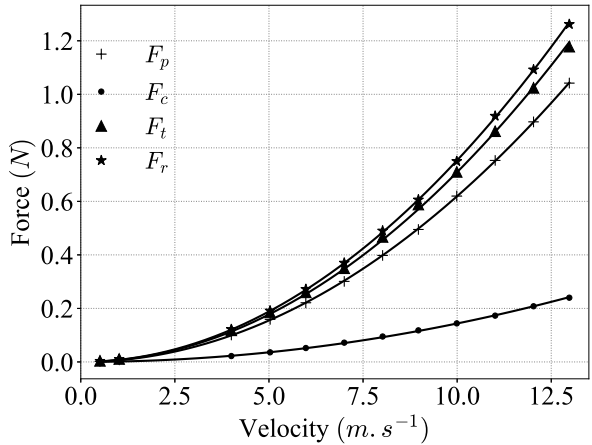}
    \caption{Drag force for the plates and the frame represented in figure \ref{fig:frame}. Values and drag coefficients are in table \ref{tab:coupling}. Data are fitted using a quadratic law with respect to the velocity $v_0$. The surface is orthogonal to the mean far field flow direction $\bm{v}_0$.}
    \label{fig:force-app-C}
\end{figure}

We notice that correcting our data with a constant coupling (interference) drag term would not change the curve shape. Doing so with a linear coupling will have a minor effect. Indeed, if the use of a frame to stretch the textiles seems to underestimate the drag coefficient, the difference for the worst case in the drag coefficient between the screen measured directly ($0.993$) and the screen measured with the frame after subtraction of its drag force ($0.939$) is closed to the order of the uncertainty calculated in the next section. Even if we added the coupling drag force estimated for solidity $1$ in all our data, this would not change our conclusions.
Therefore, it is reasonable to neglect the interference drag, and simply proceed with the subtraction of the frame drag force from the total drag force.

\subsection{Determination of the drag coefficient}
To determine the drag coefficient we calculate a non-linear regression of the corrected data with the method of the least squares. The model function is
\begin{equation}\label{eq-corr-3}
   {F_p}_i = f(v_i,C_D) = \frac{1}{2}\rho S_p C_D v_i^2,
\end{equation}
where $C_D$ is the adjustable parameter. We minimize the sum of the square residuals $S$
\begin{equation}\label{eq-corr-4}
   S = \sum_{i = 1}^n \left(y_i - f(v_i,C_D)\right)^2.
\end{equation}
This leads to
\begin{equation}\label{eq-corr-5}
   C_D = \frac{2}{\rho S}\frac{\displaystyle\sum_{i=1}^n {F_p}_i v_i^2}{\displaystyle\sum_{i=1}^n v_i^4}.
\end{equation}

\subsection{Measurement uncertainty}

In what follows, we estimate the measurement uncertainty $u$ of the physical quantities. We assume for simplicity that the parameters $\rho$, ${F_p}_i$, $v_i$ and $S$, for $i \in   \llbracket 1,n \rrbracket $ are mutually independent, and that their respective uncertainty is small compared to their value. We neglect the uncertainty on the velocity, then the uncertainty can be calculated with
\begin{equation}\label{eq-corr-6}
   u^2(C_D) = \frac{4}{\rho^2 S^2 \left(\displaystyle\sum_{i=1}^n v_i^4\right)^2}\left(\left(\displaystyle\sum_{i=1}^n {F_p}_i v_i^2\right)^2\left(\frac{u^2(\rho)}{\rho^2}+\frac{u^2(S)}{S^2}\right)+\displaystyle\sum_{i=1}^n v_i^4 u^2({F_p}_i)\right).
\end{equation}

We assume that the force uncertainty is the same for all the data, this assumption is supported by the different repeated measurements we performed for several porous screens, we take the mean value of the deviation we obtained for the force uncertainty.

\subsubsection{Solidity uncertainty}

The solidity of the porous screens with regular nylon woven mesh are determined using image of the screen at the scale of a hundred meshes, taken with a microscope. Several image analysis are used to estimate the solidity and the associated uncertainty. The uncertainty is estimated to $\pm 0.02$.

\subsubsection{Parameter and drag force uncertainties}

The uncertainty of the drag force arises from the error of the force balance. By repeating the measurement several times, for different porous screens, we estimated the drag force uncertainty to approximately $\pm 0.02 N$. The uncertainty of the air density is $\pm 0.005 \hspace{0.1cm} kg.m^{-3}$. Finally, the uncertainty on the surface of the screen is estimated to $\pm 4\times 10^{-6}\hspace{0.1cm}m^2$.

\bibliographystyle{jfm}
\bibliography{biblioscreen}

\end{document}